\documentclass{article}

\usepackage{natbib}
\usepackage{epubtk}
\usepackage{booktabs}
\usepackage{graphicx}

\begin{document}

\title{Coronal Loops: Observations and Modeling of Confined Plasma}

\author{\epubtkAuthorData{Fabio Reale}{%
Dipartimento di Scienze Fisiche \& Astronomiche,\\
Universit\`a di Palermo, Sezione di Astronomia,\\
Piazza Parlamento 1, 90134 Palermo, Italy}{%
reale@astropa.unipa.it}{%
http://www.astropa.unipa.it/~reale/}
}

\date{}
\maketitle

\begin{abstract}
Coronal loops are the building blocks of the X-ray bright solar corona. They owe their brightness to the dense confined plasma, and this review focuses on loops mostly as structures confining plasma. After a brief  historical overview, the review is divided into two separate but not independent sections: the first illustrates the observational framework, the second reviews the theoretical knowledge. Quiescent loops and their confined plasma are considered, and therefore topics such as loop oscillations and flaring loops (except for non-solar ones which provide information on stellar loops) are not specifically addressed here. 
The observational section discusses loop classification and populations, and then describes the morphology of coronal loops, its relationship with the magnetic field, and the concept of loops as multi-stranded structures. The following part of this section is devoted to the characteristics of the loop plasma, and of its thermal structure in particular, according to the classification into hot, warm and cool loops. Then, temporal analyses of loops and the observations of plasma dynamics and flows are illustrated. In the modeling section starts  some basics of loop physics are provided, supplying some fundamental scaling laws and timescales,  a useful tool for consultation. The concept of loop modeling is introduced, and  models are distinguished between those treating loops as monolithic and static, and those resolving loops into thin and dynamic strands. Then more specific discussions address modeling the loop fine structure, and the plasma flowing along the loops. Special attention is devoted to the question of loop heating, with separate discussion of wave (AC) and impulsive (DC) heating. Finally, a brief discussion about stellar X-ray emitting structures related to coronal loops is included and followed by conclusions and open questions.
\end{abstract}

\epubtkKeywords{The Sun}

\newpage


\section{Introduction}
\label{section:introduction}

The corona is the outer part of the solar atmosphere. Its name derives from the fact that, since it is extremely tenuous with respect to the lower atmosphere, it is visible in the optical band only during the solar eclipses as a faint crown (corona in Latin) around the black moon disk. When inspected through spectroscopy the corona reveals unexpected emission lines, which were first identified as due to a new element (coronium) but which were later ascertained to be due to high excitation states of iron \citep[e.g.,][]{1997soco.book.....G,2001nsss.book.....G}. It became then clear that the corona is made of very high temperature gas, hotter than 1~MK. Almost all the gas is fully ionized there and thus interacts effectively with the ambient magnetic field. It is for this reason that the corona appears so inhomogeneous when observed in the X-ray band, in which plasma at million degrees emits most of its radiation. In particular, the plasma is confined inside magnetic flux tubes which are anchored on both sides to the underlying photosphere. When the confined plasma is heated more than the surroundings, its pressure and density increase. Since the tenuous plasma is optically thin, the intensity of its radiation is proportional to the square of the density, and the tube becomes much brighter than the surrounding ones and looks like a bright closed arch: a coronal loop.

When observed in the X-ray band, the bright corona appears to be made entirely by coronal loops that can therefore be considered as the building blocks of X-ray bright corona. This review specifically addresses coronal loops as bright structures confining plasma. It first provides an observational framework that is the basis for the second part dealing with modeling and interpretation. There have been several earlier books \citep{1991plsc.book.....B,1997soco.book.....G,2004psci.book.....A} and reviews \citep{1978ARAA..16..393V,1990MmSAI..61..401P,1996Ap&SS.237...33G,2001ARAA..39..175A,2005ESASP.600E..27R}, in particular on coronal heating \citep{1993SoPh..148...43Z,1995itsa.conf...17C,2006SoPh..234...41K}, that have in general a larger or different scope but include information about coronal loops.
Interested readers are urged to survey these other reviews in order to complement and fill in any gaps in topical coverage of the present paper.

\newpage


\section{Historical Keynotes}
\label{section:histor}

First evidence of magnetic confinement came from rocket missions in the 1960s. In particular, in 1965, arcmin angular resolution was achieved with grazing incidence optics \citep{1965ApJ...142.1274G}. 
The data analysis led to the first density and temperature diagnostics with wide band filters, to derive high pressure in compact regions with intense bipolar magnetic fields and to propose the magnetic confinement \citep{1968ApJ...151..333R}. 
The first coronal loop structures were identified properly after a rocket launch in 1968 which provided for the first time an image of an X-ray flare \citep{1968Sci...161..564V}, 
with a resolution of a few arcsec. 

In the course of collecting the results of all rocket missions of the American Science and Engineering (AS\&E) program, \cite{1973SoPh...32...81V} 
proposed a classification of the morphology of the X-ray corona as fundamentally consisting of arch-like structures connecting regions of opposite magnetic polarity in the photosphere.  The classification was based on the loop size, on the physical conditions of the confined plasma, on the underlying photospheric regions. They distinguished active regions, coronal holes, active regions interconnection, filament cavities, bright points, and large-scale structures \citep{1978ARAA..16..393V,1990MmSAI..61..401P}. 

The magnetic structuring of the solar corona is evident. However, the magnetic field lines can be traced only indirectly because direct measurements are feasible generally only low in the photosphere through the Zeeman effect on spectral lines. It is anyhow possible to extrapolate the magnetic field in a volume. This was done to derive the magnetic field structure of a relatively stable active region by \cite{1975SoPh...44...83P}
using the \cite{1964NASSP..50..107S}
method, under the assumption of negligible currents in the corona. This was useful also to derive magnetic field intensities sufficient for hot plasma confinement. Later on, even more reliable magnetic field topologies were derived assuming force-free fields \citep[e.g.,][]{1981SoPh...69..343S},
i.e., with currents everywhere parallel to the magnetic field as it is expected in coronal loops.

The rocket missions lacked good time coverage and the information about the evolution of coronal loops was only limited, mostly available from the Orbiting Solar Observatory-IV (OSO-IV) mission \citep{1972SoPh...22..150K}.
This satellite had an angular resolution in the order of the arcmin and could not resolve individual loops. In 1973 the X-ray telescope S-054 on-board Skylab monitored the evolution of coronal loops for several months, taking 32,000 X-ray photographs with a maximum resolution of 2~arcsec  and an extended dynamic range. It was  possible to study the whole evolution of an active region, from the emergence as compact loops filled with dense plasma to its late spreading, a few solar rotations later, as progressively longer and longer loops filled with less and less dense plasma \citep{1982ApJ...259..359G}.
It was confirmed that the whole X-ray bright corona consists of magnetic loops, whose lifetime is typically much longer than the characteristic cooling times \citep{1978ApJ...220..643R}. No exception is made by the coronal holes where the magnetic field opens radially to the interplanetary space and the plasma streams outwards with practically no X-ray emission.

In the same mission coronal loops were detected also in the UV band at temperatures below 1~MK, by Extreme UltraViolet (EUV) telescopes S-055 \citep{1977ApOpt..16..837R}
and S-082 \citep{1977ApOpt..16..870T,1977ApOpt..16..879B}.
These loops are invisible in the X-ray band and many of them depart from sunspots, appear coaxial and are progressively thinner for progressively lower temperature ions \citep{1975SoPh...43..327F,1976ApJ...210..575F},
suggesting nested loops hotter toward outer shells, although this suggestion has never been validated. The apparent scale height of the emission is larger than that expected from a static model, but the loops appear to be steady for long times. \cite{1976ApJ...210..575F}
proposed a few explanations including siphon flows and thermal instability of the plasma at the loop apex. New observations of such cool loops were performed several years later with the SOlar and Heliospheric Observatory (SoHO) mission and provided new details and confirmations (Section~\ref{sec:obs_flo}).

A different target was addressed by the Solar Maximum Mission \citep[SMM, 1980\,--\,1989,][]{1980SoPh...65....5B,1980SoPh...65...53A},
which included high-resolution spectrometers in several X-ray lines, i.e.\ the Bent Crystal Spectrometer (BCS) and the Flat Crystal Spectrometer (FCS), mostly devoted to obtain time-resolved spectroscopy of coronal flares \citep[e.g.,][]{1985SoPh...99..167M}.
Similarly, the Hinotori mission \citep[1981\,--\,1991,][]{1983SoPh...86....3T}
was dedicated mainly to solar flare observations in the X-ray band. This was also the scope of the later Yohkoh mission,
\citep[1991\,--\,2001,][]{1991SoPh..136....1O}
by means of high resolution X-ray  spectroscopy, adding the monitoring
and imaging of the hot and flaring corona. \cite{1992PASJ...44L.135H}
found first indications of plasma at 5\,--\,6~MK in active regions with the Soft X-ray Telescope \citep[SXT,][]{1991SoPh..136...37T}.

Normal incidence optics have been developed in the late 1980s. An early experiment was the Normal Incidence X-ray Telescope \citep[NIXT,][]{1989SPIE.1160..629G},
which provided a few high resolution coronal images in the EUV band.

More recent space missions dedicated to study the corona have been 
the Solar and Heliospheric Observatory \citep[SoHO,][]{1995SoPh..162....1D}, launched in
1995 and still operative,
and the Transition
Region and Coronal Explorer \citep[TRACE,][]{1999SoPh..187..229H},
launched in 1998 and replaced in 2010 by the Solar Dynamic Observatory instruments. Both SoHO and TRACE were
tailored to observe the quiet corona (below 2~MK). SoHO images the whole
corona \citep[Extreme ultraviolet Imaging Telescope, EIT,][]{1995SoPh..162..291D}
and performs wide band spectroscopy \citep[Solar Ultraviolet Measurements of Emitted Radiation, SUMER,][]{1995SoPh..162..189W} and \citep[Coronal Diagnostic Spectrometer, CDS,][]{1995SoPh..162..233H}
in the EUV band; TRACE imaged
the EUV corona with high spatial (0.5~arcsec) and temporal (30~s) resolution. Both SoHO/EIT and TRACE are based on normal incidence optics and contain three different EUV filters that provide limited thermal diagnostics.

Thanks to their capabilities, both missions allowed to address finer diagnostics, in particular to investigate the fine transverse structuring of coronal loops, both in its geometric and thermal components, and the plasma dynamics and the heating mechanisms at a higher
level of detail. SoHO and TRACE have been complementary in many respects and several studies attempted to couple the information from them.

Among other relevant missions, we mention the CORONAS series \citep{1998SPIE.3406...20I,2002russian}, with instruments like SPectroheliographIc X-Ray Imaging Telescope \citep[SPIRIT,][]{2003MNRAS.338...67Z}, Rentgenovsky Spektrometr s Izognutymi Kristalami \citep[ReSIK,][]{1998ESASP.417..313S} and Solar Photometer in X-rays \citep[SPHINX,][]{2008JApA...29..339S} which have contributed to the investigation of coronal loops.

In late 2006, two other major solar missions started, namely Hinode \citep{2007SoPh..243....3K} and the Solar TErrestrial Relations Observatory (STEREO) \citep[e.g.,][]{2008SSRv..136....5K}.
On-board Hinode, two instruments address particularly the study of coronal loops: the X-Ray Telescope \citep[XRT,][]{2007SoPh..243...63G} and the Extreme-ultraviolet Imaging Spectrometer \citep[EIS,][]{2007PASJ...59S.751C}).
Both these instruments offer considerable improvements on previous missions. The XRT has a spatial resolution of about 1 arcsec, a very low scattering and the possibility to switch among nine filters and combinations of them. EIS combines well spectral ($\sim 2$~mA), spatial (2") and temporal ($\sim 10$ s) resolution to obtain accurate diagnostics of plasma dynamics and density. One big achievement of the STEREO mission is that, since it consists of two separate spacecrafts getting farther and farther from each other, it allows -- through for instance its Sun-Earth Connection Coronal and Heliospheric Investigation (SECCHI) package -- a first 3D reconstruction of coronal loops \citep{2009ApJ...695...12A,2009SoPh..259..109K}.

Hinode and STEREO represent the state-of-art for coronal loop observations, and the coronal loop studies including recent findings of these missions will be overviewed. In 2010, the new big solar mission Solar Dynamic Observatory (SDO) has been launched and new interesting results about the solar corona are expected soon, with instruments like the Atmospheric Imaging Assembly (AIA).

\newpage


\section{The Observational Framework}
\label{sec:obs}

\subsection{General properties}
\label{sec:obs_gen}

Although coronal loops are often well defined and studied in the EUV band, detected by many space mission spectrometers like those on board SoHO and Hinode, and by high resolution imagers such as NIXT and TRACE, the bulk of coronal loops is visible in the X-ray band (Figure~\ref{fig:x_uv}). Also, the peak of the coronal emission measure of active regions -- where the loops are brightest -- is above 2~MK, which is best observed in X-rays \citep[e.g.,][]{2000ApJ...528..537P,2009ApJ...704L..58R}.

\epubtkImage{x_vs_uv.png}{%
\begin{figure}[htbp]
  \centerline{\includegraphics[scale=0.5]{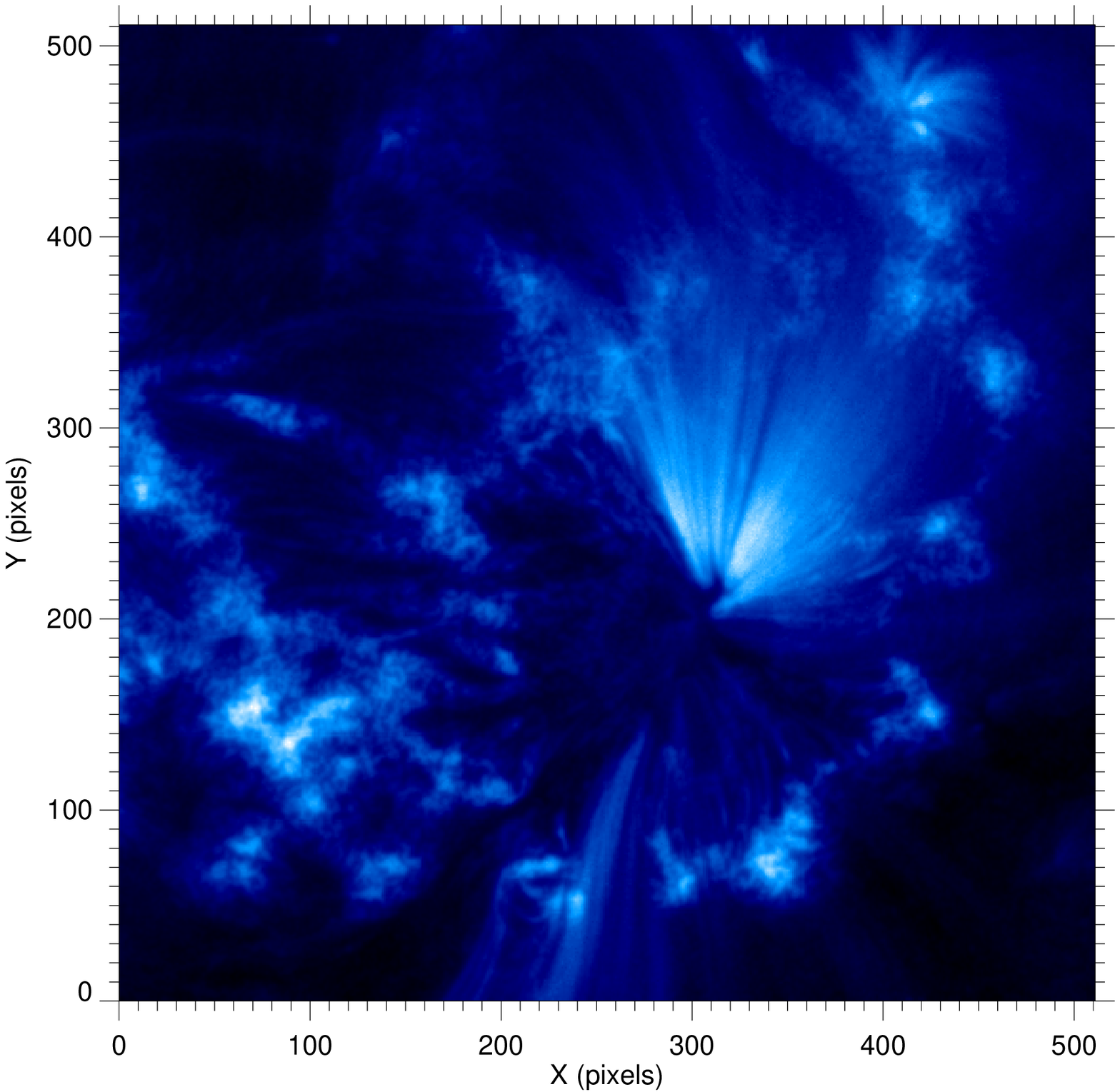}}
  \centerline{\includegraphics[scale=0.5]{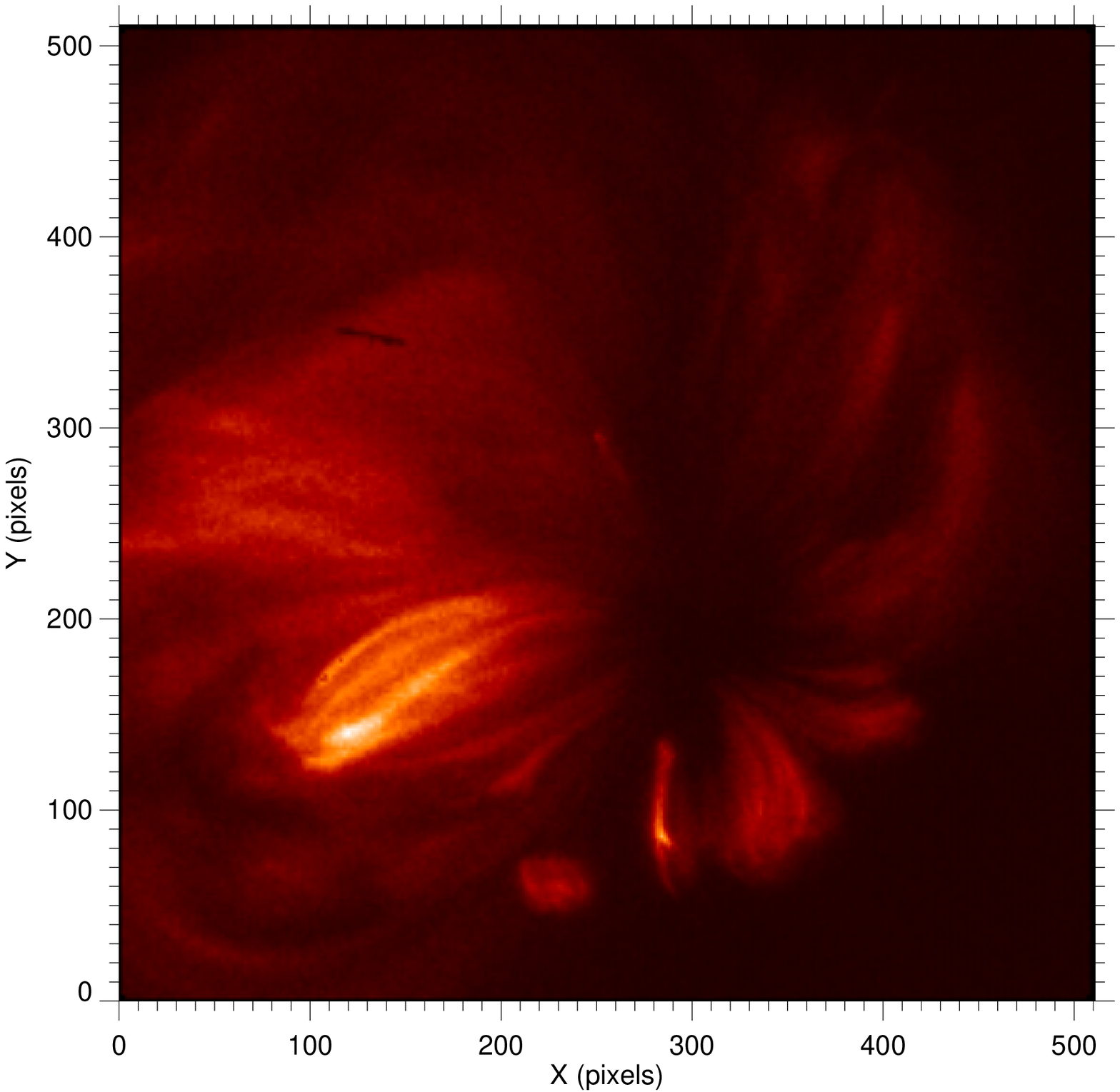}}
  \caption{Images of the same active region, taken in the EUV band
  with TRACE (top) and in the X-ray band with Hinode/XRT (bottom), on 14 november 2006. The
  X-ray image shows more clearly that the active region is densely
  populated with coronal loops.}
  \label{fig:x_uv}
\end{figure}}

Coronal loops are characterized by an arch-like shape that recalls typical magnetic field topology. This shape is replicated over a wide range of dimensions. Referring, for the moment, to the soft X-ray band, the main properties of coronal loops are listed in Table~\ref{tab:loop_par}. The length of coronal loops spans over at least 4 orders of magnitude: bright points ($\sim 10^8$~cm), small Active Region (AR) loops ($\sim 10^9$~cm), AR loops ($\sim 10^{10}$~cm), giant arches ($\sim 10^{11}$~cm) (Figure~\ref{fig:loop_types}). As already mentioned, the loops owe their high luminosity and variety to their nature of magnetic flux tubes where the plasma is confined and isolated from the surroundings. Magnetized fully-ionized plasma conducts thermal energy mostly along the magnetic field lines. Due to the high thermal insulation, coronal loops can have different temperatures, from $\sim 10^5$~K (cool loops), to a few $\sim 10^6$ (X-ray loops), up to a few $\sim 10^7$~K (flaring loops). A density of the confined plasma below $10^7 - 10^8$~cm$^{-3}$ can be difficult to detect, while more typical values of bright loops are $10^9 - 10^{10}$~cm$^{-3}$ in quiescent and active regions loops. Flaring loops can be easily a factor 10 denser. The corresponding plasma  pressure can typically vary between $10^{-3}$ and 10~dyne~cm$^{-2}$ for non-flaring loops, corresponding to confining magnetic fields of the order of 0.1\,--\,10~G. One characterizing feature of coronal loops is that typically their cross-section is constant along their length above the transition region, at variance from the topology of potential magnetic fields. There is evidence that the cross-section varies across the transition region, as documented in \cite{1976RSPTA.281..339G}.

\begin{table}[htbp]
\caption{Typical X-ray coronal loop parameters}
\label{tab:loop_par}
\centering
\begin{tabular}{l c c c c}
\toprule
Type & Length      & Temperature & Density           & Pressure \\
~    & [$10^9$~cm] & [MK]        & [$10^9$~cm$^{-3}$] & [dyne~cm$^{-2}$] \\
\midrule
Bright points & 0.1\,--\,1  & 2        & 5           & 3 \\
Active region & 1\,--\,10   & 3        & 1\,--\,10   & 1\,--\,10 \\
Giant arches  & 10\,--\,100 & 1\,--\,2 & 0.1\,--\,1  & 0.1 \\ 
Flaring loops & 1--10       & $> 10$   & $> 50$      & $> 100$ \\
\bottomrule
\end{tabular}
\end{table}


\epubtkImage{loop_types.png}{%
\begin{figure}[htbp]
  \centerline{\includegraphics[scale=0.7]{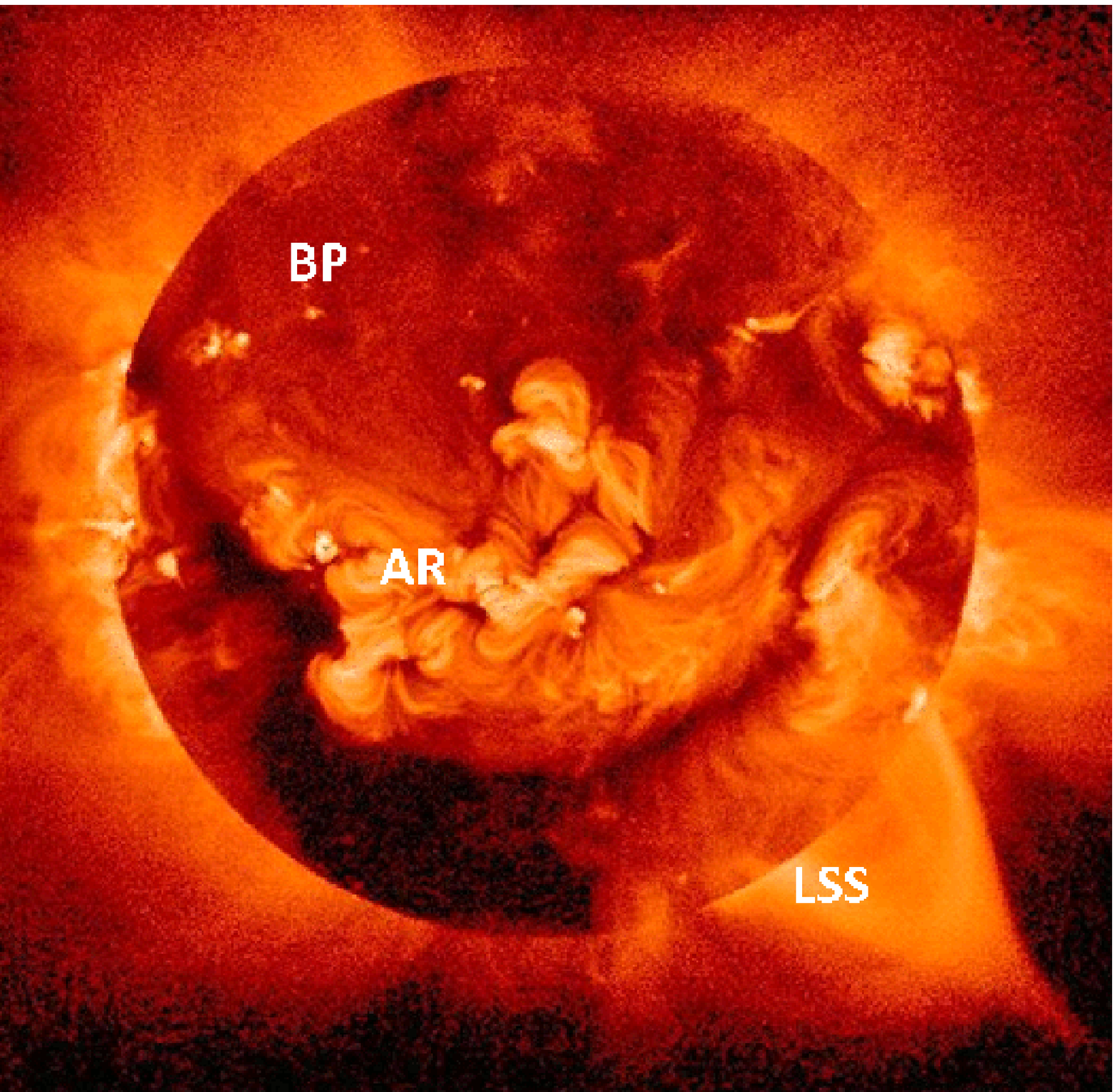}}
  \caption{The X-ray corona contains loops with different spatial
  scales, e.g., bright points (BP), active region loops (AR), large
  scale structures (LSS) (Image credit: Yohkoh mission, ISAS, Japan).}
  \label{fig:loop_types}
\end{figure}}

\subsubsection{Classification}
\label{sec:obs_class}


Myriads of loops populate the solar corona and constitute statistical ensembles. Attempts to define and classify coronal loops were never easy, and no finally established result exists to-date. Early attempts were based on morphological criteria, i.e., bright points, active region loops and large scale structures \cite{1973SoPh...32...81V}, largely observed with instruments in the X-ray band.
In addition to such classification, more recently, the observation of loops in different spectral bands and the suspicion that the difference lies not only in the band, but also in intrinsic properties, have stimulated another classification based on the temperature regime, i.e., \textit{cool, warm, hot} loops (Table~\ref{tab:loop_therm}). \textit{Cool loops} are generally detected in UV lines at temperatures between $10^5$ and $10^6$~K. They were first addressed by \cite{1976ApJ...210..575F} 
and later explored more with SoHO observations \citep{1997SoPh..175..511B}.
\textit{Warm loops} are well observed by EUV imagers such as SoHO/EIT and TRACE, and confine plasma at temperature around 1\,--\,1.5~MK \cite{1999ApJ...517L.155L}. 
\textit{Hot loops} are those typically observed in the X-ray band and hot UV lines (e.g.\ Fe\,{\sc xvi}), with temperatures around or above 2~MK (Table~\ref{tab:loop_par}). These are the coronal loops already identified, for instance, in the early rocket missions \cite{1973SoPh...32...81V}.
This distinction is not only due to observation with different instruments and in different bands, but there are hints that it may be more substantial and physical, i.e., there may be two or more classes of loops that may be governed by different regimes of physical processes. For instance, the temperature along warm loops appears to be distributed uniformly and the density to be higher than that predicted by equilibrium conditions. Does this make such loops intrinsically different from hot loops, or is it just the signature that warm loops are a transient conditions of hot loops? 

\begin{table}[htbp]
\caption{Thermal coronal loop classification}
\label{tab:loop_therm}
\centering
\begin{tabular}{l c}
\toprule
Type & Temperature [MK] \\
\midrule
Cool & 0.1\,--\,1 \\
Warm & 1\,--\,1.5 \\
Hot & $\geq 2$ \\
\bottomrule
\end{tabular}
\end{table}


%
%


A real progress in the insight into coronal loops is expected from the study of large samples of loops or of loop populations. Systematic studies of coronal loops suffer from the problem of the sample selection and loop identification, because, for instance, loops in active regions overlap along the line sight. Attempts of systematic studies have been performed in the past on Yohkoh and TRACE data \citep[e.g.,][]{1995ApJ...454..499P,2000ApJ...541.1059A}.
A large number of loops were analyzed and it was possible to obtain meaningful statistics. However,
it is difficult to generalize the results because of limited samples and/or selection effects, e.g., best observed loops, specific instrument. One basic problem for statistical studies of coronal loops is that it is very difficult to define an objective criterion for loop identification. In fact, loops are rarely isolated; they coexist with other loops which intersect or even overlap along the line of sight. This is especially true in active regions where most of the loops are found. 
In order to make a real progress along this line, we should obtain loop samples and populations selected on totally objective and unbiased criteria, which is difficult due to the problems outlined above. Some steps are coming in this direction and we will see results in the future.

\subsection{Morphology and fine structuring}
\label{sec:obs_mor}

\subsubsection{Geometry}
\label{sec:obs_mor_mag}


Coronal loops are magnetic structures and might therefore be mapped easily and safely by mapping reliably the coronal magnetic field. Unfortunately, it is well-known that it is very difficult to measure the magnetic field in the corona, and it can be done only in very special conditions, e.g. very strong local field \citep{1991ApJ...366L..43W}. 
In some cases it is possible to use coronal seismology \citep[first proposed by][]{1970PASJ...22..341U} 
to determine the average magnetic field strength in an oscillating loop first used by \cite{1999Sci...285..862N}, and \cite{2001A&A...372L..53N}
on TRACE loops, and recently investigated in a number of studies. The accuracy of this method depends on the correct detection of the temporally and spatially resolved mode of oscillation, and on the details of the loop geometry.

\epubtkImage{loop_shape.png}{%
\begin{figure}[htbp]
  \centerline{\includegraphics[scale=0.6]{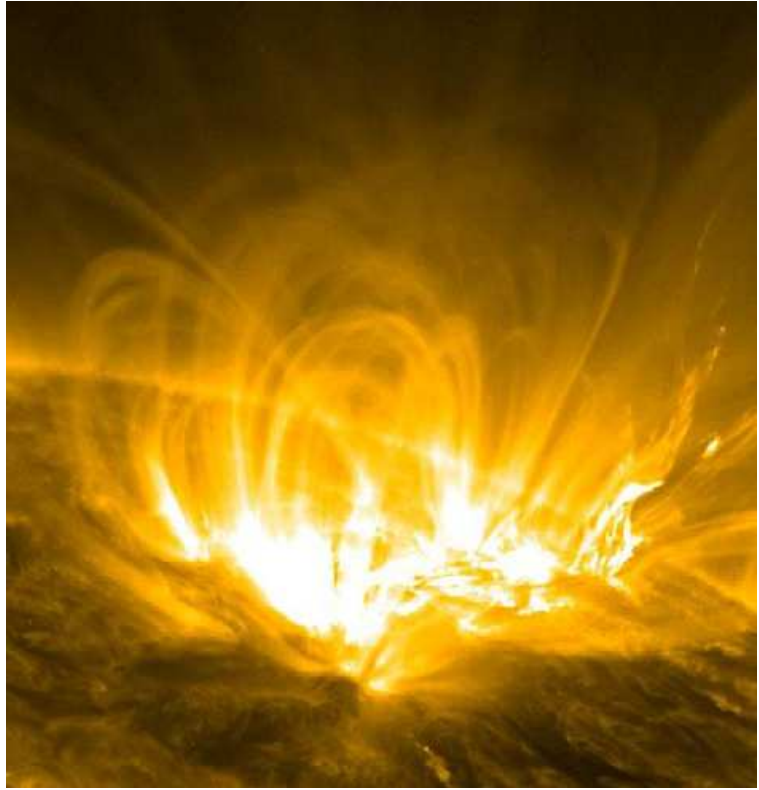}}
  \caption{Coronal loops have approximately a semicircular shape (image: SDO/AIA,  171 A filter, 9 July 2010, credit: NASA/SDO).}
  \label{fig:loop_shape}
\end{figure}}

%

Since we cannot well determine the coronal magnetic field, coronal loop geometry deserves specific analysis.
As a good approximation, loops generally have a semicircular shape (Figure~\ref{fig:loop_shape}). The loop aspect, of course, depends on the loop orientation with respect to the line of sight: loops with the footpoints on the limb more easily appear as semicircular, as well as loops very inclined on the surface near the center of the disk. The assumption of semicircular shape can be useful to measure the loop length even in the presence of important deformations due to projection effects: the de-projected distance of the loop footpoints is the diameter of the arc. However, deviations from circularity are rather common and, in general, the detailed analysis of the loop geometry is not a trivial task. The accurate determination of the loop geometry is rather important for the implications on the magnetic field topology and reconstruction. It is less important for the structure and evolution of the confined plasma, which follow the field lines whatever shape they have and change little also with moderate changes of the gravity component along the field lines. First works on the accurate determination of the loop geometry date back to the sixties \cite{1964ApJ...140..760S}. 
More specific ones take advantage of stereoscopic views allowed by huge loops during solar rotation, with the aid of magnetic field reconstruction methods. These studies find deviations from ideal circularity and symmetry, not surprising for such large structures \cite{1985SoPh...96...93B}. 
The geometry of a specific loop observed with TRACE was measured in the framework of a complete study including time-dependent hydrodynamic modeling \cite{2000ApJ...535..412R,2000ApJ...535..423R}. In that case, the discrepancy between the length derived from the distance of the footpoints taken as loop diameter and the length measured along the loop itself allowed to assess the loop as elongated.
Later, a reconstruction of loop geometry was applied to TRACE observation of medium-sized oscillating loops, to derive the properties of the oscillations. In this case, a semicircular pattern was applied \cite{2002SoPh..206...99A}.
The importance of the deviations from circularity on constraining loop oscillations was remarked later \cite{2006AA...459..241D}.

The STEREO mission is actually contributing much to the analysis of loop morphology and geometry, thanks to its unique capability to observe the Sun simultaneously from different positions. 
\cite{2007ApJ...671L.205F} presented a first stereoscopic reconstruction of the three-dimensional shape of magnetic loops in an active region from two different vantage points based on simultaneously recorded STEREO/SECCHI images. They derived parameters of five relatively long loops and constraints on the local magnetic field, and found reconstructed loops to be non-planar and more curved than field lines extrapolated from SoHO/MDI measurements, probably due to the inadequacy of the linear force-free field model used for the extrapolation.
A misalignment of 20\,--\,40~deg between theoretical model and observed loops has been quantified from STEREO results and discussed \citep{2009SoPh..259....1S,2009ApJ...696.1780D} 
\cite{2008ApJ...679..827A} presented triangulations and 3D reconstructions of 30 coronal loops, using the Extreme UltraViolet Imager (EUVI) telescopes of both STEREO spacecrafts and deriving a series of loop characteristics, such as the loop plane inclination angles, and the coplanarity and circularity. They derived the parameters of seven complete and quite inclined loops, and found deviations from circularity within 30\% and less significant from coplanarity.
\cite{2009ApJ...695...12A} applied a reconstruction method to an active region observed with STEREO by combining stereoscopic triangulation of 70 loops and addressing mainly density and temperature modeling with a filling factor equivalent to tomographic volume rendering.

%
      
Another interesting issue regarding coronal loop geometry is the analysis of the loop cross-section, which also provides information about the structure of the coronal magnetic field. \linebreak
Yohkoh/\,SXT allowed for systematic and quantitative studies of loop morphology and showed that the cross-section of coronal loops is approximately constant along their length and do not increase significantly.
More in detail, a systematic analysis of a sample of ten loops showed that the loops tend to be only slightly ($\sim$~30\%)
wider at their midpoints than at their footpoints, while for a bipolar field configuration we would expect expansions by factors. One possible explanation of this effect is the presence of significant twisting of the magnetic field lines, and therefore the development of electric currents and strong deviations from a potential field. The effect might be seen either as a twisting of a single loop or as a ``braiding'' of a bundle of unresolved thin loops. At the same time it was found that the variation of width along each loop tends to be modest, implying that the cross section has an approximately circular shape \citep{1992PASJ...44L.181K}. 
Implications of these results on the theory of coronal heating are discussed in \cite{2000SoPh..193...53K}, but the conclusion is that none of the current models alone is able to explain all observed properties.

\epubtkImage{cor_magn.png}{%
\begin{figure}[htbp]
  \centerline{\includegraphics[scale=0.6]{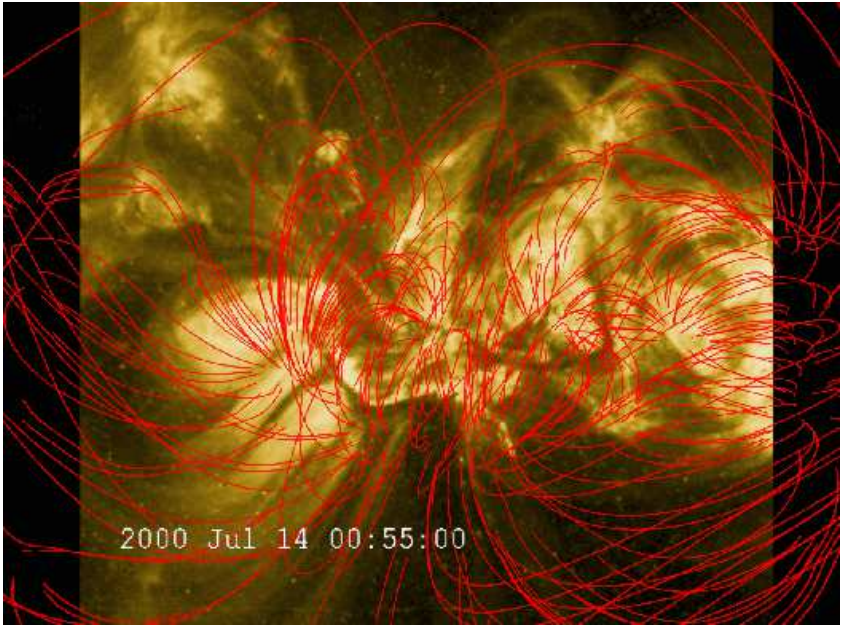}}
  \caption{Magnetic field lines extrapolated from optical magnetogram superposed on a TRACE image. Credit: NASA/ESA/LMSAL }
  \label{fig:cor_mag}
\end{figure}}

Important information about the internal structuring of coronal loops comes from the joint analysis of the photospheric and coronal magnetic field (Figure~\ref{fig:cor_mag}). 
An analysis of the magnetic field at the footpoints of hot and cool loops showed, among other results, that the magnetic filling factor is lower in hot loops (0.05\,--\,0.3 out of sunspots) than in warm loops (0.2\,--\,0.6) \citep{2005ApJ...621..498K}.
\cite{2006ApJ...639..459L} investigated the
magnetic structure of loops observed with TRACE,  applying a linear
force-free extrapolation model to SoHO/MDI data and comparing the resulting
configuration of the magnetic field directly with TRACE images of the same
region. They confirmed that, whereas the model predicts a significant expansion of the
magnetic field structures, at least a factor two, from the footpoints to the corona, and a significant asymmetry of the structures, because the magnetic field lines starting from the
same footpoint can diverge to different other end footpoints, these features
are not observed: also TRACE loops are quite symmetric and their cross-section is
constant to a good degree of approximation, as it had been already found for Yohkoh loops 
\citep{1992PASJ...44L.181K}.
The results in \cite{2006ApJ...639..459L} suggest that the tangling 
of the magnetic flux strands driven by the photospheric
convection might be very strong, and confirm, therefore, that the magnetic field structure is far more
complicated than it can be modelled even with linear force-free extrapolation.
\cite{2007ApJ...662L.119S} studied 
braiding-induced interchange reconnection of the magnetic field and the width of solar coronal loops and showed that loop width observations
support the hypothesis that granular braiding is countered statistically by frequent coronal reconnections,
which in turn explain the general absence of entangled coronal field structures in high-resolution observations of the quiescent solar corona.
\cite{2007ApJ...661..532D} addressed the apparent
uniform cross section of bright threadlike structures in the corona, long apparent scale height, and the inconsistency between loop densities derived by spectral and photometric means. They found that, if coronal loops are interpreted as a mixture of diffuse background and very dense, unresolved stranded structures, this requires a combination of high plasma density within the structures, which greatly increases the emissivity of the structures, and geometric effects that attenuate the apparent brightness of the feature at low altitudes.

New methods to improve the morphological analysis of loops are being developed. For instance, 
\cite{2007AA...466..347D} made multispectral analysis of solar EUV images and explored
the possibility of separating the different solar structures from a linear combination of images. They found source images with more contrast than the original ones.


\subsubsection{Fine structuring}
\label{sec:obs_fib}


It has been long claimed \citep[e.g.,][]{1993ApJ...405..767G} 
that coronal loops consist of bundles of thin strands, to scales below the current instrumental resolution. The task to investigate this substructuring is not easy because the thickness of the elementary components may be as small as a few km, according to some nanoflare models \citep[e.g.,][]{2009AA...499L...5V}, and the measured one goes down to the resolution limit of the most powerful imaging instruments \citep[e.g.,][]{1993ApJ...405..767G}. First limited evidence of fine structuring was the low filling factor inferred for loops observed with NIXT \citep{1999AA...342..563D} (see Section~\ref{sec:obs_warm_hot}).
The high spatial resolution achieved by the
TRACE normal incidence telescope allowed to address the
transverse structure of the imaged coronal loops. TRACE images visibly
show that coronal loops are substructured (Figure~\ref{fig:trace_strands}).

\epubtkImage{trace_strands.png}{%
\begin{figure}[htbp]
  \centerline{\includegraphics[scale=0.4]{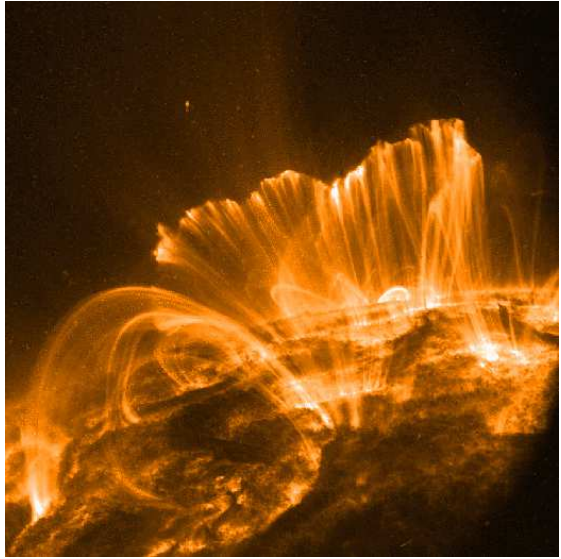}}
  \caption{TRACE image including a system of coronal loops (9 November 2000, 2 UT). Bundles of
  strands are clearly visible.}
  \label{fig:trace_strands}
\end{figure}}

There were some early attempts to study the structure along the single strands in TRACE observations \citep{2002ApJ...580.1159T}.
Later, it was shown that in many
cases, hot loop structures observed with the Yohkoh/SXT are not exactly co-spatial with warm structures observed 
with the SoHO/EIT, which is sensitive in the same bands as TRACE, nor they cool down to become visible to the EIT \citep{2003ApJ...590.1095N,2004ApJ...601..530S}.

The detailed morphological comparison of an active region showed that hot loops seen in SXT ($T > 3$~MK) and 
warm loops seen in the SoHO/EIT 195~\AA\ band ($T \sim 1 - 2$~MK) are located in almost alternating manner \citep{2003ApJ...590.1095N}.
The anti-coincidence of the hot and the warm loops is conserved for a duration longer than the estimated cooling timescale. These results suggest that loops are not isothermal in the transverse direction, rather they have a differential emission measure distribution of modest but finite width that peaks at different temperatures for different loops (see Section~\ref{sec:obs_dia}).

In apparent alternative, 
\cite{2005ApJ...626..543W} showed that hot monolithic loops visible with the
Yohkoh/SXT are later resolved as stranded cooler structures with TRACE.
However, this occurs with a time delay of 1 to 3 hours, much longer than the
plasma cooling times, and, therefore, correlation can be hardly established
between the plasma detected by Yohkoh and that detected by 
TRACE. Direct density measurements of loop plasma 
made from multi-line observations at the solar limb indicate not very dense
plasma and relatively high plasma filling factor \citep[0.2\,--\,0.9,][]{2005AA...439..351U}).
This is in the direction of a moderate structuring of the loops.

\cite{2005ApJ...633..499A} analyzed specifically 
and systematically TRACE images to search for the thinnest coherent structures 
that can be resolved with TRACE.  They found that about 10\% of the positions can
be fitted with an isothermal model and proposed that, since the corresponding
structures have a uniform thermal distribution, they should be elementary loop
components, with an average width of about 2000~km.
\cite{2007ApJ...656..577A} studied statistically a large set of coronal loops and found further evidence of elementary loop strands resolved by TRACE.

Other studies based both on models and on analysis of observations independently suggest that elementary loop components should be even finer, with typical cross-sections of the strands to be of the order of 10\,--\,100~km \citep{2003SoPh..216...27B,2004ApJ...605..911C,2009AA...499L...5V}.

\subsection{Diagnostics and thermal structuring}
\label{sec:obs_dia}
%

The investigation of the thermal structure of coronal loops is very important for their exhaustive physical comprehension and to understand the underlying heating mechanisms. For instance, one of the classifications outlined above is based on the loop thermal regime, and, we remark, it is debated whether the classification indicates a real physical difference.


Diagnostics of temperature are not trivial in the corona. No direct measurements are available. Since the plasma is optically thin, we receive information integrated on all the plasma column along the line of sight. The problem is to separate the distinct contributing thermal components and reconstruct the detailed thermal structure along the line of sight. However, even the determination of global and average values deserves great attention.

Moderate diagnostic power is allowed by imaging instruments, by means of multifilter observations. Filter ratio maps provide information about the spatial distribution of temperature and emission measure \citep[e.g.,][]{1973SoPh...32...81V}.
The emission of an optically thin isothermal plasma as measured in a $j$-th filter passband is:

\begin{equation}
I_j= EM ~~ G_j(T)
\end{equation}
where $T$ is the temperature and $EM$ is the emission measure, defined as 
\[
EM = \int_V n^2 dV
\]
where $n$ is the particle density, and $V$ the plasma volume.
The ratio $R_{ij}$ of the emission in two different filters $i,j$ is then independent of the density, and only a function of the temperature:

\begin{equation}
R_{ij} = \frac{I_i}{I_j} = \frac{G_i(T)}{G_j(T)}
\end{equation}
The inversion of this relationship provides a value of temperature.

The limitations of this method are substantial. In particular, one filter ratio value provides one temperature value for each pixel; this is a reliable measurement, within experimental errors, as long as the assumption of isothermal plasma approximately holds for the plasma column in the pixel along the line of sight. If the plasma is considerably multithermal, the temperature value is an average weighted for the instrumental response. Since the response is a highly non-linear function of the emitting plasma temperature, it is not trivial to interpret the related maps correctly. In addition, it is fundamental to know the instrument response with high precision, in order to avoid systematic errors, which propagate dangerously when filter ratios are evaluated. In this sense, broadband filters provide robust thermal diagnostics, because they are weakly dependent on the details of the atomic physics models, e.g., on the presence of unknown or not well-known spectral lines, on the choice of element abundances. Narrowband filters can show non-unique dependencies of filter ratio values on temperature \citep[e.g.,][]{2007ApJ...667..591P}, due to the presence of several important spectral lines in the bands, but a more general problem can be the bias to detect narrow ranges of temperatures forced by the specific instrument characteristics. This problem can be important especially when the distribution of the emission measure along the line of sight is not simple and highly non-linear \citep[e.g.,][]{2009ApJ...698..756R}.
The problem of diagnostics of loop plasma from filter ratios, and, more in general, the whole analysis of loop observations, are made even more difficult by the invariable presence of other structures intersecting along the line of sight. A uniform diffuse background emission also affects the
temperature diagnostics, by adding systematic offsets which alter the filter
ratio values. The task of subtracting this ``background emission'' from the measured emission is non-trivial and can affect seriously the results of the whole analysis. 
This problem emerged dramatically when the analysis of the same large
loop structure observed with Yohkoh/SXT on the solar limb led to three different results
depending mostly on the different ways to treat the background 
\citep{2000ApJ...539.1002P,2001ApJ...559L.171A,2002ApJ...580..566R}.
The amount of background depends on the instrument characteristics, such as the passband and the point response function: it is most of the signal in TRACE UV filterbands, for instance, and its subtraction becomes a very delicate issue \citep[e.g.,][]{2003AA...406.1089D,2006AA...449.1177R,2008ApJ...680.1477A,2010AA...515A...7T}. 
The problem can be mitigated if one analyzes loops as far as possible isolated from other loops, but this is not easy, for instance, in active regions. If this is not the case, broadband filters may also include contamination from many structures at relatively different temperature and make the analysis of single loops harder.
The problem of background subtraction in loop analysis has been addressed by several authors, who apply different subtraction ranging from simple offset, to emission in nearby pixels or subregions, to values interpolated between the loop sides, to whole images at times when the loop is no longer (or not yet) visible \citep{2002ApJ...580.1159T,2003AA...406.1089D,2003ApJ...599..604S,2005ApJ...633..499A,2006AA...449.1177R,2008ApJ...680.1477A,2010AA...515A...7T}.

\epubtkImage{tmap.png}{%
\begin{figure}[htbp]
  \centerline{\includegraphics[scale=0.5]{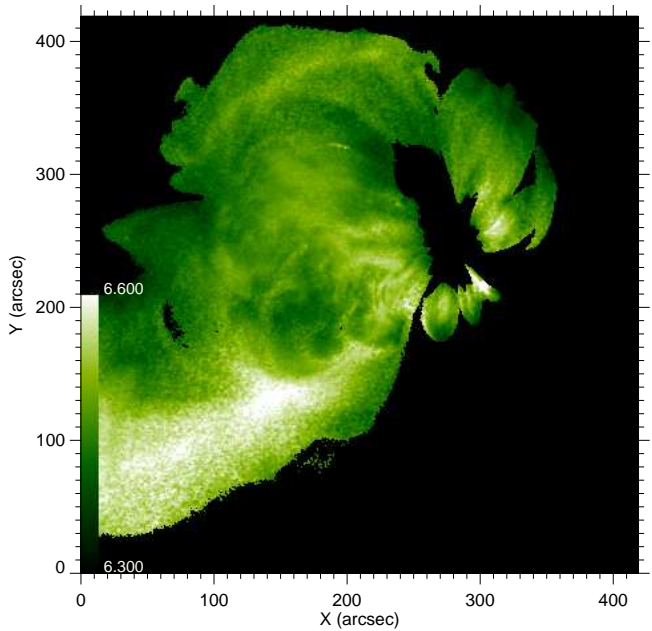}}
  \caption{Temperature map of an active region obtained from the ratio of two images in different broadband filters with Hinode/XRT (12 November 2006, 12 UT).}
  \label{fig:tmap}
\end{figure}}

More accurate diagnostics, although with less time and space resolution, is in principle provided by spectrometers and observations in temperature-sensitive spectral lines, which are being constantly improved to provide better and better spatial information. Early results from UV spectroscopy already recognized the link between transition region and coronal loops, for instance, from Skylab mission \citep{1979ApJ...229..369F,1980ApJ...240..300M}. 
Together with background subtraction, one major difficulty met by spectroscopic analysis is that, in the UV band, the density of lines is so high that they are often blended and, therefore, it is hard to separate the contribution of the single lines, especially the weak ones. Fine diagnostics, such as Doppler shifts and line broadening, can become very tricky in these conditions and results are subject to continuous revisitation and warnings from the specialized community. The problem of background subtraction is serious also for spectral data, because their lower spatial and temporal resolution determines the presence of more structures, and therefore more thermal components, along the line of sight in the same spatial element.

Care should be paid also when assembling information from many
spectral lines into a reconstruction of the global thermal structure
along the line of sight. Methods are well-established \citep[e.g.,][]{1975MNRAS.173..397G} and several approaches are
available. The so-called method of the emission measure loci
\citep{1987MNRAS.225..903J} is able to tell whether plasma is
isothermal of multithermal along the line of sight (Figure~\ref{fig:em_loci}), but less able to
add details. Detailed emission measure distributions can be obtained
from differential emission measure (DEM) reconstruction methods  \citep[e.g.,][]{1996ApJS..106..143B,1998ApJ...503..450K}, but this is an ill-posed mathematical problem, and, therefore, results are not unique and are subject to systematic and unknown errors. Forward modeling and simulations can be ways to escape from these problems, but they require non-trivial computational efforts and programming, and it is not always possible to provide accurate confidence levels. All these approaches are constantly improved and probably the best way to proceed is to combine different approaches and multiband observations and to finally obtain a global consistency.

In addition to the problems intrinsic to diagnostic techniques, we have to consider that loops appear to have different properties in different bands, as mentioned in Section~\ref{sec:obs_class}. It is still debated whether such differences originate from an observational bias due to the instruments or from intrinsic physical differences, or both. In view of this uncertainty, in the following we will make a distinction between hot and warm loops, which will generally correspond to loops observed in the X-ray (and hot UV lines, e.g., Yohkoh/SXT, Hinode/XRT, SoHO/CDS Fe\,{\sc xvi} line) and in the UV band (e.g., SoHO/EIT, TRACE), respectively. Cool loops are also observed in the UV band. The boundary between hot and warm loops is, of course, not sharp, and it is not even clear whether they are aspects of the same basic structure, or they really are physically different and are heated differently (see also Section~\ref{sec:obs_dia_tra}). We will devote attention to the comparison between hot and warm loops.


\subsubsection{Hot loops}
\label{sec:obs_dia_yoh}

After the pioneering analyses driven by the Skylab X-ray instruments (Section~\ref{section:histor}),
Yohkoh/SXT allowed to conduct large scale studies on the thermal and structure diagnostics of hot loops, and the comparison with other instruments, for instance on-board SoHO, allowed to obtain important cross-checks and additional information. Filter ratio maps of flaring loops were shown early after the mission launch \citep{1992PASJ...44L..63T}.

Systematic measurements of temperature, pressure and length of tens of quiescent and active region coronal loops were conducted on Yohkoh observations \citep{1995ApJ...454..499P} using the filter ratio method. For this sample of loops, selected to be steady and isolated, the lengths were measured with assumptions on the loop geometry and ranged in a decade between $5 \times 10^9 < 2L < 5 \times 10^{10}$ cm. The temperature measurements were averaged over about half of the loops and also ranged in a decade ($2 < T < 30$~MK), with a mean of about 6~MK. Therefore, it appears as a sample of particularly hot loops, although the uncertainties in the hot tail of the distribution are very large, probably due to the flat dependence of the temperature on the filter ratio at high temperature. Pressures were derived from the equation of state, after derivation of the density, from the emission measure and from the volume inferred from the length and assumptions on the loop aspect. They ranged in two decades ($0.1 < p < 20$ dyne cm$^{-2}$). Overall, it was shown that the temperature and length of this sample of hot loops are uncorrelated, that pressure varies inversely with length (as overall expected for a thermally homogeneous sample from loop scaling laws, see Section~\ref{sec:mod_bas_rtv}), although with a large spread. Such distributions were used as constraints on the loop heating through the derivation of the dependence of the magnetic field intensity on the loop length \citep{1995Natur.377..131K}. They also led to accurate analysis of data uncertainties \citep{1995ApJ...448..925K}. Another systematic analysis was made on a sample of about 30 bright steady Yohkoh loops located in active regions \citep{1995ApJ...454..934K}. The temperatures were measured after averaging several images and taking the value at the loop top. While this analysis confirmed some of the correlations found in \cite{1995ApJ...454..499P}, it found a correlation between the loop length and the temperature, and showed deviations from RTV scaling laws (Section~\ref{sec:mod_bas_rtv}). It cannot be excluded that correlations between parameters depend on the loop sample, as a single scaling law links three parameters.
Yohkoh/SXT loops hotter than 3~MK were found also in another study, the hottest ones with shorter lifetimes (less than few hours), and often exhibiting cusps \citep{1996ApJ...459..342Y}. 

A big effort has recently been devoted to the possible detection of
hot plasma outside of evident flares. This would be a conclusive
evidence of nanoflaring activity in coronal loop \citep[e.g.,][see Section~\ref{sec:mod_hea}]{2006SoPh..234...41K}.
Hinode instruments appear to be able to provide new interesting contributions to this topic.
The analysis of spectroscopic observations of hot lines  in solar active regions from Hinode/EIS allows to construct emission measure distributions in the 1\,--\,5~MK temperature range, and shows that the distributions are flat or slowly increasing up to approximately 3~MK and then fall off rapidly at higher temperatures \citep{2009ApJ...696..760P}.
Evidence of emission from hot lines has been found also in other Hinode/EIS observations, and in particular from the analysis of the emission from the 
Ca\,{\sc xvii} at 192.858~\AA, formed near a temperature of $6\times 10^6$~K, which has been found in several parts of active regions \citep{2009ApJ...697.1956K}.
Using Fe lines, \cite{2009AA...495..587Y}
has shown very accurate density measurements ($\approx$~5\%) across an active region, with values in the range $8.5\leq \log\,(N_{\rm e}/{\rm cm}^{-3}) \leq 11.0$. A smaller density range (from $10^{8.5}$ to $10^{9.5}$~cm$^{-3}$ has been found by \cite{2009ApJ...692.1294W} 
using the Fe\,{\sc xiii} line group, although one pair has been found to reach the high density limit. Density sensitive lines have been used to measure the filling factor of coronal structures. 
\cite{2008AA...491..561D,2009AA...497..287D} has used spectra and images obtained with EIS and comparison with TRACE to 
determine the volumetric filling factor of bright points. The emission measure and bright point widths have been compared with the electron densities and with TRACE data. The plasma-filling factor has been found to vary from $3 \times 10^{-3}$ to 0.3 with a median value of 0.04, which may indicate considerable subresolution structure, or the presence of a single completely-filled unresolved loop with subarcsec width.

Thanks to its multifilter observations, also Hinode/XRT is providing useful information about the thermal structure of the bright X-ray corona.  Temperature maps derived with combined filter ratios show fine structuring to the limit of the instrument resolution  and evidence of multithermal components \citep{2007Sci...318.1582R}.
This kind of temperature diagnostics is supported by the evidence of warm structures bright in the TRACE images. Observations including flare filters show evidence of a hot component in active regions outside of flares \citep{2009ApJ...693L.131S} 
and data in the medium thickness filters appear to constrain better this component of hot plasma as widespread, although minor, and peaking around $\log T \sim 6.8-6.9$, with a tail above 10~MK \citep{2009ApJ...698..756R}.
This may support the hypothesis that active regions are heated impulsively. Evidence of a persistent although small hot plasma component outside of flare is shown also by RHESSI data \citep{2009ApJ...697...94M}, and the comparison between RHESSI and XRT data seem to support this scenario in a consistent way \citep{2009ApJ...704L..58R}. 
The topic of coronal active region heating is debated. Evidence interpreted in the direction of  more gradual heating has been obtained by \cite{2010ApJ...711..228W} (see Section~\ref{sec:mod_hea}).

\subsubsection{Comparison of hot and warm loops}
\label{sec:obs_warm_hot}

Before the SoHO/EIT and TRACE observations, warm loops had been imaged in a similar spectral band and with similar optics by the rocket NIXT mission (see Section~\ref{section:histor}). Studies of NIXT loops including the comparison with hydrostatic loop models (Section~\ref{sec:mod_bas_rtv}) pointed out that bright spots also visible in H$\alpha$ band were the footpoints of hot high-pressure loops \citep{1994ApJ...422..412P}. This result was confirmed by 
the comparison of the temperature structure obtained from Yohkoh with NIXT data  \citep{1995PASJ...47L..15Y} (see also Section~\ref{sec:obs_dia_tra}).

Another comparison of loops imaged with NIXT and Yohkoh/SXT showed that the compact loop structures (length $\sim  10^9$~cm) have a good general morphological correspondence, while larger scale NIXT loops ($\sim 10^{10}$~cm)
have no obvious SXT counterpart \citep{1999AA...342..563D}. Comparison with static loop models (see Section~\ref{sec:mod_bas_rtv}) allowed to derive estimates of the loop filling factors, important for the loop fine structure (Section~\ref{sec:obs_fib}). In the NIXT band, the filling factor of short loops was found to be very low ($10^{-3} - 10^{-2}$), but of the order of unity in the SXT band and for the largest structure. Information about the loop filling factor was derived also from the analysis of
simultaneous SoHO and Yohkoh observations of a small solar active region, suggesting a volume filling factor decreasing with increasing density and possible differences between emitting material in active regions and the quiet Sun \citep{2000ApJ...537..481G}. 

Some similarity between loops observed in the TRACE EUV band and Yohkoh X-ray band was found out of the core of active region loops \cite{2000SoPh..195..123N} and interpreted as evidence of loops with a broad range of temperatures. Core loops were instead observed only in the X-rays and found to be variable, indicating that probably they are not steady. 

The thermal distribution across the loop structures, i.e., along the line of sight, can be investigated with the analysis of observations in several spectral lines, as obtained, for instance, with the SoHO/CDS. Information on 
the validity of the data analysis and of the loop diagnostics can be  obtained from
the comparison with simultaneous and co-spatial data from imaging instruments. 

Density and temperatures in two active regions were accurately determined from 
SOHO-CDS observations \citep{1999SoPh..189..129M} 
and it was confirmed quantitatively that the AR cores are hotter than larger loop structures extending above the limb. 
From the analysis of a single loop observed on the solar limb with
SoHO/CDS, \cite{2001ApJ...556..896S} 
found a bias to obtain flat temperature distributions along
the loop from ratios of single lines or narrow band filters (TRACE), while a careful DEM reconstruction at selected points along the loop is inconsistent with isothermal
plasma both across and along the loop. A whole line of works started from this
study reconsidering and questioning the basic validity of the
temperature diagnostics with TRACE and emphasizing the
importance of the background subtraction, but also the need to obtain
accurate spectral data \citep{2002ApJ...578L.161S,2002ApJ...577L.115M,2002ApJ...580L..79A,2003ApJ...599..604S}.
Similar results but different conclusions
were reached by \cite{2004ApJ...608.1133L,2004ApJ...611..537L}
who analyzed a loop observed with SoHO and, finding it nearly isothermal,
considered this evidence as real and invoked a non-constant cross-section to explain it. On the other hand, evidence of non-uniform temperature along loops observed with TRACE was also found  \citep{2003AA...406.1089D,2006AA...449.1177R}, emphasizing that the temperature diagnostics with narrow band instruments is a delicate issue.

\epubtkImage{em_loci}{%
\begin{figure}[htbp]
  \centerline{\includegraphics[width=11cm]{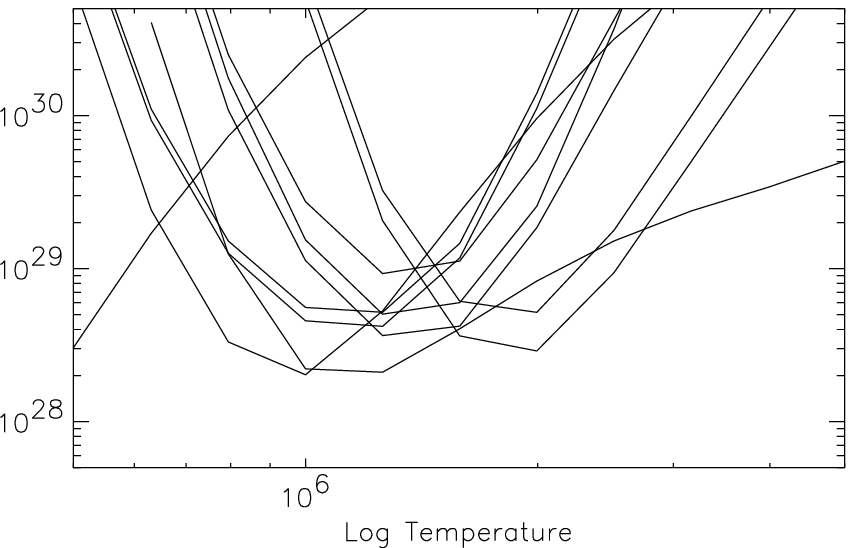}}
  \centerline{\includegraphics[width=8.8cm,angle=270]{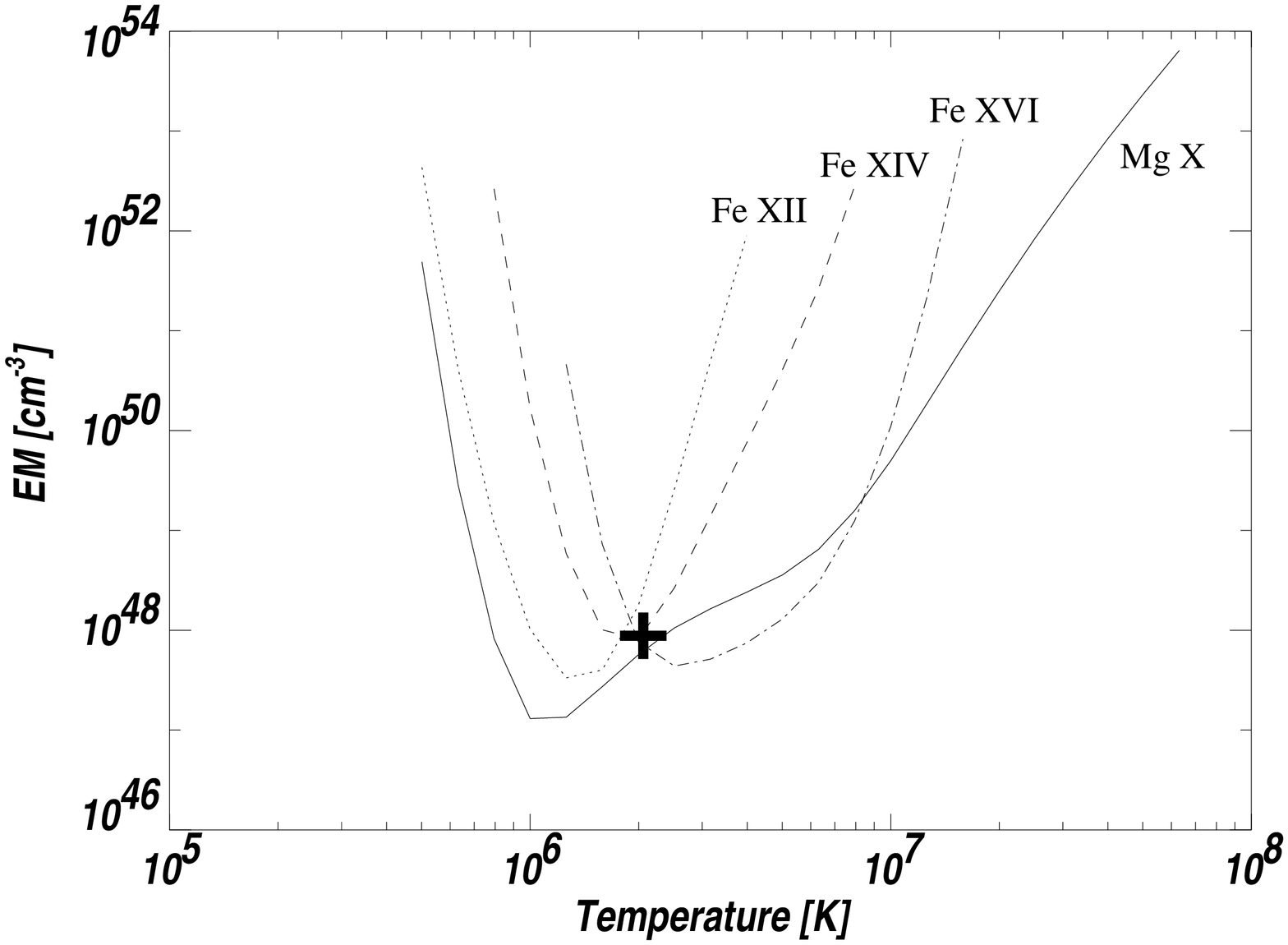}}
  \caption{EM-loci plots for two different loop systems, one showing a
    multi-thermal structure \citep[top, from][reproduced by permission
    of the AAS]{2005ApJ...627L..81S},
    the other an almost isothermal one \citep[bottom,
      from][]{2003AA...406..323D}. The EM is per unit area in the top panel.}
  \label{fig:em_loci}
\end{figure}}

An interesting debate has focussed on the question whether
the loops observed with TRACE and CDS have a uniform transverse thermal
distribution, i.e., a narrow DEM, or a multi-thermal distribution, i.e., a wide
DEM which may group together warm and hot loops. Although tackled from a different perspective, this question also concerns the fine longitudinal structuring of the loops and of their heating and is therefore strictly connected to the subject of Sections~\ref{sec:obs_fib} and ~\ref{sec:mod_hea}. 
\cite{2003AA...406.1089D} found a loop detected with TRACE to be isothermal (with temperatures below 1~MK) along the line of sight from diagnostics of spectral lines obtained with SoHO/CDS.
\cite{2005ApJ...627L..81S} analyzed another loop on the limb observed with SoHO/CDS, with a DEM
reconstruction and a careful analysis of background subtraction, and found a
multi-thermal distribution across the loop. From the comparison with the isothermal structure of hot loops derived 
from CDS data \citep{2003AA...406..323D,2004ApJ...608.1133L}, 
they concluded that there may be two
different classes of loops, multi-thermal and isothermal, which they found to be
confirmed by a systematic inspection of the CDS atlas (Figure~\ref{fig:em_loci}). 

Multiband observations allow to obtain even more information and constraints.
\cite{2006AA...449.1177R} analyzed a well-defined loop
system detected in a time-resolved observation in several spectral bands, namely three TRACE UV filters, one Yohkoh/SXT filter, two rasters taken with SoHO/CDS in twelve
relevant lines ($5.4 \leq \log T \leq 6.4$). The data
analysis supported a coherent scenario across the different bands and
instruments, indicating a globally cooling loop and the presence of thermal structuring. The study overall indicated that the loop analysis can be easily affected by a variety of instrumental biases and uncertainties, for instance due to gross background subtraction. The fact that the loop that could be well analyzed across several bands and lines is a cooling loop may not be by chance (see end of Section~\ref{sec:obs_dia_tra}). 


Specific analyses of SoHO spectrometric data have continued to contribute much to the study of the loop thermal structure up to recently. 
A Differential Emission Measure (DEM) analysis  of coronal loops using a forward-folding technique on SoHO/CDS data has shown different results for two loops, one to be isothermal and the other to have a broad DEM \citep{2007ApJ...658L.119S}. 
\cite{2008ApJ...672..674L} have analyzed an extensive active region spectrum observed by the SUMER instrument on board SOHO and found that the plasma is made of three
distinct isothermal components, whose physical properties are similar to coronal hole, quiet-Sun, and active region plasmas.

Hinode/EIS observations of active region loops certify that structures which are clearly discernible in cooler lines ($\sim 1$~MK) become fuzzy at higher temperatures \citep[$\sim 2$~MK,][]{2009ApJ...694.1256T},  Figure~\ref{fig:tripathi}),
as already pointed out by \cite{2006ApJ...636L..53B}. 

\epubtkImage{tripathi}{%
\begin{figure}[htbp]
  \centerline{\includegraphics[scale=0.8]{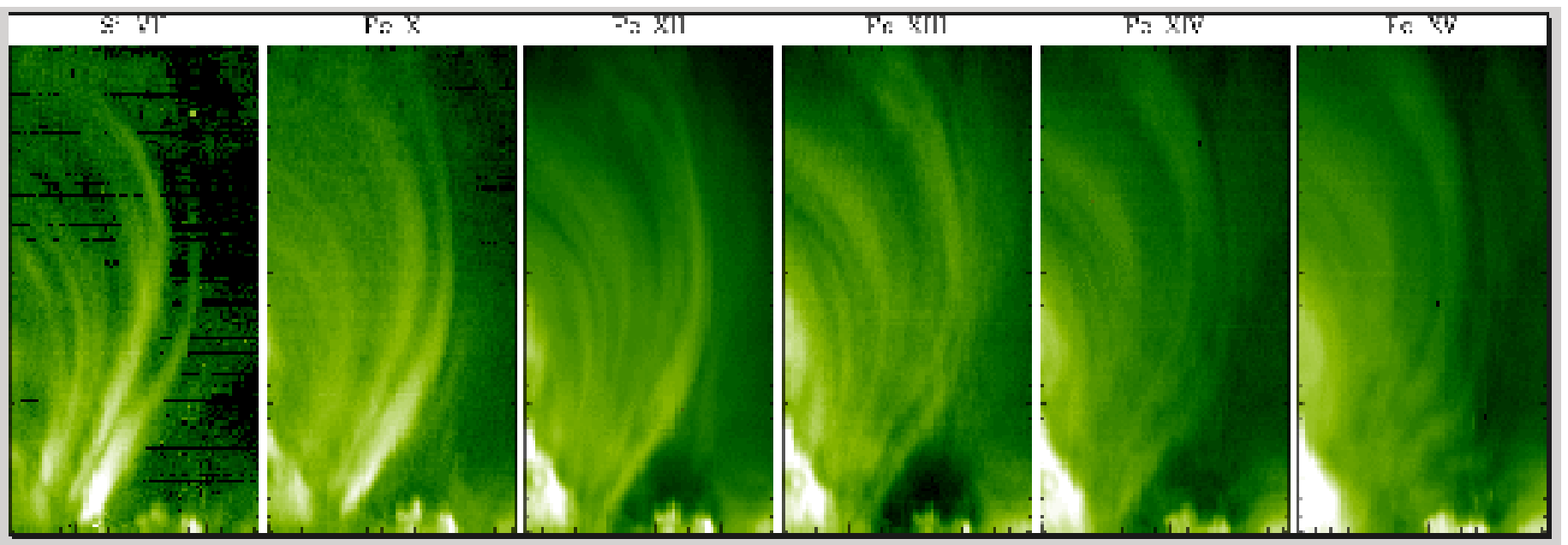}}
  \caption{Loop system observed in several EUV spectral lines with Hinode/EIS (19 May 2007, 11:41-16:35 UT). The loops become less and less contrasted, i.e., fuzzier and fuzzier, at higher and higher temperature (courtesy D.\ Tripathi).}
  \label{fig:tripathi}
\end{figure}}

Comparative studies of active region loops in the transition region and the corona \cite{2009ApJ...695..642U} 
observed with Hinode seem to point out the presence of two  dominant loop populations, i.e., core multitemperature loops that undergo a
continuous process of heating and cooling in the full observed temperature range 0.4\,--\,2.5~MK 
shown by the X-Ray Telescope, and peripheral loops which evolve mostly in the temperature range 0.4\,--\,1.3~MK.

\subsubsection{Warm loops}
\label{sec:obs_dia_tra}


The TRACE mission opened new intriguing questions because the data showed new features, e.g., stranded bright structures mostly localized in active regions, name ``the moss'', and because the narrow band filters offered some limited thermal diagnostics, but not easy to interpret. Reliable temperatures are in fact found in a very narrow range, and many coronal loops are found to be isothermal in that range.

As mentioned in Section~\ref{sec:obs_warm_hot}, first loop diagnostics with normal incidence telescopes were obtained from data collected with NIXT \cite{1994ApJ...422..412P}. 
The bright spots with H$\alpha$ counterparts were identified with the footpoints of high pressure loops, invisible with NIXT because not sensitive to plasma hotter than 1~MK. They have been later addressed as the \textit{moss} in the TRACE images, which undergo the same effect. Interactions of moss with underlying chromospheric structures were first described by \cite{1999SoPh..190..409B}. 
Comparison of SOHO/CDS and TRACE observations led to establish that the plasma responsible for
the moss emission has a temperature range of about 1~MK and is associated with hot loops at 1\,--\,2~MK, with a volume filling factor of order 0.1 \citep{1999ApJ...520L.135F}. It was also found that the path along which the emission originates is of order 1000~km long.
According to an analytical loop model, a filling factor of about 0.1 is in agreement with the hypothesis of moss emission from the legs of 3~MK loops \citep{2000ApJ...537..471M}.

As for temperature diagnostics with narrow band filters,
loops soon appeared to be mostly isothermal with ratios of TRACE filters \citep{1999ApJ...517L.155L,2000ApJ...541.1059A}. Is this a new class of loops?
Equivalent SoHO/EIT filter ratios provided analogous results \citep{1999ApJ...515..842A}.
This evidence is intriguing and many investigations have addressed it (see also Section~\ref{sec:obs_warm_hot}). From the diagnostic perspective, 
\cite{2001ApJ...556..896S} reconstructed DEM distributions along the line of sight from spectral SoHO/CDS data and synthesized EIT count rates from them, which led to almost uniform temperatures along the loop, pointing again to an instrumental bias. 
\cite{2005ApJ...635L.101W} confirmed that, provided they are flat, i.e. top-hat-shaped, even broad DEMs along the line of sight produce constant TRACE filter ratio values. However, we learn from DEM studies made both with spectrometers and from multi-wideband imagers that the DEM of coronal loops is most probably neither isothermal nor broad and flat, instead peaked with components extending both to low and high temperatures \citep[e.g.,][]{2000ApJ...528..537P,2009ApJ...698..756R}. The critical point becomes the DEM width and its range of variation.

Later, 
\cite{2007ApJ...660L.157S} found that even TRACE triple-filter data  cannot, in general, constrain the temperature distribution for plasma in warm loops. On the other hand,
\cite{2007ApJ...667..591P} studied the cross-field thermal structure of a sample of coronal loops from triple-filter TRACE observations, and found that the observations are compatible with multithermal plasma with significant emission measure throughout the range 1\,--\,3~MK.
\cite{2009ApJ...691..503S} used TRACE filter ratios, emission measure loci, and two methods of differential emission measure analysis to examine the temperature structure of three different loops. In agreement with previous studies, they found both isothermal and multithermal cases. This might not be a contradiction, in the view of the presence of at least three possible conditions of warm loops, as discussed at the end of this section.
\cite{2008ApJ...674.1191N} compared TRACE to CDS data to measure the temperature along a coronal loop in an active region on the solar limb. Their double filter ratio temperature analysis technique led to temperatures between 1.0 and 1.3~MK. Emission measure loci from CDS lines were consistent with a line-of-sight isothermal structure which increases in temperature from $\sim$~1.20 to 1.75~MK along the loop, in contrast with the nearby multithermal background.
                                                              
Another puzzling issue, certainly linked to the loop isothermal appearance, is that warm loops are often diagnosed to be overdense with respect to the equilibrium values predicted by loop scaling laws \citep[][Section~\ref{sec:mod_bas_rtv}]{1999ApJ...517L.155L,2003ApJ...587..439W}. To explain both these pieces of evidence, several authors claimed that the loops cannot be at equilibrium and that they must be filamented and cooling from a hotter state, probably continuously subject to heating episodes \citep[nanoflares,][Sections~\ref{sec:mod_fin} and ~\ref{sec:mod_hea}]{2002ApJ...579L..41W,2003ApJ...593.1174W}.
Other authors proposed that part of the effect might be due to inaccurate background subtraction \citep{2003AA...406.1089D}. 


The Hinode mission is stimulating new analyses of warm coronal structures, mostly based on its high quality EIS spectral data. Modeling observations of coronal moss with Hinode/EIS 
confirmed that the moss intensities predicted by steady, uniformly heated loop models are too intense relative to the observations \citep{2008ApJ...677.1395W}. A nonuniform filling factor is required 
and must vary inversely with the loop pressure. Observations of active region loops with EIS indicate
that isolated coronal loops that
are bright in Fe\,{\sc xii} generally have very narrow temperature distributions ($3 \times 10^5$~K), but are not properly isothermal and have a volumetric filling factors of approximately 10\% \citep{2008ApJ...686L.131W}. 

\cite{2008ApJ...684L.115S} studied temperatures of loops identified in a TRACE image in three density-sensitive line ratios. While emission measure loci plots indicated that the loop plasma is not isothermal, a more detailed differential emission measure analysis showed that two broad components can reproduce the background-subtracted intensities. They proposed that the two-component DEM distribution represents two ensembles of strands, one for each of the loops seen in the TRACE image.

Density diagnostics through density-sensitive line ratio also led to measure directly density values, for instance in active regions \citep[e.g.,][]{2007PASJ...59S.707D}. \cite{2008AA...481L..53T} found
that
the hot core of the active region is densest with values as high as $10^{10.5}$~cm$^{-3}$. The electron density estimated in specific regions in the active region moss decreases with increasing temperature. The density within the moss region was highest at $\log T = 5.8{-}6.1$, with a value around $10^{10}$~cm$^{-3}$.

In a cooler regime ($4.15 < \log T < 5.45$) observed in coordination by SOHO spectrometers and imagers, STEREO/EUVI, and Hinode/EIS, active region plasma at the limb has been found to cool down from a coronal hole status with temperatures in the $5.6 < \log T < 5.9$ range \citep{2009ApJ...695..221L}.

The loop reconstruction analysis described in \cite{2009ApJ...695...12A} was used mainly for density and temperature modeling of the warm loops of an active region observed with STEREO. The rendering reduces the problem of background subtraction. Although based on simple model assumptions, the derived density and temperature distributions are able to reproduce the total observed fluxes within 20\%. The modeling extrapolates results quite outside of the range of sensitivity of the STEREO EUV filters, anyhow finding emission measure distributions not very different from those obtained from spectroscopic observations \citep{1996ApJS..106..143B} and deviations from hydrostatic values in agreement with other previous studies \citep{1999ApJ...517L.155L,2003ApJ...587..439W}.

In summary, the current observational framework and loop analysis seems to indicate that for a coherent scenario warm loops are manifestations of at least three different loop conditions: i) in loops consisting of bundles of thin independently-heated strands, few cooling strands of steady hot X-ray loops might be detected as warm overdense loops in the UV band. These warm loops would coexist with hot loops and would show a multithermal emission measure distribution \citep{2007ApJ...667..591P,2008ApJ...686L.131W,2009ApJ...694.1256T}; ii) we might have warm loops as an obvious result of a relatively low average heating input in the loop. These loops would be much less visible in the X-rays and thus would not be cospatial with hot loops, and would also be much less multithermal \citep{2003AA...406..323D,2004ApJ...608.1133L,2005ApJ...633..499A,2008ApJ...674.1191N}; iii) warm loops might be globally cooling from a status of hot X-ray loop \citep{2006AA...449.1177R}. These loops would also be overdense and cospatial with hot loops but with a time shift of the X-ray and UV light curves, i.e., they would be bright in the X-rays before they are in the UV band. Also these loops would have a relatively narrow thermal distribution along the line of sight.


\subsection{Temporal analysis}
\label{sec:obs_tim}


The solar corona is the site of a variety of transient phenomena. Coronal loops sometimes flare in active regions \citep[see the review by][]{2008LRSP....5....1B}. However, most coronal loops are well-known to remain in a steady state for most of their life, much longer than the plasma characteristic cooling times \citep[][see Section~\ref{sec:mod_bas_rtv}]{1978ApJ...220..643R}. This is taken as an indication that a heating mechanism must be on and steady long enough to bring the loop to an equilibrium condition, and keep it there for a long time. Nevertheless, the emission of coronal loops is found to vary significantly on various timescales, and the temporal analyses of coronal loop data have been used to obtain different kinds of information, and as a help to characterize the dynamics and heating mechanisms. The time variability of loop emission is generally not trivial to interpret. The problem is that the emission is very sensitive to density and less to temperature. Therefore, variations are not direct signatures of heating episodes, not even of local compressions, because the plasma is free to flow along the magnetic field lines. Variations must therefore be explained in the light of the evolution of the whole loop. This typically needs accurate modeling, or, at least, care must be paid to many relevant and concurrent effects. 

Another important issue is the band in which we observe. The EUV bands of the normal incidence telescopes are quite narrow. Observations are then more sensitive to variations because cooling or heating plasma is seen to turn on and off rapidly as it crosses the band sensitivity. On the other hand, telescopes in the X-ray band detect hotter plasma which is expected to be more sensitive to heating and therefore to vary more promptly, but the bandwidths are large and do not take as much advantage of the temperature sensitivity as the narrow bands. Finally, spectroscopic observations are, in principle, very sensitive to temperature variations, because they observe single lines, but their time cadence is typically low and able to follow variations only on large timescales. Time analysis studies can be classified to address two main classes of phenomena: temporal variability of steady structures and single transient events, such as flare-like brightenings.

In spite of limited time coverage, the instrument S-054 on-board Skylab already allowed for early studies of variability of hot X-ray loops. Decay times were studied by \cite{1978SoPh...56..107K} 
who found evidence of continued evaporation of coronal plasma in slowly-decaying structures. \cite{1980SoPh...66...79S} and \cite{1985SoPh...98..323H}
found timescales of moderate variability of a few hours over a
substantial steadiness for observations of active region loops in 2~MK 
lines such as Fe\,{\c xv} and Si\,{\sc xii}. Substantial (but non-flaring) temporal variability was reported by \cite{1988ApJS...68..371H} 
in two active region loops observed with SMM in a few relatively hot X-ray lines ($\sim 5$~MK) on time scales of some minutes. Cooler loops ($< 1$~MK) first studied in detail by \cite{1976ApJ...210..575F}
(see Section~\ref{sec:obs_flo}) were found to be more variable and dynamic \citep[e.g.,][]{1985SoPh...98...91K}.



The high time coverage and resolution of Yohkoh triggered studies of brightenings on short time scales.
The Yohkoh/SXT resolution and dynamic range allowed to study the interaction of differently bright hot loops and to show, for instance, that X-ray bright points often involve loops considerably larger than the bright points themselves, and that they vary on timescales from minutes to hours \citep{1992PASJ...44L.161S}.
The analysis of a large set (142) of macroscopic transient X-ray brightenings indicated that they derive from the interaction of multiple loops at their footpoints \citep{1994ApJ...422..906S}.
Some other more specific loop variations were also observed, e.g., the shrinkage of large-scale non-flare loops \citep{1997ApJ...478L..41W}.
This was interpreted not as an apparent motion, but as a real contraction of coronal loops that
brighten due to heating at footpoints followed by gradual cooling.
Fine-scale motions and brightness variations of the emission were found on timescales of 1 minute or less, with dark inclusions
corresponding to jets of chromospheric plasma seen in the wings of H$\alpha$. Such small scale variations are associated with the fine structure and dynamics of the conductively
heated upper transition region between the solar chromosphere and corona \citep{1999ApJ...519L..97B}.

Loop variability was specifically studied in several UV spectral lines observed with SoHO/CDS for about 3 hours by \cite{2003AA...406..323D}. 
In the hottest lines, within the limited time resolution of about 10 min, a few brightenings of a hot loop ($\sim 2$~MK) were detected but they are minor perturbations over a steadily high emission level. The observation of the whole life of a cool loop ($\log T \sim 5.3$) on a time lapse of a few hours confirmed the highly transient nature of cool loops, probably linked to the presence of substantial flows (Section~\ref{sec:obs_flo}).

Variability analyses have been conducted also on warm loops present in TRACE data. The brightening of a single coronal loop was analyzed in detail by \cite{2000ApJ...535..412R} in an observation of more than 2 hours with a cadence of about 30 s. The loop evolves coherently in the rise phase and brightens from the footpoints to the top, allowing for detailed hydrodynamic modeling \citep{2000ApJ...535..423R} (see also Section~\ref{sec:mod_hea}).
Active region transient events  i.e., short-lived brightenings in small-scale loops, were observed with TRACE, with a high cadence of 35 s over half an hour \citep{2001ApJ...563L.173S}. 
Several brightenings detected over a neutral line in a region of emerging flux were interpreted as reconnection
events associated with flux emergence, possible EUV counterparts to active region transient brightenings.
The fast evolution probably implies high speed flows and high coronal densities.
\cite{2002SoPh..206..133S}
noticed apparent shrinking and expansion of brightening warm loops and proposed heating and cooling of different concentric strands, leading to coronal rain visible in the H$\alpha$ line.
Plasma condensations in hot and warm loops were detected also in the analysis of line intensity and velocity in temporal series data from SOHO/CDS \citep{2007AA...475L..25O}.
\cite{2003ApJ...590..547A}
found no significant variability of the moss regions observed with TRACE. This has been taken as part of the evidence toward steady coronal heating in active region cores \citep[][Section~\ref{sec:mod_hea}]{2010ApJ...711..228W}.


The analysis of temporal series from various missions has been used, more recently, to investigate the possible presence of continuous impulsive heating by nanoflares.
The temporal evolution of hot coronal loops was studied in data taken with GOES Solar X-ray Imager (SXI), an instrument with moderate spatial resolution and spectral band similar to Yohkoh/SXT \citep{2007ApJ...657.1127L}. 
The durations and characteristic timescales of the emission rise, steady and decay phases were found to be much longer than the
cooling time and indicate that the loop-averaged heating rate increases slowly, reaches a maintenance level, and then
decreases slowly (Figure~\ref{fig:xlc}), not in contradiction with the early results of Skylab (Section~\ref{section:histor}). This slow evolution is taken as an indication of a single heating mechanism operating for the entire lifetime of the loop. If so, the timescale of the loop-averaged heating rate might be roughly proportional
to the timescale of the observed intensity variations.

\epubtkImage{light_curve_sxi_jim06.png}{%
\begin{figure}[htbp]
  \centerline{\includegraphics[scale=0.5]{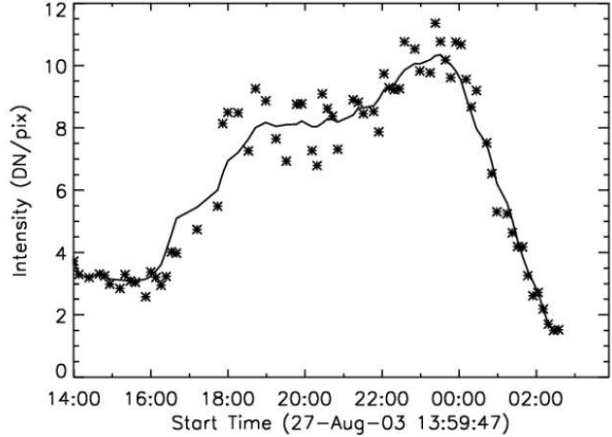}}
  \caption{X-ray light curve observed with the SXI telescope on board
  GOES. The loop lifetime is much longer than the characteristic
  cooling times (courtesy of J.A.\ Klimchuk and  M.C.\ L\`{o}pez
  Fuentes).}
  \label{fig:xlc}
\end{figure}}

Joint TRACE and SOHO/CDS observations allowed to study temperature as a function of time in active region loops \citep{2007ApJ...655..598C}.
In many locations along the loops, the emission measure loci
were found consistent with an isothermal structure, but the results also indicated significant changes in the loop temperature (between 1 and 2~MK) over the 6 hrs observing period. This was interpreted as one more indication of multistranded loops, substructured below the resolution of the imager and of the spectrometer.
Further support to fine structuring comes from the analysis of the auto-correlation
functions in SXT and TRACE loop observations \citep{2008ApJ...689.1421S}.
The duration of the intensity fluctuations for the hot SXT loops was found to be relatively short because of the significant photon noise, but that for the warm TRACE loops agrees well with the characteristic cooling timescale. This may support loops to be continuously heated by impulsive
nanoflares. The energy of nanoflares is estimated to be $10^{25}$~erg for SXT loops and $10^{23}$~erg for TRACE loops. The
occurrence rate of nanoflares is about 0.4 and 30 nanoflares $s^{-1}$ in a typical hot SXT loop and a typical warm TRACE loop, respectively.

A recent study on time series has been performed on data taken with the Hinode mission.
Hinode's Solar Optical Telescope (SOT) magnetograms and high-cadence EIS spectral data were taken to study the relationship between chromospheric, transition region, and coronal emission and the evolution of the magnetic field \citep{2008ApJ...689L..77B}.
The data have allowed to distinguish hot, relatively steadily
emitting warm coronal loops from isolated transient brightenings and to find that they are both associated with highly dynamic magnetic flux regions. Brightenings have been typically found in regions of flux
collision and cancellation, while warm loops are generally rooted in magnetic field regions that are locally
unipolar with unmixed flux. The authors suggest that the type of heating (transient vs.\ steady) is related to the structure of the magnetic field, and that the heating in transient events may be fundamentally different from that in warm coronal loops.

\subsection{Flows}
\label{sec:obs_flo}


Diagnosing the presence of significant flows in coronal loops is not an easy
task. Apparently moving brightness variations may not be a conclusive 
evidence of plasma motion, since the same effect may be produced by the
propagation of thermal fronts or waves. Conclusive evidence of plasma motion comes from
measurements of Doppler shifts in relevant spectral lines. However, the detection of
significant Doppler shifts requires several conditions to be fulfilled
altogether, e.g., significant component along the line of sight, 
amount of moving plasma larger than amount of static plasma, plasma motion
comparable to typical line broadening effects.

In general, we can distinguish two main classes of mass bulk motions inside coronal loops: siphon flows, due to a pressure difference between the footpoints, and loop filling or draining, due to transient heating and subsequent cooling, respectively. Some other evidence of bulk motions, such as systematic redshifts in UV lines, has been difficult to interpret.

Siphon flows have been mainly invoked to explain motions in cool loops.
The existence of cold loops has been known for a long time \citep{1976ApJ...210..575F} (see Section~\ref{section:histor})
and SoHO has collected
high-quality data showing the presence of dynamic cool loops \citep{1997SoPh..175..511B}. A well-identified detection was found in SoHO/SUMER data, i.e., a small loop showing a supersonic siphon-like flow  \citep{2004AA...427.1065T} and in SoHO/CDS data \citep{2003AA...406..323D}.

Redshifts in transition region UV lines have been extensively observed  on the solar disk \citep[e.g.,][]{1976ApJ...205L.177D,1981ApJ...251L.115G,1982SoPh...77...77D,1982ApJ...255..325F,1987ApJ...323..368K,1990ApJ...358..693R,1993ApJ...408..735B,1999ApJ...516..490P}.
Some mechanisms have been proposed to explain these redshifts: downward  propagating acoustic waves \citep{1993ApJ...402..741H},
downdrafts driven by radiatively-cooling condensations in the solar transition region 
\citep{1996AA...316..215R,1997AA...318..506R}, nanoflares \citep{1999AA...352L..99T};
a conclusive word is still to be given with the better and better definition of the observational framework.

Blue-shifts in the transition region are also studied but not necessarily associated with coronal loops \citep[e.g.,][]{1986ApJ...310..456D}. More localized and transient episodes of high velocity outflows, named explosive events, have been observed in the transition lines such as C\,{\sc iv}, formed at 100,000~K \citep[e.g.,][]{1989SoPh..123...41D,1998ApJ...504L.123C,1999ApJ...526..471W,2002ApJ...565.1298W,2002ApJ...570L.105W}.
However, \cite{2002AA...392..309T} found indications that such EUV explosive events are not directly relevant in heating the corona, are characteristic of structures not obviously connected with the upper corona, and have a chromospheric origin.

\cite{1998ApJ...505..957C} found Doppler shifts increasing and then decreasing with increasing temperature and explained them by the dominance of emission from plasma flowing downward from the upper hot region to the lower cool region along flux tubes with varying cross section (a factor about 30). \cite{1999AA...349..636T}
confirmed these results with the exception of blueshifts at higher temperature.

EUV spectra of coronal loops above an active region show clear
evidence of strong dynamical activity. In the O\,{\sc v}~629~\AA\ line, formed at 240,000~K, line-of-sight velocities greater than 50~km s$^{-1}$ have been measured with the shift extending over a large fraction of a loop \citep{1997SoPh..175..511B}.
Active region loop structures appear to be extremely time variable and dynamic at transition region temperatures, with large Doppler-shifts \citep{1999SoPh..190..379B}.
\cite{2003AA...406..323D} showed the direct observation and identification of the
birth, evolution and cooling of one of such transient cool loops, and measured a blue-shifted up-flow all along the loop, probably a one-direction siphon flow.

\cite{2002ApJ...567L..89W} analyzed   co-aligned TRACE and the SoHO/SUMER observations
of warm active region loops. Although these loops appear static in the TRACE images, SUMER detects line-of-sight flows along the loops of up to 40~km s$^{-1}$. Apparent motions were also detected in TRACE images \citep{2001ApJ...553L..81W}.

In SoHO/EIT high-cadence 304~\AA\ images, 
\cite{2004AA...415.1141D} analyzed systematic intensity variations along an off-limb half loop structure propagating from the top towards the footpoint and reported several arguments supporting that these intensity variations are due to flowing/falling plasma blobs and not to slow magneto-acoustic waves (Section~\ref{sec:mod_hea}). This evidence has been the object of modeling studies (see Section~\ref{sec:mod_flo}).

\epubtkImage{helen_flows.png}{%
\begin{figure}[htbp]
  \centerline{\includegraphics[scale=0.75]{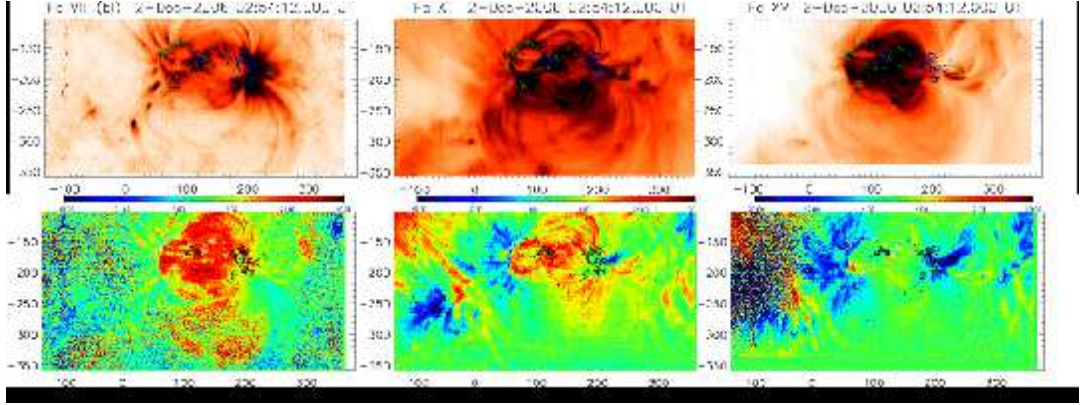}}
  \caption{Monochromatic (negative) images and dopplergrams (km
    $s^{-1}$) of an active region (NOAA 10926) observed with
    Hinode/EIS in Fe\,{\sc viii}, Fe\,{\sc xii}, Fe\,{\sc xv}
    lines. (Courtesy G.\ Del Zanna)}
  \label{fig:flows}
\end{figure}}

%

The high spectral resolution of Hinode/EIS is allowing for very detailed studies of persistent loop plasma motions, which are very important to assess whether the loops are to be treated more as static or dynamic structures as a whole.
More specifically, nonthermal velocities were detected in solar active regions \citep{2007ApJ...667L.109D}.
The largest widths seem to be located more in relatively faint zones, some of which also show Doppler outflows. Doppler flows in active region loops observed by Hinode EIS were explicitly addressed by \cite{2008AA...481L..49D}, 
who found a multifaceted scenario (Figure~\ref{fig:flows}). Persistent redshifts, stronger in
cooler lines (about 5\,--\,10~km s$^{-1}$ in Fe\,{\sc xii} and 20\,--\,30~km
s$^{-1}$ in Fe\,{\sc viii}), were observed in most loop
structures. Persistent blueshifts, stronger in the hotter lines
(typically 5\,--\,20~km s$^{-1}$ in Fe\,{\sc xii} and 10\,--\,30~km
s$^{-1}$ in Fe\,{\sc xv}), were present in areas of weak emission, in a sharp boundary between the low-lying ``hot'' 3~MK loops and the higher ``warm'' 1~MK loops.
Strong localized outflows ($\sim$~50~km/s) in a widespread downflow region were clearly visible in Doppler-shifts maps obtained with EIS \citep{2008ApJ...686.1362D}. The outflows might be tracers of long loops and/or open magnetic fields.

\cite{2008AA...482L...9O} used the high resolution Hinode SOT observations, and detected cool plasma flowing in multi-threaded coronal loops with speeds in the range 74-123 km/s. In addition to flows, the loops exhibited transverse oscillations.

Further analysis of coronal plasma motions near footpoints of active region loops showed again a strong correlation between Doppler velocity and nonthermal velocity \citep{2008ApJ...678L..67H}.
Significant deviations from a single Gaussian profile were found in the blue wing of the line profiles for the upflows. These may suggest that there are unresolved high-speed upflows.
\cite{2009ApJ...694.1256T} found that an active region was comprised of redshifted emissions (downflows) in the core and blueshifted emissions (upflows) at the boundary. All these results have to be matched with the recent finding of extensive blueshifts correlated with spicules upflows and with coronal emission intensity \citep{2009ApJ...701L...1D} (Section~\ref{sec:mod_hea}).

\newpage


\section{Loop Physics and Modeling}
\label{sec:mod}

\subsection{Basics}
\label{sec:mod_bas}


The basics of loop plasma physics are well established since the 1970s \citep[e.g.,][]{1978SoPh...58...57P}.
In typical coronal conditions, i.e., ratio of thermal and magnetic pressure $\beta \ll 1$, temperature of a few MK, density of $10^8 - 10^{10}$ cm$^{-3}$,  the plasma confined in coronal loops can be assumed as a compressible
fluid moving and transporting energy only along the magnetic field lines,
i.e., along the loop itself \citep[e.g.,][]{1978ApJ...220..643R,1979ApJ...233..987V}.
In this configuration, the magnetic field has only the role of
confining the plasma. It is also customary to assume constant loop 
cross-section (see Section~\ref{sec:obs_mor_mag}). In these conditions, neglecting gradients across the direction of the field, effects of curvature, and transverse waves, the plasma
evolution can be described by means of the one-dimensional hydrodynamic 
equations for a compressible fluid, using only the coordinate along the loop
(Figure~\ref{fig:loop_model}).

\epubtkImage{loop_model}{%
\begin{figure}[htbp]
  \centerline{\includegraphics[scale=0.6]{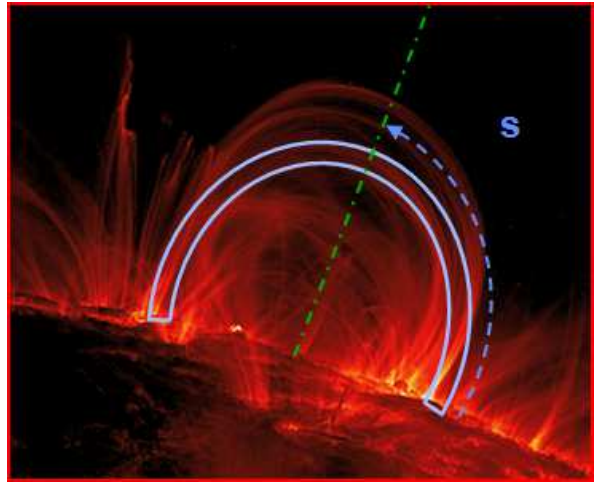}}
  \caption{The plasma confined in a loop can be described with
  one-dimensional hydrodynamic modeling, with a single coordinate ($s$)
  along the loop (image: TRACE, 6 November 1999, 2 UT).}
  \label{fig:loop_model}
\end{figure}}

The time-dependent equations of mass, momentum and energy conservation 
typically include the effects of the gravity component
along the loop, the radiative losses from an optically thin plasma, the
plasma thermal conduction, an external heating input, the plasma compressional viscosity:
 
\begin{equation}
 \label{eq:mass}
    \frac{dn}{dt}= -n \frac{\partial v}{\partial s},
\end{equation}
\begin{equation}
 \label{eq:mom}
    n m_{\rm H} \frac{dv}{dt} = - \frac{\partial p}{\partial s}
    + n m_{\rm H} g + \frac{\partial}{\partial s}
    (\mu \frac{\partial v}{\partial s}),
\end{equation}
\begin{equation}
 \label{eq:en}
    \frac{d\epsilon}{dt} +(p+\epsilon) \frac{\partial v}{\partial s} =
    H - n^2 \beta_i P(T) + \mu \left( \frac{\partial v}{\partial s} \right)^2
    + F_c,
\end{equation}
with $p$ and $\epsilon$ defined by:
\begin{equation}
 \label{eqm:pressener}
    p = (1+\beta_i) n k_{\rm B} T \hspace{1cm}
            \epsilon =\frac{3}{2} p + n \beta_i \chi,
\end{equation}
and the conductive flux:
\begin{equation}
F_c = \frac{\partial}{\partial s} \left( 
	    \kappa T^{5/2} \frac{\partial T}{\partial s} \right)
\end{equation}
where $n$ is the hydrogen number density, $s$ the spatial coordinate
along the loop, $v$ the plasma velocity, $m_{\rm H}$ the mass of
hydrogen atom, $\mu$ the effective plasma viscosity, $P(T)$ the
radiative losses function per unit emission measure \citep[e.g.,][]{1976ApJ...204..290R},
$\beta_i$ the
fractional ionization, i.e.,\ $n_{\rm e}/n_{\rm H}$, $F_c$ the conductive flux,
$\kappa$ the
thermal conductivity \citep{1962pfig.book.....S},
$k_{\rm B}$ the Boltzmann
constant, and $\chi$ the hydrogen ionization potential.  $H(s,t)$ is
a function of both space and time which describes the heat input in the
loop.

These equations can be solved numerically and several specific codes have been used
extensively to investigate the physics of coronal loops and of X-ray
flares \citep[e.g.,][]{1980SoPh...68..351N,1982ApJ...252..791P,1982ApJ...258..373D,1984ApJ...279..896N,1985ApJ...289..414Fv,1985ApJ...289..414Fvi,1985ApJ...289..414Fvii,1986SoPh..103...47M,1991AA...241..618G,1993ApJ...402..741H,1997AAS..122..585B,1999ApJ...512..985A,2003AA...411..605M,2002ApJ...580L..85O,2003AA...407.1127B,2006AA...458..987B}.

The concept of numerical loop modeling is to use simulations, first of all, to get insight into the physics of coronal loops, i.e., the reaction of confined plasma to external drivers, to describe plasma evolution, and to derive predictions to compare with observations. One major target of modeling is of course to discriminate between concurrent hypotheses, for instance, regarding the heating mechanisms, and to constrain the related parameters.

The models require to be provided with initial loop conditions and boundary conditions. In view of our ignorance of the specific heating mechanisms (see Section~\ref{sec:mod_hea}), the models require to define an input heating function, specifying its time-dependence, for instance it can be steady, slowly or impulsively changing, and its position in space. The output typically consists of distributions of temperature, density and velocity along the loop evolving with time. From simulation results, some modelers derive observables, i.e., the plasma emission, which can be compared directly to data collected with the telescopes. The model results are, in this case, to be folded with the instrumental response. This forward-modeling allows to obtain constraints on model parameters and, therefore, quantitative information about the questions to be solved, e.g., the heating rate and location \citep[e.g.,][]{2000ApJ...535..423R}. 

Loop codes are typically based on finite difference numerical methods. Although they are one-dimensional, and therefore typically less demanding than other multi-dimensional codes that study systems with more complex geometry, and although they do not include the explicit description of the magnetic field, as full MHD codes, loop codes require some special care. One of the main difficulties consists in the appropriate resolution of the steep transition region (1\,--\,100~km thick) between the chromosphere and the corona, which can easily drift up and down depending on the dynamics of the event to be simulated. The temperature gradient there is very large due to the local balance between the steep temperature dependence of the thermal conduction and the peak of the radiative losses function \citep{1981ApJ...243..288S}. The density is steep as well so to maintain the pressure balance. The transition region can become very narrow during flares.

Also a fine temporal resolution is extremely important, because the highly efficient thermal conduction in a hot magnetized plasma can lead to a very small time step and make execution times not so small even nowadays. Another important issue is the necessary presence of a relatively thick, cool and dense region under the transition region, i.e., a chromosphere, otherwise the atmosphere gets unstable. The main role of the chromosphere in this context is only that of a mass reservoir, and therefore, in several codes, it is chosen to treat it as simply as possible, e.g., an isothermal inactive layer which neither emits, nor conducts heat. In other cases, a more accurate description is chosen, e.g., to include a detailed model \citep[e.g.,][]{1981ApJS...45..635V}
and to maintain a simplified radiative emission and a detailed energy balance with an \textit{ad hoc} heat input \citep{1982ApJ...252..791P,2000ApJ...535..423R}. 

In recent years, time-dependent loop modeling has revived in the light of the observations with SoHO and TRACE for the investigation of the loop dynamics and heating. 
The upgrade driven by the higher quality of the data has consisted in the introduction of more detailed mechanisms for the heating input, for the momentum deposition, or others, e.g., the time-dependent ionization and the saturated thermal conduction \citep{2006AA...458..987B,2008ApJ...684..715R}. Some codes have been upgraded to include adaptive mesh refinement for better resolution in regions of high gradients, such as in the transition region, or during impulsive events \citep[e.g.,][]{1997AAS..122..585B}. Another form of improvement has been the description of loops as collections of thin strands. Each strand is a self-standing, isolated and independent atmosphere, to be treated exactly as a single loop. This approach has been adopted both to describe loops as static \citep{2000ApJ...528L..45R} (Figure~\ref{fig:reale00})
and as impulsively heated by nanoflares \citep{2002ApJ...579L..41W}.
On the same line, collections of loop models have been applied to describe entire active regions \citep{2006ApJ...645..711W}.

\epubtkImage{reale00.png}{%
\begin{figure}[htbp]
  \centerline{\includegraphics[scale=0.6]{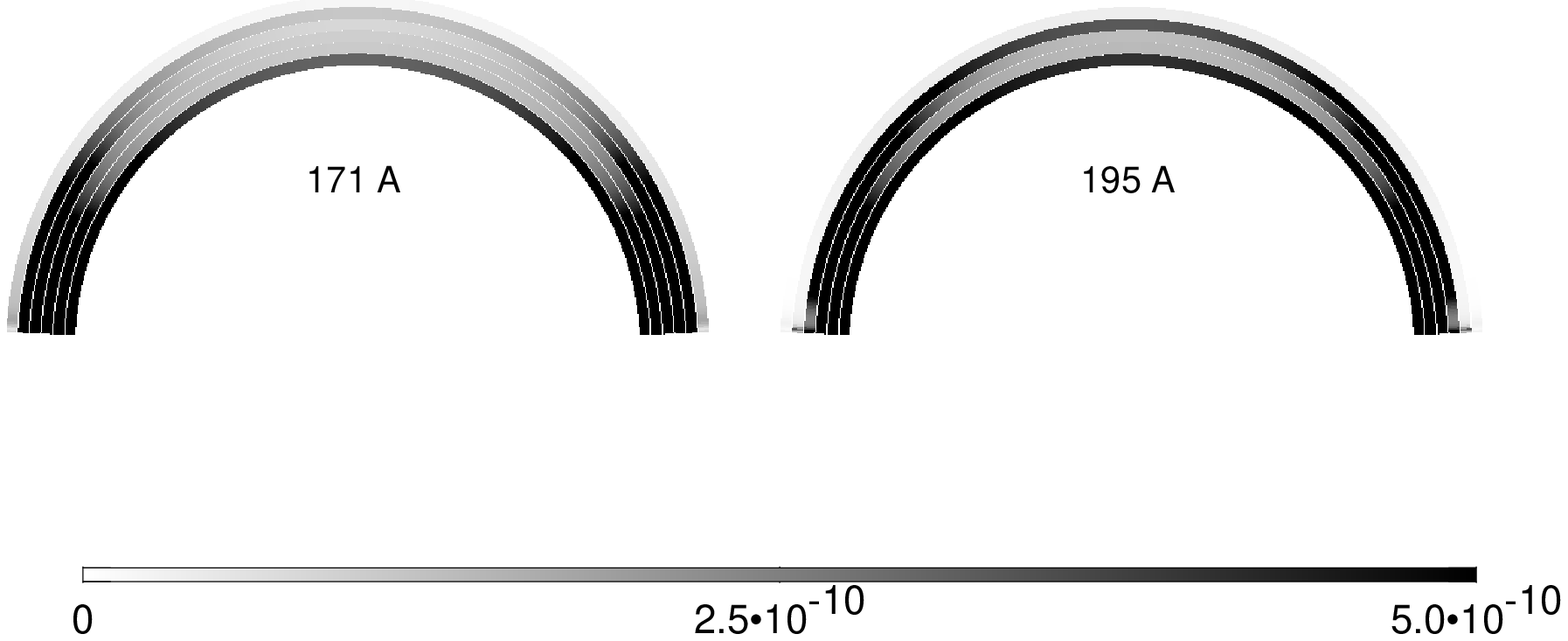}}
  \caption{Emission in two TRACE filterbands predicted by a model of
  loop made by several thin strands \citep[from][]{2000ApJ...528L..45R}.}
  \label{fig:reale00}
\end{figure}}

One limitation of current 1D loop models is that they are unable to treat conveniently the tapering expected going down from the corona to the chromosphere (or expansion upwards) through the transition region. This effect can be neglected in many circumstances, but it is becoming  increasingly important with the finer and finer level of diagnostics allowed by upcoming observational data. For instance, the presence of tapering changes considerably the predicted distribution of emission measure in the low temperature region (Section \ref{sec:mod_bas_rtv}).

Possible deviations from pure 1D evolution might be driven by intense oscillations or kinks, as described in \cite{2009ApJ...694..502O}. The effect of the three-dimensional loop structure should then be taken into account to describe the interaction with excited MHD waves \citep{2008ApJ...682.1338M,2009AA...505..319P,2009AnGeo..27.3899S}.

However, the real power of 1D loop models, that makes them still on the edge, is that they fully exploit the property of the confined plasma to evolve as a fluid and practically independently of the magnetic field, and that they can include the coronal part, the transition  region, and the photospheric footpoint in a single model with thermal conduction. In this framework, we may even simulate a multi-thread structure only by collecting many single loop models together, still with no need to include the description and interaction with the magnetic field. We should however be aware that the magnetic confinement of the loop material is not as strong and the thermal conduction is not as anisotropic below the coronal part of the loop as it is in the corona. 

\cite{2008ApJ...682.1351K} illustrate a new efficient model of dynamic coronal loops called ``Enthalpy-Based Thermal Evolution of Loops'' (EBTEL), which accurately describes the evolution of the average temperature, pressure, and density along a coronal strand with a ``0-D'', very fast approach. This model is particularly useful for the description of loops as collections of myriads of strands. In more detailed modeling, it has been recently shown that non-local thermal conduction may lengthen considerably the conduction cooling times and may enhance the chances of observing hot nanoflare-heated plasma \citep{2008SoPh..252...89W}.

Alternative approaches to single or multiple loop modeling have been
developed more recently, thanks also to the increasing availability of
high performance computing systems and resources. A global ``ab
initio'' approach was presented by \cite{2005ApJ...618.1020G} and by
\cite{2007ASPC..368..107H} \citep[see also][for the case of a flare model]{2001ApJ...549.1160Y}.
They model a small part of the solar corona in a computational box using a three-dimensional MHD code that span the entire solar atmosphere from the upper convection zone to the lower corona. These models include non-grey, non-LTE (Local Thermodynamic Equilibrium) radiative transport in the photosphere and chromosphere, optically thin radiative losses, as well as magnetic field-aligned heat conduction in the transition region and corona. Although such models still cannot resolve well fine structures, such as current sheets and the transition region, they certainly represent the first important step toward fully self-consistent modeling of the magnetized corona.

\subsubsection{Monolithic (static) loops: scaling laws}
\label{sec:mod_bas_rtv}


The Skylab mission remarked, and later missions confirmed
(Figure~\ref{fig:xlc}), that many X-ray emitting coronal loops persist
mostly unchanged for a time considerably longer than their cooling
times by radiation and/or thermal conduction \citep[][and references therein]{1978ApJ...220..643R}. This means that, for most of their lives, they can be well described as systems at equilibrium and has been the starting point for several early theoretical studies \citep{1975AA....42..213L,1976RSPTA.281..339G,1976RSPTA.281..391J,1979ApJ...233..987V,1980AA....86..355J}.
\cite{1978ApJ...220..643R} devised a model of coronal loops in hydrostatic equilibrium with several realistic simplifying assumptions: symmetry with respect to the apex, constant cross section (see Section~\ref{sec:obs_mor_mag}), length much shorter than the pressure scale height, heat deposited uniformly along the loop, low thermal flux at the base of the transition region, i.e., the lower boundary of the model. In these conditions, the pressure is uniform all along the loop, which is then described only by the energy balance between the heat input and the two main losses mentioned above. From the integration of the equation of energy conservation, one obtains the well-known scaling laws:

\begin{equation}
 \label{eq:rtv1}
    T_{0,6} = 1.4 \left( p L_9 \right)^{1/3} 
\end{equation}
and
\begin{equation}
 \label{eq:rtv2}
    H_{-3}  = 3 p^{7/6}  L_9^{-5/6} 
\end{equation}
where $T_{0,6}$, $L_9$ and $H_{-3}$ are the loop maximum temperature
$T_0$, length $L$ and heating rate per unit volume $H$, measured in
units of $10^6$~K (MK), $10^9$~cm and $10^{-3} $~erg cm$^{-3} $ s$^{-1} $ respectively. These scaling laws were found in agreement with Skylab data within a factor 2.

Analogous models were developed in the same framework \citep{1975AA....42..213L}
and equivalent scaling laws were found independently by \cite{1978AA....70....1C}
and more general ones by \cite{1979AA....77..233H}.
They have been derived with a more general formalism by \cite{1991plsc.book.....B}.
Although scaling laws could explain several observed properties, some features such as the emission measure in UV lines and the cool loops above sunspots could not be reproduced, and, although the laws have been questioned a number of times \citep[e.g.,][]{1995ApJ...454..934K}
in front of the acquisition of new data, such as those by Yohkoh and TRACE, they anyhow provide a basic physical reference frame to interpret any loop feature. For instance, they provide reference equilibrium values even for studies of transient coronal events, they have allowed to constrain that many loop structures observed with TRACE are overdense \citep[e.g.,][Section\ref{sec:mod_bas_tsc}]{1999ApJ...517L.155L,2003ApJ...587..439W}
and, as such, these loops must be cooling from hotter status \citep{2005ApJ...626..543W} (see Section~\ref{sec:obs_dia_tra},
and so on. They also are useful for density estimates when closed with the equation of state, and for coronal energy budget when integrated on relevant volumes and times.

Scaling laws have been extended to loops higher than the pressure scale height \citep{1981ApJ...243..288S}
and limited by the finding that very long loops become unstable \citep{1981SoPh...70..293W}.
According to \cite{1986ApJ...301..440A},
the cool loops belong to a different family and are low-lying, and may eventually explain an evidence of excess of emission measure at low temperature. 

The numerical solution of the complete set of hydrostatic equations allowed to obtain detailed profiles of the physical quantities along the loop, including the steep transition region. Figure~\ref{fig:stat} shows two examples of solution for different values of heating uniformly distributed along the loop.

\epubtkImage{loop_stat.png}{%
\begin{figure}[htbp]
  \centerline{\includegraphics[scale=0.6]{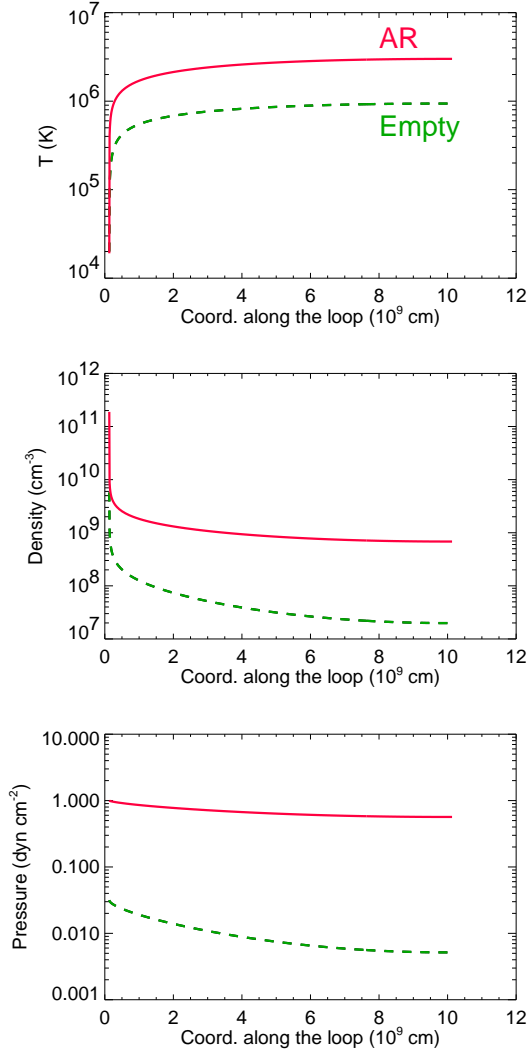}}
  \caption{Distributions of temperature, density and pressure along a
  hydrostatic loop computed from the model of
  \cite{1981ApJ...243..288S} for a high pressure loop (AR) and a low
  pressure one (Empty) with heating uniformly distributed along the loop.}
  \label{fig:stat}
\end{figure}}

\cite{1999SoPh..190..139R} and \cite{2000ApJ...535L..59A} investigated in detail the effect of hydrostatic weighting on the loop visibility and on the vertical temperature structure of the solar corona.
From the comparison of SOHO-CDS observations of active region loops with a static, isobaric loop model \citep{2002AA...383..653L}, \cite{2002AA...383..661B}
showed that a classical model is not able to reproduce the observations, but ad hoc assumptions might be needed. Further improvements of this approach including flows did not improve the agreement between the model and the observations \citep{2004ApJ...608.1133L}.
Using static models, 
\cite{2004ApJ...611..537L} found that the loop models overestimate the footpoint emission by orders of magnitude and proposed that non-uniformity in the loop cross section, more specifically a significant decrease of the cross section near the footpoints, is the most likely solution to the discrepancy (Section \ref{sec:mod_bas}).
On the same line, \cite{2008ApJ...676..672W} modelled X-ray loops and EUV moss in an active region core with steady uniform heating and found that a filling factor of 8\%
and loops that expand with height provide the best agreement with the intensity in two X-ray filters, though maintaining still some discrepancies with observations.
\cite{2008AA...489..441G} studied the distribution of coronal heating in a solar active region using a simple electrodynamic
model and attributed the observed small coronal-loop width expansion to both the preferential heating of
coronal loops of small cross-section variation, and the cross-section confinement due to the random electric currents flowing along the loops.

The strength of scaling laws is certainly their simplicity and their easy and general application, even in the wider realm of stellar coronae. However, increasing evidence of dynamically heated, fine structured loops is indicating the need for improvements.
 

\subsubsection{Structured (dynamic) loops}
\label{sec:mod_bas_tsc}


In the scenario of loops consisting of bundles of thin strands, each strand behaves as an independent atmosphere and can be described as an isolated loop itself. If the strands are numerous and heated independently, a loop can be globally maintained steady with a sequence of short heat pulses, each igniting a single or a few strands (nanoflares). In this case, although the loop remains steady on average for a long time, each strand has a continuously dynamic evolution. The evolution of a loop structure under the effect of an impulsive heating is well-known and studied from observations and from modeling \citep[e.g.,][]{1980SoPh...68..351N,1982ApJ...252..791P,1983ApJ...265.1090C,
1984ApJ...279..896N,1985ApJ...289..414Fv,1985ApJ...289..414Fvi,1985ApJ...289..414Fvii,1986SoPh..103...47M,2001AA...380..341B},
since it resembles the evolution of single coronal flaring loops. We have to mention here that there have been attempts to model even flaring loops as consisting of several flaring strands \citep{1997ApJ...489..426H,1998ApJ...500..492H,2002ApJ...578..590R,2006ApJ...637..522W}.

The evolution of single coronal loops or single loop strands subject to impulsive heating has been recently summarized in the context of the diagnostics of stellar flares \citep{2007AA...471..271R}.
A heat pulse injected in an inactive tenuous strand makes chromospheric plasma expand in the coronal section of the strand, and become hot and dense, X-ray bright, coronal plasma. After the end of the heat pulse, the plasma begins to cool slowly. In general, the plasma cooling is governed by the thermal conduction to the
cool chromosphere and by radiation from optically thin conditions.
In the following we outline the evolution of the confined heated plasma into four phases, according to \cite{2007AA...471..271R}. Figure~\ref{fig:ntx_time} tracks this evolution which maps on the path drawn in
the density-temperature diagram of Figure~\ref{fig:nt_diag_eq}
\citep[see also][]{1992AA...253..269J}.

\epubtkImage{reale07_fig1.png}{%
\begin{figure}[htbp]
  \centerline{\includegraphics[scale=0.4]{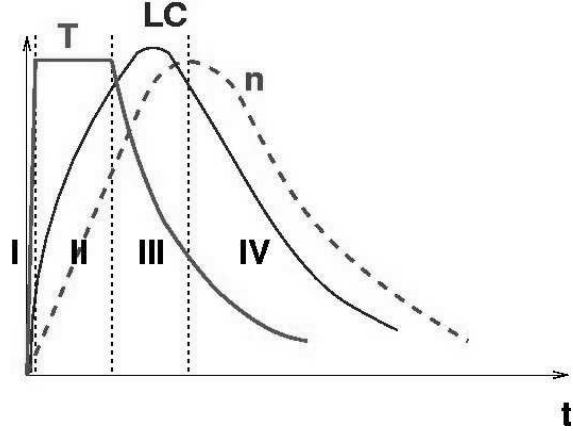}}
  \caption{Scheme of the evolution of temperature (T, thick solid
    line), X-ray emission, i.e. the light curve (LC, thinner solid line) and density (n,
    dashed line) in a loop strand ignited by a heat pulse. The strand
    evolution is divided into four phases (I, II, III, IV, see text
    for further details). From \cite{2007AA...471..271R}.}
  \label{fig:ntx_time}
\end{figure}}

\begin{description}

\item[Phase I]: from the start of the heat pulse to the temperature peak
(\textit{heating}). If the heat pulse is triggered in the coronal part of the loop, the heat is
efficiently conducted down to the much cooler and denser chromosphere.
The temperature rapidly increases in the whole loop, with a time scale
given by the conduction time in a low density plasma (see below). This evolution changes only slightly if the heat pulse is deposited near the loop footpoints: the conduction front then propagates mainly upwards and on timescales not very different from the evaporation time scales, also because the heat conduction saturates \citep[e.g.,][]{2006SoPh..234...41K,2008ApJ...684..715R}.
In this case the distinction from Phase II is not clearly marked.

\item[Phase II]: from the temperature peak to the end of the heat
pulse (\textit{evaporation}). The temperature settles to the maximum value ($T_0$).
The chromospheric plasma is strongly
heated, expands upwards, and fills the loop with much denser plasma. This occurs both if the heating is conducted from the highest parts of the corona and if it released directly near the loop footpoints.
The evaporation is explosive at first, with a timescale given by the
isothermal sound crossing time ($s$), since the temperature is approximately uniform in the highly conductive corona:

\begin{equation}
\tau_{sd} = \frac{L} {\sqrt{2 k_B T_0 / m}} \approx 80 \frac{L_9}{ \sqrt{T_{0,6}}}
\label{eq:tsound}
\end{equation}
where $m$ is the average particle mass.
After the evaporation front has reached the loop apex, the loop 
continues to fill more gently. The time scale during this
more gradual evaporation is dictated by the time taken by the cooling
rate to balance the heat input rate.

\item[Phase III]: from the end of the heat pulse to the density peak
(\textit{conductive cooling}).  When the heat pulse stops, the plasma
immediately starts to cool due to the efficient thermal conduction
\citep[e.g.,][]{2004ApJ...605..911C},
with a time scale ($s$):

\begin{equation}
\tau_c=\frac{3 n_c k_B T_0 L^2}{2/7\kappa T_0^{7/2}}
= \frac{10.5 n_c k_B L^2}{\kappa T_0^{5/2}} \approx 1500 \frac{n_{9} L_9^2}{T_{6}^{5/2}}
\label{eq:tcon}
\end{equation}
where $n_c$ ($n_{c,10}$) is the particle density ($10^{10}$~cm$^{-3}$) at
the end of the heat pulse, the thermal conductivity is
$\kappa = 9 \times 10^{-7}$ (c.g.s. units). Since the plasma is dense, we expect no saturation effects in this phase.

The heat stop time can be generally traced as the time at which the
temperature begins to decrease significantly and monotonically. While
the conduction cooling dominates, the plasma evaporation is still going on and the density increasing.  The efficiency of radiation cooling increases as well, while the efficiency of conduction cooling decreases with the temperature.

\item[Phase IV]: from the density peak afterwards (\textit{Radiative
cooling}).  As soon as the radiation cooling time becomes equal to the
conduction cooling time \citep{2004ApJ...605..911C},
the density reaches
its maximum, and the loop depletion starts, slowly at first and then
progressively faster. The pressure begins to decrease inside the loop,
and is no longer able to sustain the plasma. The radiation
becomes the dominant cooling mechanism, with the following time scale ($s$):

\begin{equation}
\tau_r = \frac{3 k_B T_M}{n_M P(T)} =  \frac{3 k_B T_M}{b T_M^\alpha n_M} \approx
3000 \frac{T_{M,6}^{3/2}} {n_{M,9}}
\label{eq:trad}
\end{equation}
where $T_M$ ($T_{M,6}$) is the temperature at the time of the density
maximum 
($10^7$~K), $n_M$ ($n_{M,9}$) the maximum density ($10^{9}$~cm$^{-3}$),
$P(T)$ the plasma emissivity per unit emission measure, expressed as:
\[
P(T) =  b T^\alpha
\]
with $b = 1.5 \times 10^{-19}$ and $\alpha = -1/2$. The density and the temperature both decrease monotonically. 

The presence of significant residual heating
could make the decay slower. In single loops, this can be diagnosed from the analysis of the
slope of the decay path in the density-temperature diagram \citep{1993AA...267..586S,1997AA...325..782R}. The free decay has a slope between 1.5 and 2 in a $\log$ density vs $\log$ temperature diagram; heated decay path is flatter down to a slope $\sim 0.5$.
In non-flaring loops, the effect of residual heating can be mimicked by the effect of a strong gravity component, as in long loops perpendicular to the solar surface. The dependence of the decay slope on the pressure scale height has been first studied in \cite{1993AA...272..486R}, and, more recently, in terms of enthalpy flux by \cite{2010ApJ...717..163B}. 

As clear from Fig.~\ref{fig:nt_diag_eq} the path in this phase is totally below, or at most approaches, the QSS curve. This means that for a given temperature value the plasma density is higher than that expected for an equilibrium loop at that temperature, i.e., the plasma is ``overdense". Evidence of such overdensity (Section~\ref{sec:obs_dia_tra}) has been taken as an important indication of steadily pulse-heated loops.

\end{description}

\epubtkImage{reale07_fig2.png}{%
\begin{figure}[htbp]
  \centerline{\includegraphics[scale=0.7]{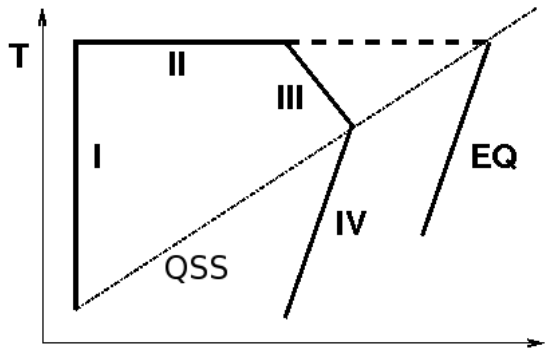}}
  \caption{Scheme of the evolution of pulse-heated loop plasma of
    Figure~\ref{fig:ntx_time} in a density-temperature diagram
    (\emph{solid line}). The four phases are labeled. The locus of
    the equilibrium loops is shown (\emph{dashed-dotted line}, marked with QSS), as
    well as the evolution path with an extremely long heat pulse
    (\emph{dashed line}) and the corresponding decay path (marked
    with EQ). Adapted from \cite{2007AA...471..271R}.}
\label{fig:nt_diag_eq}}
\end{figure}


This is the evolution of a loop strand ignited
by a transient heat pulse. Important properties of the heated plasma can be obtained from the analysis of the evolution after the heating stops, i.e., when the plasma cools down.

\cite{1991AA...241..197S} derived a global thermodynamic time scale
for the pure cooling of heated plasma confined in single coronal loops, which has been later refined to be  \citep{2007AA...471..271R} ($s$):

\begin{equation}
\tau_{s} = 4.8 \times 10^{-4} \frac{L}{\sqrt{T_0}}
= 500 \frac{L_9}{\sqrt{T_{0,6}}}
\label{eq:tserio}
\end{equation}

This decay time was obtained assuming that the decay starts from equilibrium
conditions, i.e., departing from the the locus of the equilibrium
loops with a given length \citep[hereafter QSS line,][]{1992AA...253..269J}
in Figure~\ref{fig:nt_diag_eq}. It is therefore valid as long as there is no considerable contribution from the plasma draining to the eenrgy balance. The link between the assumption of equilibrium and the
plasma evolution is shown in Figure~\ref{fig:nt_diag_eq}:
if the heat pulse lasts long enough, Phase~II extends to the right,
and the heated loop asymptotically reaches equilibrium conditions, i.e.,
the horizontal line approaches the QSS line. If the decay starts from 
equilibrium conditions, Phase~III is no longer
present, and Phase~II links directly to Phase~IV. Therefore, there is no
delay between the beginning of the temperature decay and the beginning
of the density decay: the temperature and the density start to decrease
simultaneously. Also, the decay will be dominated by radiative cooling,
except at the very beginning \citep{1991AA...241..197S}.

The presence of
Phase~III implies a delay between the temperature peak and the density
peak. This delay is often observed both in solar flares \citep[e.g.,][]{1993AA...267..586S}
and in stellar flares \citep[e.g.,][]{1988AA...205..181V,1989AA...213..245V,2000AA...353..987F,2000AA...356..627M,2002AA...392..585S}.
The presence of this delay, whenever observed, is a signature of a relatively short heat pulse,
or, in other words, of a decay starting from non-equilibrium conditions.

According to \cite{2007AA...471..271R},
the time taken by the loop to reach equilibrium conditions under
the action of a constant heating is much longer than the sound crossing time
(Equation~(\ref{eq:tsound})), which rules the very initial plasma evaporation.
As already mentioned, in the late rise phase the dynamics become much less
important and the interplay between cooling and heating processes
becomes dominant. The relevant time scale is therefore that reported
in Equation~(\ref{eq:tserio}).

\epubtkImage{reale07_fig5.png}{%
\begin{figure}[htbp]
  \centerline{\includegraphics[scale=0.7]{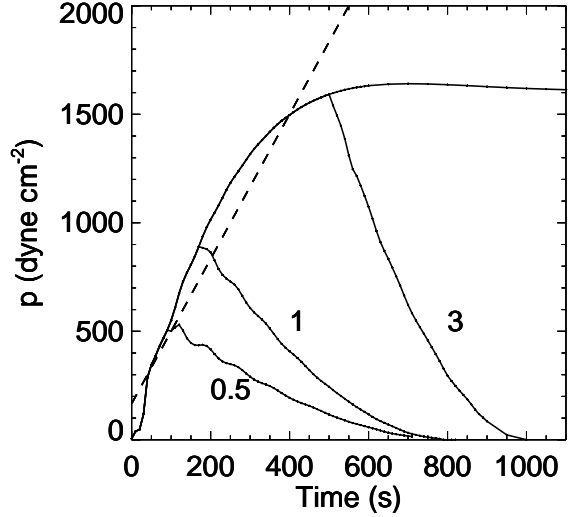}}
  \caption{Pressure evolution obtained from a hydrodynamic simulation
  of a loop strand ignited by heat pulses of different duration (0.5,
  1, 3 times the loop decay time, see text) and with a continuous
  heating. Most of the rise phase can be reasonably described with a
  linear trend ({dashed lines}). From \cite{2007AA...471..271R}.}
  \label{fig:mod_lin}
\end{figure}}

Hydrodynamic simulations confirm that
the time required to reach full equilibrium scales as the loop cooling time ($\tau_s$), and, as shown for instance in
Figure~\ref{fig:mod_lin} \citep[see also][]{1992AA...253..269J},
the time to reach flare steady-state equilibrium is:

\begin{equation}
t_{eq} \approx 2.3 \tau_s
\label{eq:tequil}
\end{equation}

For $t \geq t_{eq}$, the density asymptotically approaches the equilibrium value:

\begin{equation}
n_{0} = \frac{T_0^2}{2a^3 k_B L } = 1.3 \times 10^6 ~ \frac{T_0^2} {L}
\label{eq:n0}
\end{equation}
where $a = 1.4 \times 10^3$ (c.g.s. units), or
\begin{equation}
n_{9} = 1.3 \frac{T_{0,6}^2}{ L_9}
\end{equation}

If the heat pulse stops before the loop reaches equilibrium conditions,
the loop plasma maximum density is lower than the value at equilibrium, i.e. the plasma is underdense \citep[][Section \ref{sec:mod_hea}]{2004ApJ...605..911C}.
Figure~\ref{fig:mod_lin} shows that, after the initial impulsive evaporation
on a time scale given by Equation~(\ref{eq:tsound}), the later progressive
pressure growth can be approximated with a linear trend. Since the
temperature is almost constant in this phase,
we can approximate that the density increases linearly for most
of the time. We can then estimate the value of the maximum density at
the loop apex as:

\begin{equation}
n_{M} \approx n_0 \frac{t_{M}}{t_{eq}}
\label{eq:nmax}
\end{equation}
where $t_{M}$ is the time at which the density maximum occurs. 

Phase~III ranges between the time at which the heat pulse ends and
the time of the density maximum. 
The latter is also the time at which the decay path
crosses the locus of the equilibrium loops (QSS curve). According to \cite{2007AA...471..271R},
the temperature $T_{M}$ at which the maximum density occurs is:

\begin{equation}
T_{M} = 9 \times 10^{-4} ( n_M L )^{1/2}
\label{eq:trc}
\end{equation}
or
\[
T_{M,6} = 0.9 ( n_{M,9} L_9 )^{1/2}~~~.
\]

We can also derive the duration of Phase~III, i.e., the time from the end of the heat pulse to the density maximum, as

\begin{equation}
\Delta t_{0-M} \approx \tau_c \ln \psi 
\label{eq:dt0rc}
\end{equation}
where
\[
\psi = \frac{T_0}{T_{M}} 
\]
and $\tau_c$ (Equation~(\ref{eq:tcon})) is computed for an appropriate value of
the density $n_c$.  A good consistency with numerical simulations is obtained
for $n_c = (2/3) n_M$.

By combining Equation~(\ref{eq:dt0rc}) and Equation~(\ref{eq:nmax}) we obtain:

\begin{equation}
\frac{\Delta t_{0-M}}{t_M} \approx 1.2 \ln \psi
\label{eq:trat}
\end{equation}
which ranges between 0.2 and 0.8 for typical values of $\psi$ (1.2\,--\,2).

These scalings are related to the evolution of a single strand under the effect of a local heat pulse. The strands are below the current instrument spatial resolution and, therefore, we have to consider that, if this scenario is valid, we see the envelope of a collection of small scale events. The characteristics of the single heat pulses become,  therefore, even more difficult to diagnose, and the question of their frequency, distribution and size remains open. Also from the point of view of the modeling, a detailed description of a multistrand loop implies a much more complex and demanding effort. A possible approach is to literally build a collection of 1-D loop models, each with an independent evolution \citep{2010ApJ...719..576G}. One common approach so far has been to simulate anyhow the evolution of a single strand, and to assume that, in the presence of a multitude of such strands, in the steady state we would see at least one strand at any step of the strand evolution. In other words, a collection of nanoflare-heated strands can be described as a whole with the time-average of the evolution of a single strand \citep[][see also Section~\ref{sec:mod_fin}]{2002ApJ...579L..41W,2003ApJ...593.1174W,2003ApJ...587..439W,2003ApJ...593.1164W}. Another issue to be explored is whether it is possible, and to what extent, to describe a collection of independently-evolving strands as a single effective evolving loop. For instance,  how does the evolution of a single loop where the heating is decreasing slowly compare to the evolution of a collection of independently heated strands, with a decreasing rate of ignition? To what extent do we expect coherence and how is it connected to the degree of global coherence of the loop heating? Is there any kind of transverse coherence or ordered ignition of the strands? It is probably reasonable to describe a multi-stranded loop as a single ``effective'' loop if we can assume that the plasma loses memory of its previous history. This certainly occurs in late phases of the evolution when the cooling has been going on for a long time.


\subsection{Fine structuring}
\label{sec:mod_fin}


The description and role of fine structuring of coronal loops is certainly a challenge for coronal physics, also on the side of modeling, essentially because we have few constraints from observations (Section~\ref{sec:obs_fib}).
One of the first times that the internal structuring of coronal loops have been invoked in a modeling context was for the problem of the interpretation of the uniform filter ratio
distribution detected with TRACE \textit{along} warm loops.
\cite{1999ApJ...517L.155L} claimed that
standard hydrostatic loop models with uniform heating could not explain 
such indication of uniform temperature distribution. 
\cite{2000ApJ...528L..45R} showed that a uniform filter ratio could be
produced by the superposition of several thin strands 
described by standard static models of loops at different temperatures. 
In alternative, 
\cite{2001ApJ...559L.171A} showed that long loops heated at the footpoints result to be mostly isothermal. The problem with this model is that footpoint-heated loops (with heating scale height less than 1/3 of the loop half-length) had
been shown to be thermally unstable, and therefore they cannot be
long-lived, as instead observed. A further alternative is to explain
observations with steady non-static loops, i.e., with significant flows
inside \citep[][see below]{2001ApJ...553L..81W,2002ApJ...567L..89W}. 
Also this hypothesis does not
seem to answer the question \citep{2004ApJ...603..322P}.

A first step to modeling fine-structured loops is to use multistrand static models. Such models show some substantial inconsistencies with observations, e.g., in general they predict too large loop cross sections \citep{2000ApJ...528L..45R}.
Such strands are conceptually different from the thin strands predicted in the nanoflare scenario 
\citep{1988ApJ...330..474P}, which imply a highly dynamic evolution due to pulsed-heating. The nanoflare scenario is approached in
multi-thread loop models, convolving the independent hydrodynamic evolution 
of the plasma confined in each pulse-heated strand (see Section~\ref{sec:mod_flo}). These are able to match some more features of the evolution of warm loops observed with TRACE \citep{2002ApJ...579L..41W,2003ApJ...593.1174W,2003ApJ...587..439W,2003ApJ...593.1164W}.
By means of detailed hydrodynamic loop modeling, \cite{2002ApJ...579L..41W}
found that an ensemble of independently heated strands can be significantly brighter than a static uniformly heated loop and would have a flat filter ratio temperature when observed with TRACE. As an extension,
time-dependent hydrodynamic modeling of an evolving active region
loop observed with TRACE showed that a loop made as a set of small-scale, impulsively heated strands can generally reproduce the spatial and temporal properties of the observed loops, such as a delay between the appearance of the loop in different filters \citep{2003ApJ...593.1174W}.
As an evolution of this approach, 
\cite{2006ApJ...645..711W} modelled an entire active region for comparison with a SoHO/EIT observation. They made potential field extrapolations to compute
magnetic field lines and populate these field lines with solutions to the hydrostatic loop equations assuming steady, uniform heating. As a result, they constrained the link between the heating rate and the magnetic field and size of the structures. However, they also found significant discrepancies with the observed EIT emission.

More recently modeling a loop system as a collection of thin unresolved strand with pulsed heating has been used to explain why active regions look fuzzier in harder energy bands, i.e. X-rays, and/or hotter spectral lines, e.g. Fe XVI \citep[][Section~\ref{sec:obs_warm_hot}]{2009ApJ...694.1256T}. 
The basic reason is that in the dynamic evolution of each strand, the plasma spends a relatively longer time and with a high emission measure at temperature about 3 MK \citep{2010ApJ...719..576G}.

Although multistrand models appear much more complex than single loop models and need further refinements to match all the observational constraints, as mentioned in Section~\ref{sec:mod_bas_tsc}, they certainly represent an important issue for the future of coronal loop comprehension.

\subsection{Flows}
\label{sec:mod_flo}



A generalization of static models of loops (Section~\ref{sec:mod_bas_rtv}) is represented by models of loops with
stationary flows, driven by a pressure imbalance between the footpoints (siphon flows). The properties of siphon
flows have been studied by several authors \citep{1980SoPh...65..251C,1981sars.work..213P,1981SoPh...69...63N,1982SoPh...77..153B,1984ApJ...280..416A,1988ApJ...333..407T,1989ApJ...337..977M,1989ApJ...338.1131N,1990ApJ...359..550T,1990ApJ...355..342S,1991ApJ...375..404T,1992ApJ...389..777P,1993ApJ...402..314M}.
\cite{1995AA...294..861O} developed a complete detailed model of loop siphon flows and used it to explore the space of the solutions and to derive an extension of RTV scaling laws to loops containing subsonic flows.


\epubtkImage{orlando95b_shock.png}{%
\begin{figure}[htbp]
  \centerline{\includegraphics[scale=0.4]{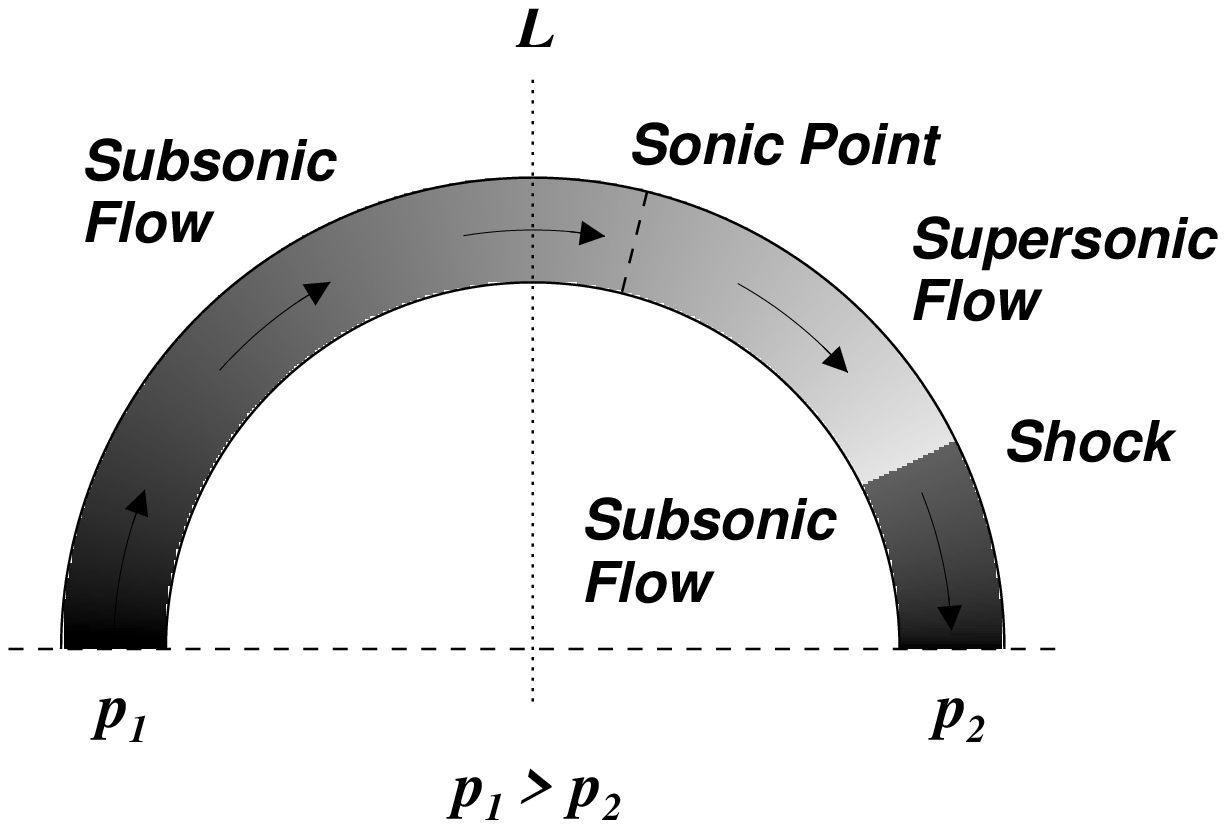}}
  \centerline{\includegraphics[scale=0.5]{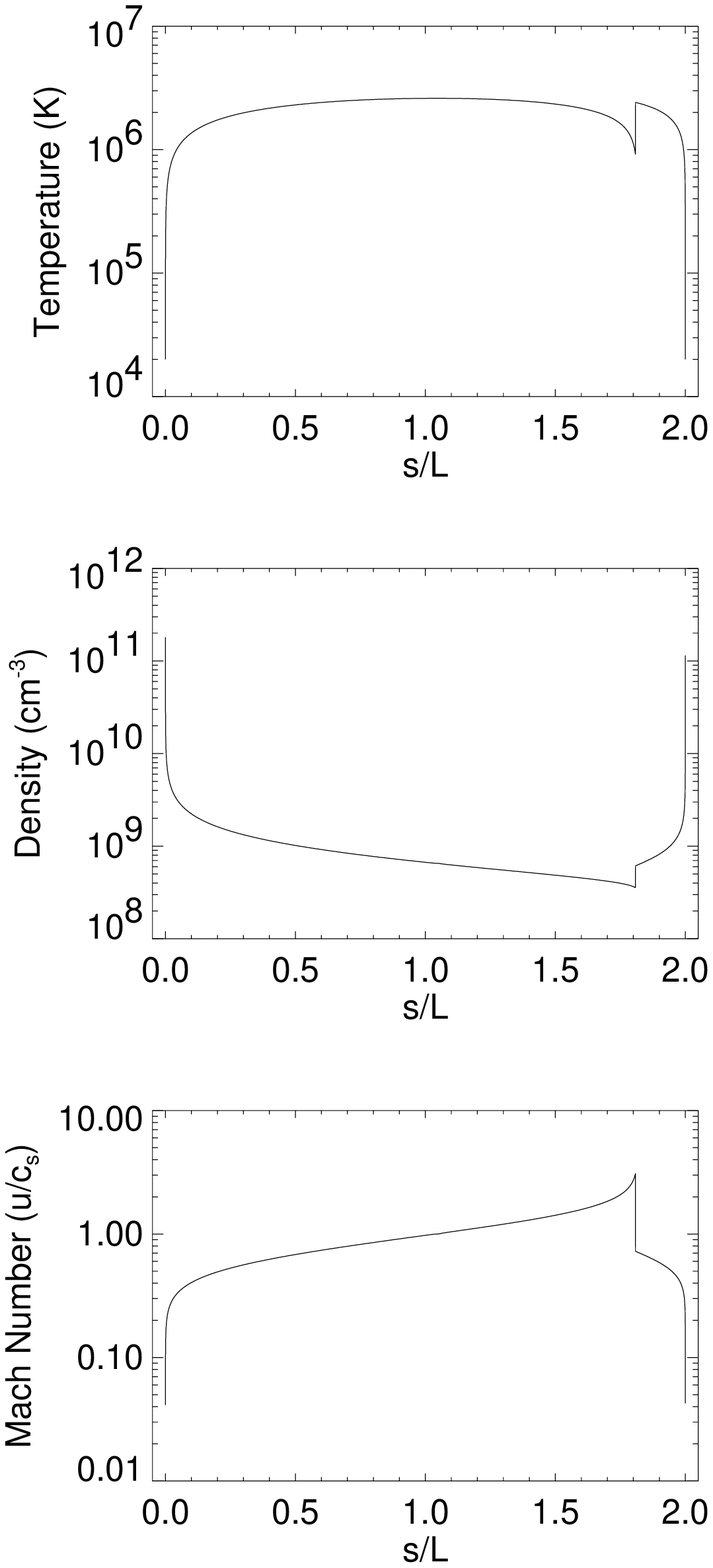}}
  \caption{Example of solutions of a siphon flow loop model including
  a shock. From \cite{1999PCEC...24..401O}.}
  \label{fig:siph}
\end{figure}}

\cite{1995AA...300..549O} explored the conditions for the presence of stationary shocks in critical and supersonic siphon flows in coronal loops (Figure~\ref{fig:siph}), finding that the shock position depends on the volumetric heating rate of the loop, and devising related scaling laws.
The presence of massive flows may alter the line emission with respect to static plasma, because of the delay of the moving plasma to settle to ionization equilibrium \citep{1989SoPh..122..245G}.
\cite{1990ApJ...355..342S} modelled that, even including the effect of ionization non-equilibrium, the UV lines are predicted to be blue-shifted by loop models. So non-equilibrium emission from flows cannot explain the observed dominant redshifts (Section~\ref{sec:obs_flo}). The effects of non-equilibrium of ionization in UV line emission from shocked siphon flows are further discussed in \cite{1999PCEC...24..401O}.

Modeling efforts were devoted in the 1990s to explain specifically the extensive evidence of redshifted UV lines on the solar disk.
\cite{1993ApJ...402..741H} used a hydrodynamic loop model including the effects of non-equilibrium of ionization to show that the redshifts might be produced by downward propagating acoustic waves, possibly stimulated by nanoflares. By means of two-dimensional hydrodynamic simulations, \cite{1996AA...316..215R,1997AA...318..506R} 
proposed that the UV redshifts might be due to downdrafts driven by radiatively-cooling condensations in the solar transition region.
In the exploration of the parameter space, they found redshifted components at speeds of several 
km/s for ambient pressure values ranging from those typical of quiet Sun to   
active regions and predicted that redshifts may occur more easily in the higher pressure plasma, typical of active regions.

\cite{1999AA...352L..99T} explored the idea that the occurrence of
nanoflares in a magnetic loop around the O\,{\sc vi} formation temperature
could explain the observed redshift of mid-low transition region
lines as well as the blueshift observed in low coronal lines ($T > 6
\times 10^5$~K). Observations were compared to numerical simulations of
the response of the solar atmosphere to an energy perturbation of $4
\times 10^{24}$~erg, including non-equilibrium of
ionization. Performing an integration over the entire period of
simulations, they found a redshift in C\,{\sc iv},  and a blueshift in O\,{\sc vi} and Ne\,{\sc viii}, of a few km/s, in reasonable agreement with observations. A similar idea was applied by \cite{2006ApJ...647.1452P} to make predictions about the presence or absence of non-thermal broadening in several spectral lines (e.g., Ne\,{\sc viii}, Mg\,{\sc x}, Fe\,{\sc xvii}) due to nanoflare-driven chromospheric evaporation. 
Clearly, the occurrence of such effects in the lines depends considerably on the choice of the heat pulse parameters. Therefore, more constraints are needed to make the whole model more consistent. In other words, modeling should address specific observations to provide more conclusive results.

Theoretical reasons indicate that flows should be invariably present in
coronal loop systems, although they may not be necessarily important in the
global loop momentum and energy budget. For instance, it has been shown that the
presence of at least moderate flows is necessary to explain why we actually see
the loops \citep{2004ApJ...604..433L}.
The loop emission and detection is in fact due to the emission
from heavy ions, like Fe. In hydrostatic equilibrium conditions, gravitational
settlement should keep the emitting elements low on the solar surface, and we
should not be able to see but the loop footpoints. Instead, detailed modeling
shows that flows of few km/s
are enough to drag ions high in the corona by Coulomb coupling and to enhance
coronal ion abundances by orders of magnitudes. Incidentally, the same modeling
shows that, for the same mechanisms, no chemical fractionation of coronal plasma with respect to photospheric composition as a function of the element First Ionization Potential (FIP)
should be present in coronal loops.

Other studies point instead to the relative unimportance of flows in coronal
loops. In particular, as already mentioned in Section~\ref{sec:obs_fib},
by means of steady hydrodynamic loop modeling (i.e.,
assuming equilibrium condition, and therefore dropping the time-dependent terms
in Equations~(\ref{eq:mass}), (\ref{eq:mom}), (\ref{eq:en})), \cite{2004ApJ...603..322P}
showed that 
flows may not be able to explain the evidence of isothermal loops, 
as instead proposed by \cite{2002ApJ...567L..89W}.
They found that 
a heating deposited asymmetrically in a
loop is able to drive significant flows in the loop and to enhance its density
to the levels typically diagnosed from TRACE observations, but it also produces
an inversion of the temperature distribution and, consequently, a highly
structured distribution of the relevant filter ratio along the loop, which is
not observed.

Plasma cooling is a mechanism that may drive significant downflows in a
loop \citep[e.g.,][]{2005AA...437..311B,2010ApJ...717..163B}.
As an extension of studies on modeling
catastrophic cooling in loops \citep{2004AA...424..289M},
\cite{2005AA...436.1067M} tried to explain the
evidence of propagating intensity variations observed in the He\,{\sc
  ii}~304~\AA\ line with SoHO/EIT 
\citep[][Section~\ref{sec:obs_flo}]{2004AA...415.1141D}. Two possible driving mechanisms had been
proposed: slow magnetoacoustic waves or blobs of cool downfalling plasma. A model of cool
downfalling blob triggered in a thermally-unstable loop heated at 
the footpoints gave a qualitative agreement with measured speeds and predicted
a significant braking in the high-pressure transition region, to be checked in
future high cadence observations in cool lines.

Plasma waves have been more recently proposed to have an important role in driving
flows within loops. Acoustic waves excited by heat pulses at the chromospheric
loop footpoints and damped by thermal conduction in corona are possible 
candidates \citep{2005AA...438..713T}.
Even more attention received the
propagation of Alfv\'en waves in coronal loops. Hydrodynamic loop modeling 
\cite{2005AA...435.1159O} showed that Alfv\'en waves deposit significant momentum in the 
plasma, and that steady state conditions with significant flows and
relatively high density can be reached. Analogous results were independently
obtained with a different approach: considering a wind-like model to
describe a long isothermal loop, \cite{2003AIPC..679..750G,2005AA...437.1081G} 
showed that the
waves can drive pressure variations along the loop which trigger siphon flows. Alfv\'en disturbances have been recently shown to be amplified by the presence of loop flows 
\citep{2009ApJ...694...69T}.

As listed above, models predict the development of flows inside coronal loops in a wide variety of situations, namely evaporation, draining, siphon flows, waves. The challenge will be to distinguish clearly among them and to assess the appropriate weight and importance to them both in the spatial and temporal distribution.

\subsection{Heating}
\label{sec:mod_hea}


The problem of what heats coronal loops is essentially the problem of coronal heating, and is a central issue in the whole solar physics. Although the magnetic origin of coronal heating has been well-established since the very first X-ray observations of the corona, the detailed mechanism of conversion of magnetic energy into thermal energy is still under intense debate, because a series of physical effects conspire to make the mechanism intrinsically elusive.

\cite{2006SoPh..234...41K} splits the heating problem into six steps: the identification of the source of energy, its conversion into heat, the plasma response to the heating, the spectrum of the emitted radiation, the final signature in observables. Outside of analytical approaches, the source and conversion of energy are typically studied in detail by means of multi-dimensional full MHD models \citep[e.g.,][]{2005ApJ...618.1020G}, which, however, are still not able to provide exhaustive predictions on the plasma response and complete diagnostics on observables. On the other hand, the plasma response is the main target of loop hydrodynamic models, which, instead, are not able to treat the heating problem in a self-consistent way (Section~\ref{sec:mod_bas}).

In the investigation of the source of energy,
\cite{1980ApJ...238..343G} already  pointed out that the magnetic field plays an active role in heating the coronal loops. They assumed that the field lines are wound continuously by the photospheric convective motions and the generated non-potential component is dissipated into heating. Several following studies were devoted to the connection and scaling of the magnetic energy to the coronal energy content \citep{1982ApJ...259..359G}
and to the rate of energy release through reconnection \citep{1981ApJ...243..301G}. 
The photospheric motions are therefore the ultimate energy source and stress the field or generate waves depending
on whether the timescale of the motion is long or short compared to the end-to-end
Alfv\'en travel time. Following \cite{2006SoPh..234...41K},
dissipation of magnetic stresses can be referred to as Direct Current (DC) heating, and dissipation of waves as Alternating Current (AC) heating. 


The question of the conversion of the magnetic energy into heat is also challenging, because dissipation is predicted to occur on very small scales or large gradients in the corona by classical theory, unless anomalous dissipation coefficients are invoked. As reviewed by \cite{2006SoPh..234...41K}, large gradients may be produced in various ways,
involving either magnetic field patterns and their evolution, magnetic instabilities such as the kink instability, or velocity pattern, such as turbulence. For waves, resonance absorption and phase mixing may be additional viable mechanisms. 

The problem of plasma response to heating has been kept historically well separated from the primary heating origin, although some attempts have been made to couple them. For instance, in \cite{2005ApJ...633..489R}
the time-dependent distribution of
energy dissipation along the loop obtained from a hybrid shell model was used as heating input of a time-dependent hydrodynamic loop model (see below). A similar concept was applied to search for signatures of turbulent heating in UV spectral lines \citep{2006ApJ...651.1219P}.

As already mentioned, studies using steady-state or time-dependent purely hydrodynamic loop modeling have addressed primarily the plasma response to heating, and also its radiative emission and the detailed comparison with observations. A forward-modeling including all these steps was performed by
\cite{2000ApJ...535..412R,2000ApJ...535..423R}. They first analyzed a TRACE observation of a brightening coronal loop (see also Section~\ref{sec:obs_tim}). The analysis was used to set up the parameters for the forward modeling, and to run loop hydrodynamic simulations with various assumptions on the heating location and time dependence. The comparison of the TRACE emission predicted by the simulations with the measured one constrained the heat pulse to be short, much less than the observed loop rise phase, and intense, appropriate for a 3~MK loop, and its location to be probably midway between the apex and one of the footpoints.

The investigation of the heating mechanisms through the plasma response is made difficult by a variety of reasons. For instance, the problem of background subtraction can be crucial in the comparison with observations, as shown by the three analyses of the same large loop structure observed with Yohkoh/SXT on the solar limb, mentioned in Section~\ref{sec:obs_dia}. More specifically,
\cite{2000ApJ...539.1002P} tried to deduce the form of the heating from Yohkoh observed temperature profiles and found that a uniform heating best describes the data, if the temperature is obtained from the ratio of the total filter intensities, with no background subtraction.
\cite{2001ApJ...559L.171A} splitted the measured emission into two components and found a better agreement with heating deposited at the loop footpoints. 
\cite{2002ApJ...580..566R} revisited the analysis of the
same loop system, considering conventional hydrostatic single-loop models and accounting accurately for
an unstructured background contribution. With forward-modeling, i.e., synthesizing from the model observable quantities to be compared directly with the data, background-subtracted data are fitted 
with acceptable statistical significance
by a model of relatively hot loop ($\sim 3.7$~MK) heated at the apex, but it was pointed out the importance of  background subtraction and the necessity of more specialized observations to address this question. More diagnostic techniques to compare models with observations were proposed afterwards \citep[e.g.,][]{2005ApJ...618.1039L}.

Independently of the adopted numerical or theoretical tool, many studies have been addressing the mechanisms of coronal loop heating clearly distinguishing between the two main classes, i.e., DC heating through moderate and frequent
explosive events, named nanoflares \citep[e.g.,][]{1988ApJ...330..474P}
and AC heating via Alfv\'en waves \citep[e.g.,][]{1998ApJ...499..945L}.

\subsubsection{DC heating}

Heating by nanoflares has a long story as a possible candidate to explain the
heating of the solar corona, and, in particular, of the coronal loops 
\citep[e.g.,][]{1993pssc.symp..151P,1993SoPh..147..263C,1993ApJ...418..496K,1995PASJ...47..251S,1998ApJ...502..981J,2001AA...373..318M,2001ApJ...557..343K,2002ApJ...579L..41W,2003ApJ...593.1174W,2003ApJ...582..486S,1997ApJ...478..799C,2004ApJ...605..911C,2004AA...424..289M,2005ApJ...622..695T,2005ApJ...633..489R,2009AA...499L...5V}.

More specifically, \cite{1994ApJ...422..381C} provided detailed predictions from a model of loops made of thousands of nanoflare-heated strands. In particular, whereas the loop total emission measure distribution should steepen above the canonical $T^{1.5}$ 
\citep{1980AA....86..355J,2000ApJ...528..524O,2001ApJ...563.1045P} dependence for temperature above 1~MK. Moreover, it was stressed the importance of the dependence of effects such as the plasma dynamics (filling and draining) on the loop filling factor driven by the elemental heat pulse size (Section~\ref{sec:mod_bas_tsc}).
In \cite{1997ApJ...478..799C} the nanoflare model was applied to the heating of coronal loops observed
by Yohkoh. A good match was found only for hot (4~MK) loops, with filling factors less than 0.1, so that it was hypothesized the existence of two distinct classes of hot loops.

Although there is evidence of intermittent heating episodes, it has been questioned whether
and to what extent nanoflares are able to provide enough energy to heat
the corona \citep[e.g.,][]{1999SoPh..190..233A}.
On the other hand, loop models with
nanoflares, and, in particular, those considering a prescribed random time
distribution of the pulses deposited at the footpoints of multi-stranded loops
have been able to explain several features of loop observations, for instance, of warm loops from TRACE \citep{2002ApJ...579L..41W,2003ApJ...593.1174W},
(see Section~\ref{sec:obs_fib}). 

Hydrodynamic loop modeling showed also that different distributions of the
heat pulses along the loop have limited effects on the observable quantities \citep{2005ApJ...628.1023P},
because most of the differences occur at the
beginning of the heat deposition, when the emission measure is low, while
later the loop loses memory of the heat distribution \citep[see also][]{2004ApJ...610L.129W}.
\cite{2008ApJ...689.1406P} applied both static and impulsive models to solar active regions and showed 
that the latter ones are able to simultaneously reproduce EUV and SXR loops in active regions, and to predict radial intensity variations consistent with the localized core and extended emissions.
\cite{2004ApJ...605..911C} showed with a semi-analytical loop model that
the cycle of loop heating/cooling naturally leads to
hot-underdense/warm-overdense loop (Section~\ref{sec:mod_bas_tsc}), as observed \citep[][Section~\ref{sec:obs_dia_tra}]{2003ApJ...593.1164W}, 
and that the width of the DEM of a nanoflare-heated
loop can depend on the number of strands which
compose the loop: a relatively flat DEM or a peaked (isothermal) DEM are
obtained with strands of diameter about 15~km or about 
150~km, respectively. This is of relevance for the diagnostics both of the loop fine
structure (Section~\ref{sec:obs_fib}) and of the DEM reconstruction
(Section~\ref{sec:obs_dia}).
As a further improvement, 
\cite{2007ApJ...666.1245W} added an impulsive
heating model to the simulation of an entire active region and found that it is possible to reproduce the total observed
soft X-ray emission in all of the Yohkoh/SXT filters. However, once again, at EUV wavelengths the agreement between the simulation and the observation is only partial.

Nanoflares have been studied also in the framework of stellar coronae. 
\cite{2005ApJ...622..695T} showed that intermittent heating by relatively intense
nanoflares deposited at the loop footpoints make the
loop stable on long time scales (loops continuously heated at the footpoints
are unstable), and, on the other hand, produces a
well-defined peak in the average DEM of the loop, similar to that derived from
the DEM reconstruction of active stars, and also to those shown in \cite{1994ApJ...422..381C}. Therefore, this is an alternative way to obtain a steep temperature dependence of the loop emission measure distribution in the low temperature range.

An alternative approach to study nanoflare heating is to analyze intensity fluctuations \citep{1997ApJ...486.1045S,2000ApJ...541.1096V,2001ApJ...557..343K,2003PPCF...45..535V}
and to derive their occurrence distribution
\citep{2008ApJ...689.1421S,2009ApJ...703.2118S}. From the width of the distributions and autocorrelation functions, it has been suggested that nanoflare signatures are more easily found in observations of warm TRACE loops than of hot Yohkoh/SXT loops. It is to be investigated whether the results change after relaxing the assumption of temperature-independent distribution widths. 
Also other variability analysis of TRACE observations was found able to put constraints on loop heating. In particular, according to \cite{2003ApJ...590..547A},
in TRACE observations, the lack of observable warm loops and of significant variations in the moss regions implies that the heating in the hot moss loops should not be
truly flare-like, but instead quasi-steady and that the heating magnitude is only weakly varying. Further evidence in this direction has been found more recently by \cite{2010ApJ...711..228W}.

An analogous approach is to analyze the intensity distributions. The distribution of impulsive events vs their number in the solar and stellar corona is typically described with a power law. The slope of the power law is a critical parameter to establish weather such events are able to heat the solar corona \citep{1991SoPh..133..357H}. In particular, a  power law index of 2 is the critical value above or below which flare-like events may be able or unable, respectively, to power the whole corona \citep[e.g.,][]{1999SoPh..190..233A}.
Unfortunately, due to the faintness of the events, the distribution of weak events is particularly difficult to derive and might even be separate from that of proper flares and microflares.
\cite{2008AA...492..857P} used a hydrodynamic model to simulate the UV emission of a loop system heated by nanoflares on small, spatially
unresolved scales. The simulations confirm previous results that several spectral lines have an intensity distribution that follows a power-law, in a
similar way to the heating function \citep{1991SoPh..133..357H}. However, only the high temperature lines best preserve the heating function's power law index (especially Fe\,{\sc xix}).


\subsubsection{AC heating}

Loop oscillations, modes and wave propagation deserve a review by themselves, and are outside of the scope of the present one. Here we account for some aspects which are relevant for the loop heating. 
A recent review of coronal waves and oscillations can be found in \cite{2005LRSP....2....3N}.
New observations from SDO AIA provide ample evidence of wave activity in the solar corona
\citep{2010AAS...21630803T}. These observations are currently the subject of intensive analysis and will be reported on in the future.  

As reviewed by  \cite{2006SoPh..234...41K},
MHD waves of many types are generated in the photosphere, e.g. acoustic, Alfv\'en, fast and slow magnetosonic waves. Propagating upwards, the waves may transfer energy to the coronal part of the loops. The question is what fraction 
of the wave flux is able to pass through the very steep density and temperature gradients in the transition region. Acoustic and slow-mode waves form shocks and are strongly damped, fast-mode waves are strongly refracted and reflected
\citep{1996SSRv...75..453N}.

\cite{1978ApJ...226..650I,1982ApJ...254..318I,1983ApJ...271..778I}
devised an LRC equivalent circuit to show the potential importance of AC processes to heat the corona. 
\cite{1984ApJ...277..392H} used a dissipation length formalism to propose resonance absorption of Alfv\'en waves as a potential coronal heating mechanism.
A loop may be considered as a high-quality resonance cavity for hydromagnetic waves. Turbulent photospheric motions can excite small-scale waves. Most Alfv\'en waves are strongly reflected in the chromosphere and transition region, where the Alfv\'en speed increases dramatically with height. Significant transmission is possible only within narrow frequency bands centered on discrete values where loop resonance conditions are satisfied \citep{1981SoPh...70...25H,1984ApJ...277..392H,1982ApJ...254..318I}.
The waves resonate as a global mode and dissipate efficiently when their frequency is near the local Alfv\'en waves frequency $\omega_A \approx 2 \pi v_A/L$. By solving the linearized MHD equations 
\cite{1987ApJ...317..514D} showed that this mechanism can potentially heat the corona, as further supported by numerical solution of MHD equations for low beta plasma \citep{1993ApJ...415..354S},
and although 
\cite{1991ApJ...372..719P} argued that Alfv\'en waves are difficult to be generated by solar convection.

Evidence for photospheric Alfv\'en waves was obtained from magnetic and velocity fluctuations in regions of strong magnetic field \citep{1996ApJ...465..436U}
and from granular motions in the quiet Sun \citep{1994AA...283..232M}
with fluxes of the order of $10^7$ erg cm$^{-2}$ s$^{-1}$, which might contribute to heating if transmitted efficiently to the corona.


\cite{1985JGR....90.7620H} estimated that enough flux may pass through the base of long ($>10^{10}$ cm) active region loops to provide their heating, but shorter loops are a problem, since they have higher resonance frequencies and the photospheric power spectrum is believed to decrease exponentially with frequency in this range.
\cite{1998ApJ...499..945L} suggested that short loops may transmit waves with low frequencies, as long as the field is sufficiently twisted. 
\cite{1988JGR....93.5423H} proposed that Alfv\'en resonance can pump energy out of the surface wave into thin layers surrounding the resonant field lines and that the energy can be distributed by an eddy viscosity throughout large portions of coronal active region loops. 

Waves may be generated directly in the corona, and evidence for their presence was found \citep[e.g.,][]{1999Sci...285..862N,1999ApJ...520..880A,1999SoPh..186..207B,2002SoPh..209...61D}.
It is unclear whether coronal waves carry a sufficient energy flux to heat the plasma \citep{2007Sci...317.1192T}.
\cite{1995ApJ...444..471O} studied the dependence on the wavenumber for comparison with observations of loop oscillations and found partial agreement with velocity amplitudes measured from non-thermal broadening of soft X-ray lines. 
The observed nonthermal broadening of transition region and coronal spectral lines implies fluxes that may be sufficient to heat both the quiet Sun and active regions, but it is unclear whether the waves are efficiently dissipated \citep{1994ApJ...435..482P}.
Furthermore, the nonthermal line broadening could be produced by unresolved loop flows that are unrelated to waves \citep{2006ApJ...647.1452P}.
\cite{1998ApJ...493..474O} included inhomogeneous density structure and found that a broadband wave spectrum becomes necessary for efficient resonance and that it fragments the loop into many density layers that resemble multistrand concept. 
The heat deposition by the resonance of Alfv\'en waves in a loop was investigated by \cite{2005AA...435.1159O} 
By assuming a functional
form first proposed by \cite{1986JGR....91.4111H},
hydrodynamic loop modeling showed that,
depending on the model parameters, heating by Alfv\'en waves leads to different classes of loop solutions, such as the isothermal cool loops indicated by
TRACE, or the hot loops observed with Yohkoh/SXT. Specific diagnostics are still to be defined for the comparison with observations.

Efficient wave dissipation may be allowed by enhanced dissipation coefficients inferred from fast damping of flaring loop oscillations in the corona 
\citep{1999Sci...285..862N},
but the same effect may also favor efficient magnetic reconnection in nanoflares. Alfv\'en waves required for resonant absorption are relatively high frequency waves. Evidence for lower frequency Alfv\'en waves has been found in the chromosphere with the Hinode SOT 
\citep{2007Sci...318.1574D}.
Such waves may supply energy in the corona even outside of resonance with different mechanisms to be explored with modeling. Among dissipation mechanisms phase mixing with enhanced resistivity was suggested by \cite{2002ApJ...576L.153O}
and supported by the analysis of \cite{2008AA...482L...9O}.
Also multistrand structure has been recognized to be important in possible wave dissipation and loop twisting, as recently modelled by \cite{2009ApJ...694..502O}.


Intensity disturbances propagating along active region loops at speeds above 100~km/s were detected with TRACE and interpreted as slow magnetoacoustic waves \citep{2000AA...362.1151N}.
These waves probably originate from the underlying oscillations, i.e., the 3-minute \linebreak chromospheric/\,transition-region oscillations in
sunspots and the 5-minute solar global oscillations (p-modes). Slow magnetosonic waves might be good candidates as coronal heating sources according to quite a detailed model by \cite{1999ApJ...526..478B},
including the effect of chromosphere and transition region and of the radiative losses in the corona. Such waves might be generated directly from upward propagating Alfv\'en waves. Contrary conclusions, in favor of fast magnetosonic waves, have been also obtained, but with much simpler modeling \citep{2001MNRAS.326..675P}.
Slow magnetosonic waves with periods of about 5 minutes have been more recently detected in the transition region and coronal emission lines by Hinode/EIS
at the footpoint of a coronal loop rooted at plage, but found to carry not enough energy to heat the corona \citep{2009ApJ...696.1448W}.

Investigation of AC heating has been made also through comparison with DC heating.
\cite{2008ApJ...688..669A} compared observational signatures of coronal heating by Alfv\'en waves and nanoflares using two coronal loop models and found that Hinode XRT intensity histograms display power-law distributions whose indices differ considerably, to be checked against observations.
\cite{2008ApJS..179..509L,2008ApJ...689.1388L} applied a method for predicting active region coronal emissions using magnetic field measurements and a chosen heating relationship to 10 active regions. With their forward-modeling, they found a volumetric coronal heating rate proportional to magnetic field and inversely proportional to field-line loop length, which seems to point to, although not conclusively, the steady-state
scaling of two heating mechanisms: van Ballegooijen's current layers theory \citep{1986ApJ...311.1001V}, taken in the AC limit, and Parker's critical angle mechanism \citep{1988ApJ...330..474P}, in the case where the angle of misalignment is a twist angle.


\subsubsection{Modeling including the magnetic field}

Some studies have investigated loop heating by including also the magnetic field in the analysis and modeling, and have tried to discriminate between different mechanisms using global approaches and scalings inferred from modeling.
Based on a previous study of the plasma parameters and the magnetic flux density \citep{2000ApJ...530..999M},
\cite{2003ApJ...586..592D} derived the dependence of the mean coronal heating rate on the magnetic flux density from the analysis of an active region.
By using the scaling laws of coronal loops, they found that models based on the dissipation
of stressed, current-carrying magnetic fields are in better agreement with the observations than models that attribute coronal heating to the dissipation of MHD waves injected at the base of the corona.
\cite{2004ApJ...615..512S} considered a similar approach applied to the whole corona, by populating magnetic field lines taken from observed magnetograms with quasi-static loop atmospheres and obtaining the best match to X-ray and EUV observation with a heating that scales as it would be expected from DC reconnection at tangential discontinuities.
These approaches will certainly be very useful when they will provide more detailed predictions and constraints.

Recent modeling has been able to explain the ignition of warm loops from primary energy release mechanisms, although it is remains unclear how the same mechanisms could produce hot loops. A large scale approach (see also Section~\ref{sec:mod_bas}) is by ``ab initio'' modeling, i.e., with full MHD modeling of an entire coronal region \citep{2005ApJ...618.1020G}.
Observed solar granular velocity pattern, a potential extrapolation of a SOHO/MDI magnetogram, and a standard stratified atmosphere were used as initial conditions. The simulation showed that, at steady state, the magnetic field is able to dissipate $(3-4) \times 10^6$~erg cm$^{-2}$ s$^{-1}$ in a highly intermittent corona, at an average temperature of $\sim 10^6$~K, adequate to reproduce typical warm loop populations observed in TRACE images. Warm loops were also obtained with time-dependent loop modeling including the
intermittent magnetic dissipation in MHD turbulence due to loop footpoint
motions \citep{2005ApJ...633..489R}.
The dissipation rate along a loop predicted with a hybrid-shell model \citep{2004PhRvL..92s4501N} 
was used as heating input (see Equation~(\ref{eq:en}))
in a proper time-dependent loop model, the Palermo-Harvard code \citep{1982ApJ...252..791P}.
It was shown that the most intense nanoflares excited in an ambient
magnetic field of about 10 G can produce warm loops with temperatures of 1\,--\,1.5~MK in the corona of a 30,000~km long loop. 

More recently,
\cite{2007ApJ...657L..47R} used reduced MHD (rMHD) to identify MHD anisotropic turbulence as the physical mechanism responsible for the transport of energy from the large scales, where energy is injected by photospheric motions, to the small scales, where it is dissipated. Strong turbulence was found for weak axial magnetic fields and long loops. The predicted heating rate is appropriate for warm loops, in agreement with \cite{2005ApJ...633..489R}.
\cite{2007ApJ...662..701B} also used shell models of rMHD turbulence to analyze the case of a coronal loop heated by photospheric turbulence and found that the Alfv\'en waves interact non-linearly and form turbulent spectra. They derived an intermittent heating function, on average able to sustain the corona and proportional to the aspect ratio of the loop to the $\sim 1.5$ power. 
\cite{2007AA...469..347B} added in the modeling a profile of density and/or magnetic field along the loop, showing that differences are found in the heat deposition, in particular in the low part of the loop. 

There are new efforts to include magnetic effects in the loop modeling.
\cite{2008AA...479..235H} studied observational properties of a kink unstable coronal loop, using a fluid code and finding potentially observable density effects.
\cite{2008AA...485..837B} studied coronal heating by nanoflares triggered by a kink instability
using three-dimensional magnetohydrodynamic numerical simulations of energy release for a cylindrical coronal loop model. Magnetic energy is
dissipated, leading to large or small heating events according to the initial current profile.

Interesting perspectives are developing from models in which self-organized criticality  triggers loop coronal heating.
For \cite{2007ApJ...671.2139U} and \cite{2008ApJ...676L..69C} coronal heating is self-regulating and keeps the coronal plasma roughly marginally collisionless. In the long run, the coronal heating process may be represented by repeating cycles that consist of fast reconnection events (i.e., nanoflares), followed by rapid evaporation episodes, followed by relatively long periods ($\sim 1$~hr) during which magnetic stresses build up and the plasma simultaneously cools down and precipitates. 
\cite{2008ApJ...682..654M} proposed an avalanche model for solar flares, based on an idealized representation of a coronal loop as a bundle of magnetic flux strands wrapping around one another. The system is driven by random deformation of the strands, and a form of
reconnection is assumed to take place when the angle subtended by two strands crossing at the same lattice site exceeds
some preset threshold. For a generic coronal loop of length $10^{10}$~cm and diameter $10^8$~cm the mechanism leads to flare energies
ranging between $10^{23}$ and $10^{29}$~erg, for an instability threshold angle of 11 degrees between contiguous magnetic flux strands.



\subsubsection{New hints}

Given the difficulty to find a conclusive answer about the heating of coronal loops, even in the presence of considerable observational and theoretical efforts, there has been recently the attempt to propose different or radically alternative scenarios. For instance, it has been suggested that warm and hot loops may be heated by different mechanisms, impulsive the former, much more steady the latter \citep{2010ApJ...711..228W}. 
New evidence from Hinode satellite indicates the presence of significant upflows in the form of widespread spicules which have correspondence in coronal observations \citep{2009ApJ...701L...1D},
This evidence suggests that the interaction between the chromosphere and the corona in the heating processes might be important and may receive support also by some modeling approaches \citep{2005ApJ...618.1020G}. 
These scenarios are intriguing but need additional investigation and support both from the observational and theoretical point of view.

\newpage


\section{Stellar Coronal Loops}

Non-solar X-ray missions since \textit{Einstein} and European X-ray Observatory SATellite \textit{EXOSAT} have established that most other stars have a confined corona, often much more active that the solar one. The level of activity is ruled by several factors, but, first of all, the age of the star is important: young fast-rotating stars are more active \citep[e.g.,][]{2005ApJ...622..653T}. 
The topic of stellar coronal loops deserves a review by itself \citep[e.g.,][]{1985ARAA..23..413R}
and here only a few relevant issues are discussed. A complete and more recent review of stellar coronae, with an extensive part regarding loops, is by \cite{2004AARv..12...71G}. In the framework of the solar-stellar connection, it is very important the comparison of what we know about the spatially resolved but single solar corona and what about the unresolved but numerous stellar coronae, which offer a variety of different environments. The lack of spatial resolution inhibits to obtain direct information about the size and appearance of the loops, and the general aspect of the corona. We therefore have to rely on indirect evidence. One possible approach to get information is to benefit from transient X-ray events, such as flares, which provide estimations of the loop scale length from their dependence on the decay and rise timescales \citep[][and Section~\ref{sec:mod_bas_tsc} for reviews] {2002ASPC..277..103R,2003AdSpR..32.1057R}.
Detailed hydrodynamic modeling can provide even more constraints, for instance, on the heat deposition \citep[e.g.,][]{1988ApJ...328..256R,2004AA...416..733R}.
The study of stellar X-ray flares allowed, for instance, to constrain that most stellar flares involve plasma confined in closed structures \citep{2002AA...383..952R},
and to infer the presence both of loops with size similar to those observed on the Sun \citep[e.g.,][]{1988ApJ...328..256R} and of giant loops \citep{2005ApJS..160..469F,2008ApJ...688..437G}, with length exceeding the stellar radius.

Another approach is to use the entire solar X-ray corona as a template and ``Rosetta stone'' to interpret stellar coronae. A detailed implementation of this approach was devised and applied extensively using Yohkoh data over its entire life, which covers a whole solar cycle
\citep{2000ApJ...528..524O,2000ApJ...528..537P}.
It was shown that the solar corona indeed provides a pattern of components, i.e., quiet structures, active regions, active region cores, flares, which can be identified in stellar coronal data and which can explain stellar activity giving different weights to the components \citep{2001ApJ...560..499O,2004ApJ...612..472P}.
The method was also applied to describe stellar coronae in terms of loop populations and to extract general information and constraints on coronal heating \citep{2004ApJ...612..472P}.
It was applied to flares 
\citep{2001ApJ...557..906R} and to describe the evolution of active regions \citep{2004AA...424..677O}.
More recently it was shown that a continuous unresolved flaring activity may explain the most active coronae, but also that the coronal heating appears to follow different scaling for quiet regions and for active and flaring regions across the cycle \citep{2008AA...488.1069A}.

\cite{2006ApJ...643..438C} realized that the strong hot peaks in the emission measure-temperature distributions in the coronae of some binary stars \citep{2003ApJS..145..147S} are similar to those expected for an impulsively-heated solar corona. A coronal model comprised of many impulsively heated strands shows that the evidence may be compatible with coronae made of many very small loops (length under $10^3$~km) heated by microflares.

The recent deeper investigation of solar coronal heating mechanisms through the evidence of hot plasma and of variability makes even more important 
future tighter links to the study of stellar coronae, which show very strong evidence of very hot steady components \citep[e.g.,][]{1990ApJ...365..704S,2005AA...432..671S}.

\newpage


\section{Conclusions and Perspectives}

Coronal loops have been the subject of in-depth studies for about 50 years. Since they owe their identity to the brightness of the confined plasma, most of the studies have addressed the physics of the confined plasma, i.e., its structure, dynamics and evolution. Most of the basic laws that rule the confined plasma, such as scaling laws, were developed early after the discovery of loops and are now well-established. 
Although the observational scenario is ever-enriching with the progress of the solar coronal missions, a variety of questions remain still open, at all levels, starting from loop identification itself. The lack of operative definitions and automatic identification tools, and the difficulty to isolate loops from other surrounding and intersecting structures have prevented systematic studies on large and unbiased loop populations. Detailed morphological and theoretical analyses converge to a fine loop substructuring, which is critical to understand the basic mechanisms but is below the resolution limit of present-day instruments. Also the large scale structure of loops leaves room for further developments, and in particular the link with the confining magnetic field, which is difficult to measure in the corona. One specific question to be addressed is the role of the loop tapering in the transition region. 

%
%
%

Looking inside the loops, open questions involve the detailed thermal structure of the confined plasma. This issue is important to assess the basic loop heating mechanism: a broad multi-component thermal distribution would indicate a structured heating, a simpler distribution a monolithic mechanism. This question is still open for many reasons; spectroscopic methods provide very detailed thermal information, but mostly concentrated in the regime of warm loops. Moreover, the methods of analysis do not appear to provide unique answers yet. Also filter ratio diagnostics from broad-band X-ray and UV imaging telescopes have not been able to provide conclusive results so far. As a consequence, we are unable, right now, to assess the problem of the apparently different nature of warm (TRACE, SoHO/EIT) and hot (Yohkoh/SXT, Hinode/XRT) loops, and whether it is simply a matter of different heating rate or the heating mechanisms are radically different. 


The role of the dynamics of plasma confined inside loops is also still under investigation. The measurement of plasma motions is made difficult by the possible ambiguity with the apparent motion of thermal fronts. It will be important in the future to evaluate the relative weight of the different flows, i.e., downflows, evaporation and draining, siphon flows, and their influence on the overall loop budget.


Also the investigation of temporal variations deserves attention. In particular, in narrow band instruments thermal variations might be confused with intensity variations, and make the interpretation difficult. On the other hand, the analysis of emission variations is very important, because it can potentially shed light on heating mechanisms based on short impulsive events (nanoflares) or on wave-like phenomena (Alfv\'en waves).

%

On the theoretical side, 1-D loop models are well-established and have provided a wealth of sound and important results. Today, their evolution consists essentially in their transformation into ``strand'' models, and so a collection of them describes a proper loop. Moreover, loops are now seen much more dynamic than they were in the past, either as the site of flows, or of wave perturbations, or of heat pulses. So they need more and more time-dependent modeling to address, for instance, the importance of flows, and the relative weight between evaporation-draining flows and siphon flows. The modeling is therefore becoming more and more demanding, although computing resources are now much more powerful than in the past and although some approaches are able to squeeze the spatial dimension, yet maintaining the temporal description. At the same time, new data seem to require other model refinements, such as the description of loop expansion in the transition region. The improvement of numerical and computing resources is also allowing more complex and complete modeling of whole loop regions including the 3-D magnetic field structure ``ab initio''. This approach is very promising and will surely provide important results in the next future, to complement those obtained with basic single loop models. We all look forward self-consistent descriptions including the generation of heat from magnetic field rearrangements and perturbations.

%
%
%
%
%
%

Special attention still deserves the problem of what heats coronal loops, which means basically what heats the whole corona. This problem has revealed to be particularly difficult, essentially because of intrinsic physical reasons, and, in particular, i) the highly effective thermal conduction, which inhibits the identification of the heating site, and ii) the expected small scales of the heating processes, which require so far prohibitive spatial/temporal resolution. Nevertheless, coronal loops remain an obvious excellent laboratory to investigate coronal heating mechanisms, because the dense plasma confined therein make them bright and easy to observe. The most recent challenge offered by coronal loops is probably the increasing evidence that the thin strands they are made of are ignited by small scale, rapid, but intense pulses. Most current efforts are devoted to study this aspect both from the theoretical/modeling and from the observational point of view. However, alternative explanations are actively explored and strongly proposed lately. 

The study of coronal loops is very alive and is the subject of Coronal Loop Workshops, taking place every two years, which are site of debate, inspiration of new investigations, and school for young investigators. We look forward further great improvements in our knowledge of coronal loops from the new mission Solar Dynamic Observatory.

\bigskip

\section{Acknowledgements}
\label{section:acknowledgements}

The author thanks P.\ Testa, S.\ Orlando, G.\ Peres, J.\ Klimchuk and the anonymous referees for suggestions.
The author acknowledges support from the Italian Ministero dell'Universit\`a e della Ricerca. The Transition Region and Coronal Explorer, TRACE, is a mission of the Stanford-Lockheed Institute for Space Research, and part of the NASA Small Explorer program. The solar X-ray images of Figure~\ref{fig:loop_types} is from the Yohkoh mission of ISAS, Japan. The X-ray telescope was prepared by the Lockheed-Martin Solar and Astrophysics Laboratory, the National Astronomical Observatory of Japan, and the University of Tokyo with the support of NASA and ISAS.



\newpage

\bibliography{refs}

\begin{thebibliography}{396}
\expandafter\ifx\csname natexlab\endcsname\relax\def\natexlab#1{#1}\fi
\expandafter\ifx\csname url\endcsname\relax
  \def\url#1{{\tt #1}}\fi
\expandafter\ifx\csname urlprefix\endcsname\relax\def\urlprefix{URL }\fi
\providecommand{\eprint}[2][]{\url{#2}}
\catcode`\% 12

\bibitem[Acton {\it et~al.\/}(1980)]{1980SoPh...65...53A}
Acton, L.W., Finch, M.L., Gilbreth, C.W., Culhane, J.L., Bentley, R.D., Bowles,
  J.A., Guttridge, P., Gabriel, A.H., Firth, J.G. and Hayes, R.W., 1980, ``The
  soft X-ray polychromator for the Solar Maximum Mission'', {\it Solar
  Phys.\/}, {\bf 65}, 53--71.
  {\small[\href{http://dx.doi.org/10.1007/BF00151384}{DOI}]},
  {\small[\href{http://adsabs.harvard.edu/abs/1980SoPh...65...53A}{ADS}]}

\bibitem[Antiochos(1984)]{1984ApJ...280..416A}
Antiochos, S.K., 1984, ``A dynamic model for the solar transition region'',
  {\it Astrophys. J.\/}, {\bf 280}, 416--422.
  {\small[\href{http://dx.doi.org/10.1086/162007}{DOI}]},
  {\small[\href{http://adsabs.harvard.edu/abs/1984ApJ...280..416A}{ADS}]}

\bibitem[Antiochos and Noci(1986)]{1986ApJ...301..440A}
Antiochos, S.K. and Noci, G., 1986, ``The structure of the static corona and
  transition region'', {\it Astrophys. J.\/}, {\bf 301}, 440--447.
  {\small[\href{http://dx.doi.org/10.1086/163912}{DOI}]},
  {\small[\href{http://adsabs.harvard.edu/abs/1986ApJ...301..440A}{ADS}]}

\bibitem[Antiochos {\it et~al.\/}(1999)]{1999ApJ...512..985A}
Antiochos, S.K., MacNeice, P.J., Spicer, D.S. and Klimchuk, J.A., 1999, ``The
  Dynamic Formation of Prominence Condensations'', {\it Astrophys. J.\/}, {\bf
  512}, 985--991. {\small[\href{http://dx.doi.org/10.1086/306804}{DOI}]},
  {\small[\href{http://adsabs.harvard.edu/abs/1999ApJ...512..985A}{ADS}]},
  {\small[\href{http://arxiv.org/abs/arXiv:astro-ph/9808199}{{arXiv:astro-ph/9%
808199}}]}

\bibitem[Antiochos {\it et~al.\/}(2003)]{2003ApJ...590..547A}
Antiochos, S.K., Karpen, J.T., DeLuca, E.E., Golub, L. and Hamilton, P., 2003,
  ``Constraints on Active Region Coronal Heating'', {\it Astrophys. J.\/}, {\bf
  590}, 547--553. {\small[\href{http://dx.doi.org/10.1086/375003}{DOI}]},
  {\small[\href{http://adsabs.harvard.edu/abs/2003ApJ...590..547A}{ADS}]}

\bibitem[Antolin {\it et~al.\/}(2008)]{2008ApJ...688..669A}
Antolin, P., Shibata, K., Kudoh, T., Shiota, D. and Brooks, D., 2008,
  ``Predicting Observational Signatures of Coronal Heating by Alfv{\'e}n Waves
  and Nanoflares'', {\it Astrophys. J.\/}, {\bf 688}, 669--682.
  {\small[\href{http://dx.doi.org/10.1086/591998}{DOI}]},
  {\small[\href{http://adsabs.harvard.edu/abs/2008ApJ...688..669A}{ADS}]}

\bibitem[Argiroffi {\it et~al.\/}(2008)]{2008AA...488.1069A}
Argiroffi, C., Peres, G., Orlando, S. and Reale, F., 2008, ``The flaring and
  quiescent components of the solar corona'', {\it Astron. Astrophys.\/}, {\bf
  488}, 1069--1077.
  {\small[\href{http://dx.doi.org/10.1051/0004-6361:200809355}{DOI}]},
  {\small[\href{http://adsabs.harvard.edu/abs/2008A&A...488.1069A}{ADS}]},
  {\small[\href{http://arxiv.org/abs/0805.2685}{{arXiv:0805.2685}}]}

\bibitem[{Aschwanden} {\it et~al.\/}(1999)]{1999ApJ...520..880A}
{Aschwanden}, M.~J., {Fletcher}, L., {Schrijver}, C.~J. and {Alexander}, D.,
  1999, ``{Coronal Loop Oscillations Observed with the Transition Region and
  Coronal Explorer}'', {\it Astrophys. J.\/}, {\bf 520}, 880--894.
  {\small[\href{http://dx.doi.org/10.1086/307502}{DOI}]},
  {\small[\href{http://adsabs.harvard.edu/abs/1999ApJ...520..880A}{ADS}]}

\bibitem[Aschwanden(1999)]{1999SoPh..190..233A}
Aschwanden, M.J., 1999, ``Do EUV Nanoflares Account for Coronal Heating?'',
  {\it Solar Phys.\/}, {\bf 190}, 233--247.
  {\small[\href{http://dx.doi.org/10.1023/A:1005288725034}{DOI}]},
  {\small[\href{http://adsabs.harvard.edu/abs/1999SoPh..190..233A}{ADS}]}

\bibitem[Aschwanden(2001)]{2001ApJ...559L.171A}
Aschwanden, M.J., 2001, ``Revisiting the Determination of the Coronal Heating
  Function from Yohkoh Data'', {\it Astrophys. J. Lett.\/}, {\bf 559},
  L171--L174. {\small[\href{http://dx.doi.org/10.1086/323788}{DOI}]},
  {\small[\href{http://adsabs.harvard.edu/abs/2001ApJ...559L.171A}{ADS}]}

\bibitem[Aschwanden(2002)]{2002ApJ...580L..79A}
Aschwanden, M.J., 2002, ``The Differential Emission Measure Distribution in the
  Multiloop Corona'', {\it Astrophys. J. Lett.\/}, {\bf 580}, L79--L83.
  {\small[\href{http://dx.doi.org/10.1086/345469}{DOI}]},
  {\small[\href{http://adsabs.harvard.edu/abs/2002ApJ...580L..79A}{ADS}]}

\bibitem[Aschwanden(2004)]{2004psci.book.....A}
Aschwanden, M.J., 2004, {\it Physics of the Solar Corona. An Introduction\/},
  Praxis Publishing Ltd.
  {\small[\href{http://adsabs.harvard.edu/abs/2004psci.book.....A}{ADS}]}

\bibitem[Aschwanden and Nightingale(2005)]{2005ApJ...633..499A}
Aschwanden, M.J. and Nightingale, R.W., 2005, ``Elementary Loop Structures in
  the Solar Corona Analyzed from TRACE Triple-Filter Images'', {\it Astrophys.
  J.\/}, {\bf 633}, 499--517.
  {\small[\href{http://dx.doi.org/10.1086/452630}{DOI}]},
  {\small[\href{http://adsabs.harvard.edu/abs/2005ApJ...633..499A}{ADS}]}

\bibitem[Aschwanden and Nitta(2000)]{2000ApJ...535L..59A}
Aschwanden, M.J. and Nitta, N., 2000, ``The Effect of Hydrostatic Weighting on
  the Vertical Temperature Structure of the Solar Corona'', {\it Astrophys. J.
  Lett.\/}, {\bf 535}, L59--L62.
  {\small[\href{http://dx.doi.org/10.1086/312695}{DOI}]},
  {\small[\href{http://adsabs.harvard.edu/abs/2000ApJ...535L..59A}{ADS}]},
  {\small[\href{http://arxiv.org/abs/arXiv:astro-ph/0004093}{{arXiv:astro-ph/0%
004093}}]}

\bibitem[Aschwanden {\it et~al.\/}(1999)]{1999ApJ...515..842A}
Aschwanden, M.J., Newmark, J.S., Delaboudini{\`e}re, J.-P., Neupert, W.M.,
  Klimchuk, J.A., Gary, G.A., Portier-Fozzani, F. and Zucker, A., 1999,
  ``Three-dimensional Stereoscopic Analysis of Solar Active Region Loops. I.
  SOHO/EIT Observations at Temperatures of $(1.0-1.5) \times 10^{6}$ K'', {\it
  Astrophys. J.\/}, {\bf 515}, 842--867.
  {\small[\href{http://dx.doi.org/10.1086/307036}{DOI}]},
  {\small[\href{http://adsabs.harvard.edu/abs/1999ApJ...515..842A}{ADS}]}

\bibitem[Aschwanden {\it et~al.\/}(2000)]{2000ApJ...541.1059A}
Aschwanden, M.J., Nightingale, R.W. and Alexander, D., 2000, ``Evidence for
  Nonuniform Heating of Coronal Loops Inferred from Multithread Modeling of
  TRACE Data'', {\it Astrophys. J.\/}, {\bf 541}, 1059--1077.
  {\small[\href{http://dx.doi.org/10.1086/309486}{DOI}]},
  {\small[\href{http://adsabs.harvard.edu/abs/2000ApJ...541.1059A}{ADS}]}

\bibitem[Aschwanden {\it et~al.\/}(2001)]{2001ARAA..39..175A}
Aschwanden, M.J., Poland, A.I. and Rabin, D.M., 2001, ``The New Solar Corona'',
  {\it Annu. Rev. Astron. Astrophys.\/}, {\bf 39}, 175--210.
  {\small[\href{http://dx.doi.org/10.1146/annurev.astro.39.1.175}{DOI}]},
  {\small[\href{http://adsabs.harvard.edu/abs/2001ARA&A..39..175A}{ADS}]}

\bibitem[Aschwanden {\it et~al.\/}(2002)]{2002SoPh..206...99A}
Aschwanden, M.J., {de Pontieu}, B., Schrijver, C.J. and Title, A.M., 2002,
  ``Transverse Oscillations in Coronal Loops Observed with TRACE II.
  Measurements of Geometric and Physical Parameters'', {\it Solar Phys.\/},
  {\bf 206}, 99--132.
  {\small[\href{http://dx.doi.org/10.1023/A:1014916701283}{DOI}]},
  {\small[\href{http://adsabs.harvard.edu/abs/2002SoPh..206...99A}{ADS}]}

\bibitem[Aschwanden {\it et~al.\/}(2007)]{2007ApJ...656..577A}
Aschwanden, M.J., Nightingale, R.W. and Boerner, P., 2007, ``A Statistical
  Model of the Inhomogeneous Corona Constrained by Triple-Filter Measurements
  of Elementary Loop Strands with TRACE'', {\it Astrophys. J.\/}, {\bf 656},
  577--597. {\small[\href{http://dx.doi.org/10.1086/510232}{DOI}]},
  {\small[\href{http://adsabs.harvard.edu/abs/2007ApJ...656..577A}{ADS}]}

\bibitem[Aschwanden {\it et~al.\/}(2008{\natexlab{a}})]{2008ApJ...680.1477A}
Aschwanden, M.J., Nitta, N.V., Wuelser, J.-P. and Lemen, J.R.,
  2008{\natexlab{a}}, ``First 3D Reconstructions of Coronal Loops with the
  STEREO A+B Spacecraft. II. Electron Density and Temperature Measurements'',
  {\it Astrophys. J.\/}, {\bf 680}, 1477--1495.
  {\small[\href{http://dx.doi.org/10.1086/588014}{DOI}]},
  {\small[\href{http://adsabs.harvard.edu/abs/2008ApJ...680.1477A}{ADS}]}

\bibitem[Aschwanden {\it et~al.\/}(2008{\natexlab{b}})]{2008ApJ...679..827A}
Aschwanden, M.J., W{\"u}lser, J.-P., Nitta, N.V. and Lemen, J.R.,
  2008{\natexlab{b}}, ``First Three-Dimensional Reconstructions of Coronal
  Loops with the STEREO A and B Spacecraft. I. Geometry'', {\it Astrophys.
  J.\/}, {\bf 679}, 827--842.
  {\small[\href{http://dx.doi.org/10.1086/529542}{DOI}]},
  {\small[\href{http://adsabs.harvard.edu/abs/2008ApJ...679..827A}{ADS}]}

\bibitem[Aschwanden {\it et~al.\/}(2009)]{2009ApJ...695...12A}
Aschwanden, M.J., Wuelser, J.-P., Nitta, N.V., Lemen, J.R. and Sandman, A.,
  2009, ``First Three-Dimensional Reconstructions of Coronal Loops with the
  STEREO A+B Spacecraft. III. Instant Stereoscopic Tomography of Active
  Regions'', {\it Astrophys. J.\/}, {\bf 695}, 12--29.
  {\small[\href{http://dx.doi.org/10.1088/0004-637X/695/1/12}{DOI}]},
  {\small[\href{http://adsabs.harvard.edu/abs/2009ApJ...695...12A}{ADS}]}

\bibitem[Bartoe {\it et~al.\/}(1977)]{1977ApOpt..16..879B}
Bartoe, J.-D.F., Brueckner, G.E., Purcell, J.D. and Tousey, R., 1977, ``Extreme
  ultraviolet spectrograph ATM experiment S082B'', {\it Appl. Optics\/}, {\bf
  16}, 879--886.
  {\small[\href{http://adsabs.harvard.edu/abs/1977ApOpt..16..879B}{ADS}]}

\bibitem[Beli{\"e}n {\it et~al.\/}(1999)]{1999ApJ...526..478B}
Beli{\"e}n, A.J.C., Martens, P.C.H. and Keppens, R., 1999, ``Coronal Heating by
  Resonant Absorption: The Effects of Chromospheric Coupling'', {\it Astrophys.
  J.\/}, {\bf 526}, 478--493.
  {\small[\href{http://dx.doi.org/10.1086/307980}{DOI}]},
  {\small[\href{http://adsabs.harvard.edu/abs/1999ApJ...526..478B}{ADS}]}

\bibitem[Benz(2008)]{2008LRSP....5....1B}
Benz, A.O., 2008, ``Flare Observations'', {\it Living Reviews in Solar
  Physics\/}, {\bf 5}, 1--+.
  {\small[\href{http://adsabs.harvard.edu/abs/2008LRSP....5....1B}{ADS}]}

\bibitem[Berger {\it et~al.\/}(1999{\natexlab{a}})]{1999SoPh..190..409B}
Berger, T.E., {de Pontieu}, B., Fletcher, L., Schrijver, C.J., Tarbell, T.D.
  and Title, A.M., 1999{\natexlab{a}}, ``What is Moss?'', {\it Solar Phys.\/},
  {\bf 190}, 409--418.
  {\small[\href{http://dx.doi.org/10.1023/A:1005286503963}{DOI}]},
  {\small[\href{http://adsabs.harvard.edu/abs/1999SoPh..190..409B}{ADS}]}

\bibitem[Berger {\it et~al.\/}(1999{\natexlab{b}})]{1999ApJ...519L..97B}
Berger, T.E., {de Pontieu}, B., Schrijver, C.J. and Title, A.M.,
  1999{\natexlab{b}}, ``High-resolution Imaging of the Solar
  Chromosphere/Corona Transition Region'', {\it Astrophys. J. Lett.\/}, {\bf
  519}, L97--L100. {\small[\href{http://dx.doi.org/10.1086/312088}{DOI}]},
  {\small[\href{http://adsabs.harvard.edu/abs/1999ApJ...519L..97B}{ADS}]}

\bibitem[Berghmans and Clette(1999)]{1999SoPh..186..207B}
Berghmans, D. and Clette, F., 1999, ``Active region EUV transient brightenings
  - First Results by EIT of SOHO JOP80'', {\it Solar Phys.\/}, {\bf 186},
  207--229.
  {\small[\href{http://adsabs.harvard.edu/abs/1999SoPh..186..207B}{ADS}]}

\bibitem[Berton and Sakurai(1985)]{1985SoPh...96...93B}
Berton, R. and Sakurai, T., 1985, ``Stereoscopic determination of the
  three-dimensional geometry of coronal magnetic loops'', {\it Solar Phys.\/},
  {\bf 96}, 93--111.
  {\small[\href{http://dx.doi.org/10.1007/BF00239795}{DOI}]},
  {\small[\href{http://adsabs.harvard.edu/abs/1985SoPh...96...93B}{ADS}]}

\bibitem[Betta {\it et~al.\/}(1997)]{1997AAS..122..585B}
Betta, R., Peres, G., Reale, F. and Serio, S., 1997, ``An adaptive grid code
  for high resolution 1-D hydrodynamics of the solar and stellar transition
  region and corona'', {\it Astron. Astrophys. Suppl. Ser.\/}, {\bf 122},
  585--592. {\small[\href{http://dx.doi.org/10.1051/aas:1997157}{DOI}]},
  {\small[\href{http://adsabs.harvard.edu/abs/1997A&AS..122..585B}{ADS}]}

\bibitem[Betta {\it et~al.\/}(2001)]{2001AA...380..341B}
Betta, R.M., Peres, G., Reale, F. and Serio, S., 2001, ``Coronal loop
  hydrodynamics. The solar flare observed on November {12}, 1980 revisited: The
  UV line emission'', {\it Astron. Astrophys.\/}, {\bf 380}, 341--346.
  {\small[\href{http://dx.doi.org/10.1051/0004-6361:20011428}{DOI}]},
  {\small[\href{http://adsabs.harvard.edu/abs/2001A&A...380..341B}{ADS}]},
  {\small[\href{http://arxiv.org/abs/arXiv:astro-ph/0110514}{{arXiv:astro-ph/0%
110514}}]}

\bibitem[Beveridge {\it et~al.\/}(2003)]{2003SoPh..216...27B}
Beveridge, C., Longcope, D.W. and Priest, E.R., 2003, ``A model for elemental
  coronal flux loops'', {\it Solar Phys.\/}, {\bf 216}, 27--40.
  {\small[\href{http://dx.doi.org/10.1023/A:1026102820634}{DOI}]},
  {\small[\href{http://adsabs.harvard.edu/abs/2003SoPh..216...27B}{ADS}]}

\bibitem[Bohlin {\it et~al.\/}(1980)]{1980SoPh...65....5B}
Bohlin, J.D., Frost, K.J., Burr, P.T., Guha, A.K. and Withbroe, G.L., 1980,
  ``Solar Maximum Mission'', {\it Solar Phys.\/}, {\bf 65}, 5--14.
  {\small[\href{http://dx.doi.org/10.1007/BF00151380}{DOI}]},
  {\small[\href{http://adsabs.harvard.edu/abs/1980SoPh...65....5B}{ADS}]}

\bibitem[Borrini and Noci(1982)]{1982SoPh...77..153B}
Borrini, G. and Noci, G., 1982, ``Non-equilibrium ionization in coronal
  loops'', {\it Solar Phys.\/}, {\bf 77}, 153--166.
  {\small[\href{http://dx.doi.org/10.1007/BF00156101}{DOI}]},
  {\small[\href{http://adsabs.harvard.edu/abs/1982SoPh...77..153B}{ADS}]}

\bibitem[{Bradshaw} and {Cargill}(2010)]{2010ApJ...717..163B}
{Bradshaw}, S.~J. and {Cargill}, P.~J., 2010, ``{The Cooling of Coronal
  Plasmas. III. Enthalpy Transfer as a Mechanism for Energy Loss}'', {\it
  Astrophys. J.\/}, {\bf 717}, 163--174.
  {\small[\href{http://dx.doi.org/10.1088/0004-637X/717/1/163}{DOI}]},
  {\small[\href{http://adsabs.harvard.edu/abs/2010ApJ...717..163B}{ADS}]}

\bibitem[Bradshaw and Cargill(2005)]{2005AA...437..311B}
Bradshaw, S.J. and Cargill, P.J., 2005, ``The cooling of coronal plasmas'',
  {\it Astron. Astrophys.\/}, {\bf 437}, 311--317.
  {\small[\href{http://dx.doi.org/10.1051/0004-6361:20042405}{DOI}]},
  {\small[\href{http://adsabs.harvard.edu/abs/2005A&A...437..311B}{ADS}]}

\bibitem[Bradshaw and Cargill(2006)]{2006AA...458..987B}
Bradshaw, S.J. and Cargill, P.J., 2006, ``Explosive heating of low-density
  coronal plasma'', {\it Astron. Astrophys.\/}, {\bf 458}, 987--995.
  {\small[\href{http://dx.doi.org/10.1051/0004-6361:20065691}{DOI}]},
  {\small[\href{http://adsabs.harvard.edu/abs/2006A&A...458..987B}{ADS}]}

\bibitem[Bradshaw and Mason(2003)]{2003AA...407.1127B}
Bradshaw, S.J. and Mason, H.E., 2003, ``The radiative response of solar loop
  plasma subject to transient heating'', {\it Astron. Astrophys.\/}, {\bf 407},
  1127--1138.
  {\small[\href{http://dx.doi.org/10.1051/0004-6361:20030986}{DOI}]},
  {\small[\href{http://adsabs.harvard.edu/abs/2003A&A...407.1127B}{ADS}]}

\bibitem[Bray {\it et~al.\/}(1991)]{1991plsc.book.....B}
Bray, R.J., Cram, L.E., Durrant, C. and Loughhead, R.E., 1991, {\it Plasma
  Loops in the Solar Corona\/}.
  {\small[\href{http://adsabs.harvard.edu/abs/1991plsc.book.....B}{ADS}]}

\bibitem[Brekke(1993)]{1993ApJ...408..735B}
Brekke, P., 1993, ``Observed redshifts in O V and downflows in the solar
  transition region'', {\it Astrophys. J.\/}, {\bf 408}, 735--743.
  {\small[\href{http://dx.doi.org/10.1086/172633}{DOI}]},
  {\small[\href{http://adsabs.harvard.edu/abs/1993ApJ...408..735B}{ADS}]}

\bibitem[Brekke(1999)]{1999SoPh..190..379B}
Brekke, P., 1999, ``Observations of Transition Region Plasma'', {\it Solar
  Phys.\/}, {\bf 190}, 379--408.
  {\small[\href{http://dx.doi.org/10.1023/A:1005224709046}{DOI}]},
  {\small[\href{http://adsabs.harvard.edu/abs/1999SoPh..190..379B}{ADS}]}

\bibitem[Brekke {\it et~al.\/}(1997)]{1997SoPh..175..511B}
Brekke, P., Kjeldseth-Moe, O. and Harrison, R.A., 1997, ``High-Velocity Flows
  in an Active Region Loop System Observed with the Coronal Diagnostic
  Spectrometer (CDS) on SOHO'', {\it Solar Phys.\/}, {\bf 175}, 511--521.
  {\small[\href{http://dx.doi.org/10.1023/A:1004950330900}{DOI}]},
  {\small[\href{http://adsabs.harvard.edu/abs/1997SoPh..175..511B}{ADS}]}

\bibitem[Brickhouse and Schmelz(2006)]{2006ApJ...636L..53B}
Brickhouse, N.S. and Schmelz, J.T., 2006, ``The Transparency of Solar Coronal
  Active Regions'', {\it Astrophys. J. Lett.\/}, {\bf 636}, L53--L56.
  {\small[\href{http://dx.doi.org/10.1086/500045}{DOI}]},
  {\small[\href{http://adsabs.harvard.edu/abs/2006ApJ...636L..53B}{ADS}]},
  {\small[\href{http://arxiv.org/abs/arXiv:astro-ph/0511683}{{arXiv:astro-ph/0%
511683}}]}

\bibitem[Brkovi{\'c} {\it et~al.\/}(2002)]{2002AA...383..661B}
Brkovi{\'c}, A., Landi, E., Landini, M., R{\"u}edi, I. and Solanki, S.K., 2002,
  ``Models for solar magnetic loops. II. Comparison with SOHO-CDS observations
  on the solar disk'', {\it Astron. Astrophys.\/}, {\bf 383}, 661--677.
  {\small[\href{http://dx.doi.org/10.1051/0004-6361:20011760}{DOI}]},
  {\small[\href{http://adsabs.harvard.edu/abs/2002A&A...383..661B}{ADS}]}

\bibitem[Brooks {\it et~al.\/}(2008)]{2008ApJ...689L..77B}
Brooks, D.H., Ugarte-Urra, I. and Warren, H.P., 2008, ``The Role of Transient
  Brightenings in Heating the Solar Corona'', {\it Astrophys. J. Lett.\/}, {\bf
  689}, L77--L80. {\small[\href{http://dx.doi.org/10.1086/595745}{DOI}]},
  {\small[\href{http://adsabs.harvard.edu/abs/2008ApJ...689L..77B}{ADS}]}

\bibitem[Brosius {\it et~al.\/}(1996)]{1996ApJS..106..143B}
Brosius, J.W., Davila, J.M., Thomas, R.J. and Monsignori-Fossi, B.C., 1996,
  ``Measuring Active and Quiet-Sun Coronal Plasma Properties with
  Extreme-Ultraviolet Spectra from SERTS'', {\it Astrophys. J. Suppl. Ser.\/},
  {\bf 106}, 143. {\small[\href{http://dx.doi.org/10.1086/192332}{DOI}]},
  {\small[\href{http://adsabs.harvard.edu/abs/1996ApJS..106..143B}{ADS}]}

\bibitem[Browning {\it et~al.\/}(2008)]{2008AA...485..837B}
Browning, P.K., Gerrard, C., Hood, A.W., Kevis, R. and van~der Linden, R.A.M.,
  2008, ``Heating the corona by nanoflares: simulations of energy release
  triggered by a kink instability'', {\it Astron. Astrophys.\/}, {\bf 485},
  837--848. {\small[\href{http://dx.doi.org/10.1051/0004-6361:20079192}{DOI}]},
  {\small[\href{http://adsabs.harvard.edu/abs/2008A&A...485..837B}{ADS}]}

\bibitem[Buchlin and Velli(2007)]{2007ApJ...662..701B}
Buchlin, E. and Velli, M., 2007, ``Shell Models of RMHD Turbulence and the
  Heating of Solar Coronal Loops'', {\it Astrophys. J.\/}, {\bf 662}, 701--714.
  {\small[\href{http://dx.doi.org/10.1086/512765}{DOI}]},
  {\small[\href{http://adsabs.harvard.edu/abs/2007ApJ...662..701B}{ADS}]},
  {\small[\href{http://arxiv.org/abs/arXiv:astro-ph/0606610}{{arXiv:astro-ph/0%
606610}}]}

\bibitem[Buchlin {\it et~al.\/}(2007)]{2007AA...469..347B}
Buchlin, E., Cargill, P.J., Bradshaw, S.J. and Velli, M., 2007, ``Profiles of
  heating in turbulent coronal magnetic loops'', {\it Astron. Astrophys.\/},
  {\bf 469}, 347--354.
  {\small[\href{http://dx.doi.org/10.1051/0004-6361:20077111}{DOI}]},
  {\small[\href{http://adsabs.harvard.edu/abs/2007A&A...469..347B}{ADS}]},
  {\small[\href{http://arxiv.org/abs/arXiv:astro-ph/0702748}{{arXiv:astro-ph/0%
702748}}]}

\bibitem[Cargill(1993)]{1993SoPh..147..263C}
Cargill, P.J., 1993, ``The Fine Structure of a Nanoflare-Heated Corona'', {\it
  Solar Phys.\/}, {\bf 147}, 263--268.
  {\small[\href{http://dx.doi.org/10.1007/BF00690717}{DOI}]},
  {\small[\href{http://adsabs.harvard.edu/abs/1993SoPh..147..263C}{ADS}]}

\bibitem[Cargill(1994)]{1994ApJ...422..381C}
Cargill, P.J., 1994, ``Some implications of the nanoflare concept'', {\it
  Astrophys. J.\/}, {\bf 422}, 381--393.
  {\small[\href{http://dx.doi.org/10.1086/173733}{DOI}]},
  {\small[\href{http://adsabs.harvard.edu/abs/1994ApJ...422..381C}{ADS}]}

\bibitem[Cargill(1995)]{1995itsa.conf...17C}
Cargill, P.J., 1995, ``Diagnostics of Coronal Heating'', in {\it Infrared tools
  for solar astrophysics: What's next?\/}, (Ed.) M.J.Penn, J.R.Kuhn~\&,
  {\small[\href{http://adsabs.harvard.edu/abs/1995itsa.conf...17C}{ADS}]}

\bibitem[Cargill and Klimchuk(1997)]{1997ApJ...478..799C}
Cargill, P.J. and Klimchuk, J.A., 1997, ``A Nanoflare Explanation for the
  Heating of Coronal Loops Observed by YOHKOH'', {\it Astrophys. J.\/}, {\bf
  478}, 799. {\small[\href{http://dx.doi.org/10.1086/303816}{DOI}]},
  {\small[\href{http://adsabs.harvard.edu/abs/1997ApJ...478..799C}{ADS}]}

\bibitem[Cargill and Klimchuk(2004)]{2004ApJ...605..911C}
Cargill, P.J. and Klimchuk, J.A., 2004, ``Nanoflare Heating of the Corona
  Revisited'', {\it Astrophys. J.\/}, {\bf 605}, 911--920.
  {\small[\href{http://dx.doi.org/10.1086/382526}{DOI}]},
  {\small[\href{http://adsabs.harvard.edu/abs/2004ApJ...605..911C}{ADS}]}

\bibitem[Cargill and Klimchuk(2006)]{2006ApJ...643..438C}
Cargill, P.J. and Klimchuk, J.A., 2006, ``On the Temperature-Emission Measure
  Distribution in Stellar Coronae'', {\it Astrophys. J.\/}, {\bf 643},
  438--443. {\small[\href{http://dx.doi.org/10.1086/501446}{DOI}]},
  {\small[\href{http://adsabs.harvard.edu/abs/2006ApJ...643..438C}{ADS}]}

\bibitem[Cargill and Priest(1980)]{1980SoPh...65..251C}
Cargill, P.J. and Priest, E.R., 1980, ``Siphon flows in coronal loops. I -
  Adiabatic flow'', {\it Solar Phys.\/}, {\bf 65}, 251--269.
  {\small[\href{http://dx.doi.org/10.1007/BF00152793}{DOI}]},
  {\small[\href{http://adsabs.harvard.edu/abs/1980SoPh...65..251C}{ADS}]}

\bibitem[Cassak {\it et~al.\/}(2008)]{2008ApJ...676L..69C}
Cassak, P.A., Mullan, D.J. and Shay, M.A., 2008, ``From Solar and Stellar
  Flares to Coronal Heating: Theory and Observations of How Magnetic
  Reconnection Regulates Coronal Conditions'', {\it Astrophys. J. Lett.\/},
  {\bf 676}, L69--L72. {\small[\href{http://dx.doi.org/10.1086/587055}{DOI}]},
  {\small[\href{http://adsabs.harvard.edu/abs/2008ApJ...676L..69C}{ADS}]},
  {\small[\href{http://arxiv.org/abs/0710.3399}{{arXiv:0710.3399}}]}

\bibitem[Chae {\it et~al.\/}(1998{\natexlab{a}})]{1998ApJ...505..957C}
Chae, J., Sch{\"u}hle, U. and Lemaire, P., 1998{\natexlab{a}}, ``SUMER
  Measurements of Nonthermal Motions: Constraints on Coronal Heating
  Mechanisms'', {\it Astrophys. J.\/}, {\bf 505}, 957--973.
  {\small[\href{http://dx.doi.org/10.1086/306179}{DOI}]},
  {\small[\href{http://adsabs.harvard.edu/abs/1998ApJ...505..957C}{ADS}]}

\bibitem[Chae {\it et~al.\/}(1998{\natexlab{b}})]{1998ApJ...504L.123C}
Chae, J., Wang, H., Lee, C.-Y., Goode, P.R. and Schuhle, U.,
  1998{\natexlab{b}}, ``Chromospheric Upflow Events Associated with Transition
  Region Explosive Events'', {\it Astrophys. J. Lett.\/}, {\bf 504}, L123+.
  {\small[\href{http://dx.doi.org/10.1086/311583}{DOI}]},
  {\small[\href{http://adsabs.harvard.edu/abs/1998ApJ...504L.123C}{ADS}]}

\bibitem[Cheng {\it et~al.\/}(1983)]{1983ApJ...265.1090C}
Cheng, C.-C., Oran, E.S., Doschek, G.A., Boris, J.P. and Mariska, J.T., 1983,
  ``Numerical simulations of loops heated to solar flare temperatures. I'',
  {\it Astrophys. J.\/}, {\bf 265}, 1090--1119.
  {\small[\href{http://dx.doi.org/10.1086/160751}{DOI}]},
  {\small[\href{http://adsabs.harvard.edu/abs/1983ApJ...265.1090C}{ADS}]}

\bibitem[Cirtain {\it et~al.\/}(2007)]{2007ApJ...655..598C}
Cirtain, J.W., Del~Zanna, G., DeLuca, E.E., Mason, H.E., Martens, P.C.H. and
  Schmelz, J.T., 2007, ``Active Region Loops: Temperature Measurements as a
  Function of Time from Joint TRACE and SOHO CDS Observations'', {\it
  Astrophys. J.\/}, {\bf 655}, 598--605.
  {\small[\href{http://dx.doi.org/10.1086/509769}{DOI}]},
  {\small[\href{http://adsabs.harvard.edu/abs/2007ApJ...655..598C}{ADS}]}

\bibitem[Craig {\it et~al.\/}(1978)]{1978AA....70....1C}
Craig, I.J.D., McClymont, A.N. and Underwood, J.H., 1978, ``The Temperature and
  Density Structure of Active Region Coronal Loops'', {\it Astron.
  Astrophys.\/}, {\bf 70}, 1.
  {\small[\href{http://adsabs.harvard.edu/abs/1978A&A....70....1C}{ADS}]}

\bibitem[Culhane {\it et~al.\/}(2007)]{2007PASJ...59S.751C}
Culhane, L., Harra, L.K., Baker, D., van Driel-Gesztelyi, L., Sun, J., Doschek,
  G.A., Brooks, D.H., Lundquist, L.L., Kamio, S., Young, P.R. and Hansteen,
  V.H., 2007, ``Hinode EUV Study of Jets in the Sun's South Polar Corona'',
  {\it Publ. Astron. Soc. Japan\/}, {\bf 59}, 751.
  {\small[\href{http://adsabs.harvard.edu/abs/2007PASJ...59S.751C}{ADS}]}

\bibitem[Davila(1987)]{1987ApJ...317..514D}
Davila, J.M., 1987, ``Heating of the solar corona by the resonant absorption of
  Alfven waves'', {\it Astrophys. J.\/}, {\bf 317}, 514--521.
  {\small[\href{http://dx.doi.org/10.1086/165295}{DOI}]},
  {\small[\href{http://adsabs.harvard.edu/abs/1987ApJ...317..514D}{ADS}]}

\bibitem[De~Groof {\it et~al.\/}(2004)]{2004AA...415.1141D}
De~Groof, A., Berghmans, D., van Driel-Gesztelyi, L. and Poedts, S., 2004,
  ``Intensity variations in EIT shutterless mode: Waves or flows?'', {\it
  Astron. Astrophys.\/}, {\bf 415}, 1141--1151.
  {\small[\href{http://dx.doi.org/10.1051/0004-6361:20034252}{DOI}]},
  {\small[\href{http://adsabs.harvard.edu/abs/2004A&A...415.1141D}{ADS}]}

\bibitem[{De Moortel} {\it et~al.\/}(2002)]{2002SoPh..209...61D}
{De Moortel}, I., Ireland, J., Walsh, R.W. and Hood, A.W., 2002, ``Longitudinal
  intensity oscillations in coronal loops observed with TRACE I. Overview of
  Measured Parameters'', {\it Solar Phys.\/}, {\bf 209}, 61--88.
  {\small[\href{http://dx.doi.org/10.1023/A:1020956421063}{DOI}]},
  {\small[\href{http://adsabs.harvard.edu/abs/2002SoPh..209...61D}{ADS}]}

\bibitem[De~Pontieu {\it et~al.\/}(2007)]{2007Sci...318.1574D}
De~Pontieu, B., McIntosh, S.W., Carlsson, M., Hansteen, V.H., Tarbell, T.D.,
  Schrijver, C.J., Title, A.M., Shine, R.A., Tsuneta, S., Katsukawa, Y.,
  Ichimoto, K., Suematsu, Y., Shimizu, T. and Nagata, S., 2007, ``Chromospheric
  Alfv{\'e}nic Waves Strong Enough to Power the Solar Wind'', {\it Science\/},
  {\bf 318}, 1574--.
  {\small[\href{http://dx.doi.org/10.1126/science.1151747}{DOI}]},
  {\small[\href{http://adsabs.harvard.edu/abs/2007Sci...318.1574D}{ADS}]}

\bibitem[De~Rosa {\it et~al.\/}(2009)]{2009ApJ...696.1780D}
De~Rosa, M.L., Schrijver, C.J., Barnes, G., Leka, K.D., Lites, B.W.,
  Aschwanden, M.J., Amari, T., Canou, A., McTiernan, J.M., R{\'e}gnier, S.,
  Thalmann, J.K., Valori, G., Wheatland, M.S., Wiegelmann, T., Cheung, M.C.M.,
  Conlon, P.A., Fuhrmann, M., Inhester, B. and Tadesse, T., 2009, ``A Critical
  Assessment of Nonlinear Force-Free Field Modeling of the Solar Corona for
  Active Region 10953'', {\it Astrophys. J.\/}, {\bf 696}, 1780--1791.
  {\small[\href{http://dx.doi.org/10.1088/0004-637X/696/2/1780}{DOI}]},
  {\small[\href{http://adsabs.harvard.edu/abs/2009ApJ...696.1780D}{ADS}]},
  {\small[\href{http://arxiv.org/abs/0902.1007}{{arXiv:0902.1007}}]}

\bibitem[DeForest(2007)]{2007ApJ...661..532D}
DeForest, C.E., 2007, ``On the Size of Structures in the Solar Corona'', {\it
  Astrophys. J.\/}, {\bf 661}, 532--542.
  {\small[\href{http://dx.doi.org/10.1086/515561}{DOI}]},
  {\small[\href{http://adsabs.harvard.edu/abs/2007ApJ...661..532D}{ADS}]},
  {\small[\href{http://arxiv.org/abs/arXiv:astro-ph/0610178}{{arXiv:astro-ph/0%
610178}}]}

\bibitem[Del~Zanna(2008)]{2008AA...481L..49D}
Del~Zanna, G., 2008, ``Flows in active region loops observed by Hinode EIS'',
  {\it Astron. Astrophys.\/}, {\bf 481}, L49--L52.
  {\small[\href{http://dx.doi.org/10.1051/0004-6361:20079087}{DOI}]},
  {\small[\href{http://adsabs.harvard.edu/abs/2008A&A...481L..49D}{ADS}]}

\bibitem[{Del Zanna} and Mason(2003)]{2003AA...406.1089D}
{Del Zanna}, G. and Mason, H.E., 2003, ``Solar active regions: SOHO/CDS and
  TRACE observations of quiescent coronal loops'', {\it Astron. Astrophys.\/},
  {\bf 406}, 1089--1103.
  {\small[\href{http://dx.doi.org/10.1051/0004-6361:20030791}{DOI}]},
  {\small[\href{http://adsabs.harvard.edu/abs/2003A&A...406.1089D}{ADS}]}

\bibitem[Delaboudini{\`e}re {\it et~al.\/}(1995)]{1995SoPh..162..291D}
Delaboudini{\`e}re, J.-P., Artzner, G.E., Brunaud, J., Gabriel, A.H., Hochedez,
  J.F., Millier, F., Song, X.Y., Au, B., Dere, K.P., Howard, R.A., Kreplin, R.,
  Michels, D.J., Moses, J.D., Defise, J.M., Jamar, C., Rochus, P., Chauvineau,
  J.P., Marioge, J.P., Catura, R.C., Lemen, J.R., Shing, L., Stern, R.A.,
  Gurman, J.B., Neupert, W.M., Maucherat, A., Clette, F., Cugnon, P. and van
  Dessel, E.L., 1995, ``EIT: Extreme-Ultraviolet Imaging Telescope for the SOHO
  Mission'', {\it Solar Phys.\/}, {\bf 162}, 291--312.
  {\small[\href{http://dx.doi.org/10.1007/BF00733432}{DOI}]},
  {\small[\href{http://adsabs.harvard.edu/abs/1995SoPh..162..291D}{ADS}]}

\bibitem[D{\'e}moulin {\it et~al.\/}(2003)]{2003ApJ...586..592D}
D{\'e}moulin, P., van Driel-Gesztelyi, L., Mandrini, C.H., Klimchuk, J.A. and
  Harra, L., 2003, ``The Long-Term Evolution of AR 7978: Testing Coronal
  Heating Models'', {\it Astrophys. J.\/}, {\bf 586}, 592--605.
  {\small[\href{http://dx.doi.org/10.1086/367634}{DOI}]},
  {\small[\href{http://adsabs.harvard.edu/abs/2003ApJ...586..592D}{ADS}]}

\bibitem[DePontieu {\it et~al.\/}(2009)]{2009ApJ...701L...1D}
DePontieu, B., McIntosh, S.W., Hansteen, V.H. and Schrijver, C.J., 2009,
  ``Observing the Roots of Solar Coronal Heating -- in the Chromosphere'', {\it
  Astrophys. J. Lett.\/}, {\bf 701}, L1--L6.
  {\small[\href{http://dx.doi.org/10.1088/0004-637X/701/1/L1}{DOI}]},
  {\small[\href{http://adsabs.harvard.edu/abs/2009ApJ...701L...1D}{ADS}]},
  {\small[\href{http://arxiv.org/abs/0906.5434}{{arXiv:0906.5434}}]}

\bibitem[Dere(1982)]{1982SoPh...77...77D}
Dere, K.P., 1982, ``Extreme ultraviolet spectra of solar active regions and
  their analysis'', {\it Solar Phys.\/}, {\bf 77}, 77--93.
  {\small[\href{http://dx.doi.org/10.1007/BF00156097}{DOI}]},
  {\small[\href{http://adsabs.harvard.edu/abs/1982SoPh...77...77D}{ADS}]}

\bibitem[Dere(2008)]{2008AA...491..561D}
Dere, K.P., 2008, ``The plasma filling factor of coronal bright points. Coronal
  bright points'', {\it Astron. Astrophys.\/}, {\bf 491}, 561--566.
  {\small[\href{http://dx.doi.org/10.1051/0004-6361:200810000}{DOI}]},
  {\small[\href{http://adsabs.harvard.edu/abs/2008A&A...491..561D}{ADS}]}

\bibitem[Dere(2009)]{2009AA...497..287D}
Dere, K.P., 2009, ``The plasma filling factor of coronal bright points. II.
  Combined EIS and TRACE results'', {\it Astron. Astrophys.\/}, {\bf 497},
  287--290.
  {\small[\href{http://dx.doi.org/10.1051/0004-6361/200811329}{DOI}]},
  {\small[\href{http://adsabs.harvard.edu/abs/2009A&A...497..287D}{ADS}]}

\bibitem[Dere {\it et~al.\/}(1986)]{1986ApJ...310..456D}
Dere, K.P., Bartoe, J.-D.F. and Brueckner, G.E., 1986, ``Outflows and ejections
  in the solar transition zone'', {\it Astrophys. J.\/}, {\bf 310}, 456--462.
  {\small[\href{http://dx.doi.org/10.1086/164698}{DOI}]},
  {\small[\href{http://adsabs.harvard.edu/abs/1986ApJ...310..456D}{ADS}]}

\bibitem[Dere {\it et~al.\/}(1989)]{1989SoPh..123...41D}
Dere, K.P., Bartoe, J.-D.F. and Brueckner, G.E., 1989, ``Explosive events in
  the solar transition zone'', {\it Solar Phys.\/}, {\bf 123}, 41--68.
  {\small[\href{http://dx.doi.org/10.1007/BF00150011}{DOI}]},
  {\small[\href{http://adsabs.harvard.edu/abs/1989SoPh..123...41D}{ADS}]}

\bibitem[Di~Giorgio {\it et~al.\/}(2003)]{2003AA...406..323D}
Di~Giorgio, S., Reale, F. and Peres, G., 2003, ``CDS/SoHO multi-line
  observation of a solar active region: Detection of a hot stable loop and of a
  cool dynamic loop'', {\it Astron. Astrophys.\/}, {\bf 406}, 323--335.
  {\small[\href{http://dx.doi.org/10.1051/0004-6361:20030492}{DOI}]},
  {\small[\href{http://adsabs.harvard.edu/abs/2003A&A...406..323D}{ADS}]}

\bibitem[Di~Matteo {\it et~al.\/}(1999)]{1999AA...342..563D}
Di~Matteo, V., Reale, F., Peres, G. and Golub, L., 1999, ``Analysis and
  comparison of loop structures imaged with NIXT and Yohkoh/SXT'', {\it Astron.
  Astrophys.\/}, {\bf 342}, 563--574.
  {\small[\href{http://adsabs.harvard.edu/abs/1999A&A...342..563D}{ADS}]}

\bibitem[Domingo {\it et~al.\/}(1995)]{1995SoPh..162....1D}
Domingo, V., Fleck, B. and Poland, A.I., 1995, ``The SOHO Mission: an
  Overview'', {\it Solar Phys.\/}, {\bf 162}, 1--37.
  {\small[\href{http://dx.doi.org/10.1007/BF00733425}{DOI}]},
  {\small[\href{http://adsabs.harvard.edu/abs/1995SoPh..162....1D}{ADS}]}

\bibitem[Doschek {\it et~al.\/}(1976)]{1976ApJ...205L.177D}
Doschek, G.A., Bohlin, J.D. and Feldman, U., 1976, ``Doppler wavelength shifts
  of transition zone lines measured in SKYLAB solar spectra'', {\it Astrophys.
  J. Lett.\/}, {\bf 205}, L177--L180.
  {\small[\href{http://dx.doi.org/10.1086/182118}{DOI}]},
  {\small[\href{http://adsabs.harvard.edu/abs/1976ApJ...205L.177D}{ADS}]}

\bibitem[Doschek {\it et~al.\/}(1982)]{1982ApJ...258..373D}
Doschek, G.A., Boris, J.P., Cheng, C.C., Mariska, J.T. and Oran, E.S., 1982,
  ``A Numerical Simulation of Cooling Coronal Flare Plasma'', {\it Astrophys.
  J.\/}, {\bf 258}, 373.
  {\small[\href{http://dx.doi.org/10.1086/160086}{DOI}]},
  {\small[\href{http://adsabs.harvard.edu/abs/1982ApJ...258..373D}{ADS}]}

\bibitem[Doschek {\it et~al.\/}(2007{\natexlab{a}})]{2007ApJ...667L.109D}
Doschek, G.A., Mariska, J.T., Warren, H.P., Brown, C.M., Culhane, J.L., Hara,
  H., Watanabe, T., Young, P.R. and Mason, H.E., 2007{\natexlab{a}},
  ``Nonthermal Velocities in Solar Active Regions Observed with the
  Extreme-Ultraviolet Imaging Spectrometer on Hinode'', {\it Astrophys. J.
  Lett.\/}, {\bf 667}, L109--L112.
  {\small[\href{http://dx.doi.org/10.1086/522087}{DOI}]},
  {\small[\href{http://adsabs.harvard.edu/abs/2007ApJ...667L.109D}{ADS}]}

\bibitem[Doschek {\it et~al.\/}(2007{\natexlab{b}})]{2007PASJ...59S.707D}
Doschek, G.A., Mariska, J.T., Warren, H.P., Culhane, L., Watanabe, T., Young,
  P.R., Mason, H.E. and Dere, K.P., 2007{\natexlab{b}}, ``The Temperature and
  Density Structure of an Active Region Observed with the Extreme-Ultraviolet
  Imaging Spectrometer on Hinode'', {\it Publ. Astron. Soc. Japan\/}, {\bf 59},
  707. {\small[\href{http://adsabs.harvard.edu/abs/2007PASJ...59S.707D}{ADS}]}

\bibitem[Doschek {\it et~al.\/}(2008)]{2008ApJ...686.1362D}
Doschek, G.A., Warren, H.P., Mariska, J.T., Muglach, K., Culhane, J.L., Hara,
  H. and Watanabe, T., 2008, ``Flows and Nonthermal Velocities in Solar Active
  Regions Observed with the EUV Imaging Spectrometer on Hinode: A Tracer of
  Active Region Sources of Heliospheric Magnetic Fields?'', {\it Astrophys.
  J.\/}, {\bf 686}, 1362--1371.
  {\small[\href{http://dx.doi.org/10.1086/591724}{DOI}]},
  {\small[\href{http://adsabs.harvard.edu/abs/2008ApJ...686.1362D}{ADS}]},
  {\small[\href{http://arxiv.org/abs/0807.2860}{{arXiv:0807.2860}}]}

\bibitem[Dudok~de Wit and Auch{\`e}re(2007)]{2007AA...466..347D}
Dudok~de Wit, T. and Auch{\`e}re, F., 2007, ``Multispectral analysis of solar
  EUV images: linking temperature to morphology'', {\it Astron. Astrophys.\/},
  {\bf 466}, 347--355.
  {\small[\href{http://dx.doi.org/10.1051/0004-6361:20066764}{DOI}]},
  {\small[\href{http://adsabs.harvard.edu/abs/2007A&A...466..347D}{ADS}]},
  {\small[\href{http://arxiv.org/abs/arXiv:astro-ph/0702052}{{arXiv:astro-ph/0%
702052}}]}

\bibitem[Dymova and Ruderman(2006)]{2006AA...459..241D}
Dymova, M.V. and Ruderman, M.S., 2006, ``The geometry effect on transverse
  oscillations of coronal loops'', {\it Astron. Astrophys.\/}, {\bf 459},
  241--244. {\small[\href{http://dx.doi.org/10.1051/0004-6361:20065929}{DOI}]},
  {\small[\href{http://adsabs.harvard.edu/abs/2006A&A...459..241D}{ADS}]}

\bibitem[Favata {\it et~al.\/}(2000)]{2000AA...353..987F}
Favata, F., Reale, F., Micela, G., Sciortino, S., Maggio, A. and Matsumoto, H.,
  2000, ``An extreme X-ray flare observed on EV Lac by ASCA in July 1998'',
  {\it Astron. Astrophys.\/}, {\bf 353}, 987--997.
  {\small[\href{http://adsabs.harvard.edu/abs/2000A&A...353..987F}{ADS}]},
  {\small[\href{http://arxiv.org/abs/arXiv:astro-ph/9909491}{{arXiv:astro-ph/9%
909491}}]}

\bibitem[Favata {\it et~al.\/}(2005)]{2005ApJS..160..469F}
Favata, F., Flaccomio, E., Reale, F., Micela, G., Sciortino, S., Shang, H.,
  Stassun, K.G. and Feigelson, E.D., 2005, ``Bright X-Ray Flares in Orion Young
  Stars from COUP: Evidence for Star-Disk Magnetic Fields?'', {\it Astrophys.
  J. Suppl. Ser.\/}, {\bf 160}, 469--502.
  {\small[\href{http://dx.doi.org/10.1086/432542}{DOI}]},
  {\small[\href{http://adsabs.harvard.edu/abs/2005ApJS..160..469F}{ADS}]},
  {\small[\href{http://arxiv.org/abs/arXiv:astro-ph/0506134}{{arXiv:astro-ph/0%
506134}}]}

\bibitem[Feldman {\it et~al.\/}(1979)]{1979ApJ...229..369F}
Feldman, U., Doschek, G.A. and Mariska, J.T., 1979, ``On the structure of the
  solar transition zone and lower corona'', {\it Astrophys. J.\/}, {\bf 229},
  369--374. {\small[\href{http://dx.doi.org/10.1086/156962}{DOI}]},
  {\small[\href{http://adsabs.harvard.edu/abs/1979ApJ...229..369F}{ADS}]}

\bibitem[Feldman {\it et~al.\/}(1982)]{1982ApJ...255..325F}
Feldman, U., Doschek, G.A. and Cohen, L., 1982, ``Doppler wavelength shifts of
  ultraviolet spectral lines in solar active regions'', {\it Astrophys. J.\/},
  {\bf 255}, 325--328. {\small[\href{http://dx.doi.org/10.1086/159833}{DOI}]},
  {\small[\href{http://adsabs.harvard.edu/abs/1982ApJ...255..325F}{ADS}]}

\bibitem[Feng {\it et~al.\/}(2007)]{2007ApJ...671L.205F}
Feng, L., Inhester, B., Solanki, S.K., Wiegelmann, T., Podlipnik, B., Howard,
  R.A. and Wuelser, J.-P., 2007, ``First Stereoscopic Coronal Loop
  Reconstructions from STEREO SECCHI Images'', {\it Astrophys. J. Lett.\/},
  {\bf 671}, L205--L208.
  {\small[\href{http://dx.doi.org/10.1086/525525}{DOI}]},
  {\small[\href{http://adsabs.harvard.edu/abs/2007ApJ...671L.205F}{ADS}]},
  {\small[\href{http://arxiv.org/abs/0802.0773}{{arXiv:0802.0773}}]}

\bibitem[Fisher {\it et~al.\/}(1985{\natexlab{a}})]{1985ApJ...289..414Fv}
Fisher, G.H., Canfield, R.C. and McClymont, A.N., 1985{\natexlab{a}}, ``Flare
  loop radiative hydrodynamics. V. Response to thick-target heating'', {\it
  Astrophys. J.\/}, {\bf 289}, 414--424.
  {\small[\href{http://dx.doi.org/10.1086/162901}{DOI}]},
  {\small[\href{http://adsabs.harvard.edu/abs/1985ApJ...289..414F}{ADS}]}

\bibitem[Fisher {\it et~al.\/}(1985{\natexlab{b}})]{1985ApJ...289..414Fvi}
Fisher, G.H., Canfield, R.C. and McClymont, A.N., 1985{\natexlab{b}}, ``Flare
  loop radiative hydrodynamics. VI. Chromospheric evaporation due to heating by
  nonthermal electrons'', {\it Astrophys. J.\/}, {\bf 289}, 425--433.
  {\small[\href{http://dx.doi.org/10.1086/162901}{DOI}]},
  {\small[\href{http://adsabs.harvard.edu/abs/1985ApJ...289..414F}{ADS}]}

\bibitem[Fisher {\it et~al.\/}(1985{\natexlab{c}})]{1985ApJ...289..414Fvii}
Fisher, G.H., Canfield, R.C. and McClymont, A.N., 1985{\natexlab{c}}, ``Flare
  loop radiative hydrodynamics. VII. Dynamics of the thick-target heated
  chromosphere'', {\it Astrophys. J.\/}, {\bf 289}, 434--441.
  {\small[\href{http://dx.doi.org/10.1086/162901}{DOI}]},
  {\small[\href{http://adsabs.harvard.edu/abs/1985ApJ...289..414F}{ADS}]}

\bibitem[Fletcher and de~Pontieu(1999)]{1999ApJ...520L.135F}
Fletcher, L. and de~Pontieu, B., 1999, ``Plasma Diagnostics of Transition
  Region `Moss' using SOHO/CDS and TRACE'', {\it Astrophys. J. Lett.\/}, {\bf
  520}, L135--L138. {\small[\href{http://dx.doi.org/10.1086/312157}{DOI}]},
  {\small[\href{http://adsabs.harvard.edu/abs/1999ApJ...520L.135F}{ADS}]}

\bibitem[Foukal(1975)]{1975SoPh...43..327F}
Foukal, P., 1975, ``The temperature structure and pressure balance of magnetic
  loops in active regions'', {\it Solar Phys.\/}, {\bf 43}, 327--336.
  {\small[\href{http://dx.doi.org/10.1007/BF00152357}{DOI}]},
  {\small[\href{http://adsabs.harvard.edu/abs/1975SoPh...43..327F}{ADS}]}

\bibitem[Foukal(1976)]{1976ApJ...210..575F}
Foukal, P.V., 1976, ``The pressure and energy balance of the cool corona over
  sunspots'', {\it Astrophys. J.\/}, {\bf 210}, 575--581.
  {\small[\href{http://dx.doi.org/10.1086/154862}{DOI}]},
  {\small[\href{http://adsabs.harvard.edu/abs/1976ApJ...210..575F}{ADS}]}

\bibitem[Gabriel(1976)]{1976RSPTA.281..339G}
Gabriel, A.H., 1976, ``A magnetic model of the solar transition region'', {\it
  Royal Society of London Philosophical Transactions Series A\/}, {\bf 281},
  339--352.
  {\small[\href{http://adsabs.harvard.edu/abs/1976RSPTA.281..339G}{ADS}]}

\bibitem[Gabriel and Jordan(1975)]{1975MNRAS.173..397G}
Gabriel, A.H. and Jordan, C., 1975, ``Analysis of EUV observations of regions
  of the quiet and active corona at the time of the 1970 March 7 eclipse'',
  {\it Mon. Not. R. Astron. Soc.\/}, {\bf 173}, 397--418.
  {\small[\href{http://adsabs.harvard.edu/abs/1975MNRAS.173..397G}{ADS}]}

\bibitem[Galeev {\it et~al.\/}(1981)]{1981ApJ...243..301G}
Galeev, A.A., Rosner, R., Serio, S. and Vaiana, G.S., 1981, ``Dynamics of
  coronal structures - Magnetic field-related heating and loop energy
  balance'', {\it Astrophys. J.\/}, {\bf 243}, 301--308.
  {\small[\href{http://dx.doi.org/10.1086/158598}{DOI}]},
  {\small[\href{http://adsabs.harvard.edu/abs/1981ApJ...243..301G}{ADS}]}

\bibitem[Gan {\it et~al.\/}(1991)]{1991AA...241..618G}
Gan, W.Q., Zhang, H.Q. and Fang, C., 1991, ``A hydrodynamic model of the
  impulsive phase of a solar flare loop'', {\it Astron. Astrophys.\/}, {\bf
  241}, 618--624.
  {\small[\href{http://adsabs.harvard.edu/abs/1991A&A...241..618G}{ADS}]}

\bibitem[Gebbie {\it et~al.\/}(1981)]{1981ApJ...251L.115G}
Gebbie, K.B., Hill, F., November, L.J., Gurman, J.B., Shine, R.A., Woodgate,
  B.E., Athay, R.G., Tandberg-Hanssen, E.A., Toomre, J. and Simon, G.W., 1981,
  ``Steady flows in the solar transition region observed with SMM'', {\it
  Astrophys. J. Lett.\/}, {\bf 251}, L115--L118.
  {\small[\href{http://dx.doi.org/10.1086/183705}{DOI}]},
  {\small[\href{http://adsabs.harvard.edu/abs/1981ApJ...251L.115G}{ADS}]}

\bibitem[Getman {\it et~al.\/}(2008)]{2008ApJ...688..437G}
Getman, K.V., Feigelson, E.D., Micela, G., Jardine, M.M., Gregory, S.G. and
  Garmire, G.P., 2008, ``X-Ray Flares in Orion Young Stars. II. Flares,
  Magnetospheres, and Protoplanetary Disks'', {\it Astrophys. J.\/}, {\bf 688},
  437--455. {\small[\href{http://dx.doi.org/10.1086/592034}{DOI}]},
  {\small[\href{http://adsabs.harvard.edu/abs/2008ApJ...688..437G}{ADS}]},
  {\small[\href{http://arxiv.org/abs/0807.3007}{{arXiv:0807.3007}}]}

\bibitem[Giacconi {\it et~al.\/}(1965)]{1965ApJ...142.1274G}
Giacconi, R., Reidy, W.P., Zehnpfennig, T., Lindsay, J.C. and Muney, W.S.,
  1965, ``Solar X-Ray Image Obtained Using Grazing-Incidence Optics.'', {\it
  Astrophys. J.\/}, {\bf 142}, 1274--1278.
  {\small[\href{http://dx.doi.org/10.1086/148404}{DOI}]},
  {\small[\href{http://adsabs.harvard.edu/abs/1965ApJ...142.1274G}{ADS}]}

\bibitem[Golub(1996)]{1996Ap&SS.237...33G}
Golub, L., 1996, ``The Solar X-Ray Corona'', {\it Astrophys. J. Suppl. Ser.\/},
  {\bf 237}, 33--48.
  {\small[\href{http://dx.doi.org/10.1007/BF02424425}{DOI}]},
  {\small[\href{http://adsabs.harvard.edu/abs/1996Ap&SS.237...33G}{ADS}]}

\bibitem[Golub and Herant(1989)]{1989SPIE.1160..629G}
Golub, L. and Herant, M., 1989, ``Analysis of the 23 June 1988 flare using NIXT
  multilayer X-ray images'', in {\it Society of Photo-Optical Instrumentation
  Engineers (SPIE) Conference Series\/}, (Ed.) Hoover, R.B., Presented at the
  Society of Photo-Optical Instrumentation Engineers (SPIE) Conference, vol.
  1160,
  {\small[\href{http://adsabs.harvard.edu/abs/1989SPIE.1160..629G}{ADS}]}

\bibitem[{Golub} and {Pasachoff}(2001)]{2001nsss.book.....G}
{Golub}, L. and {Pasachoff}, J.~M., 2001, {\it {Nearest star : the surprising
  science of our sun}\/}.
  {\small[\href{http://adsabs.harvard.edu/abs/2001nsss.book.....G}{ADS}]}

\bibitem[Golub and Pasachoff(1997)]{1997soco.book.....G}
Golub, L. and Pasachoff, J.M., 1997, {\it The Solar Corona\/}.
  {\small[\href{http://adsabs.harvard.edu/abs/1997soco.book.....G}{ADS}]}

\bibitem[Golub {\it et~al.\/}(1980)]{1980ApJ...238..343G}
Golub, L., Maxson, C., Rosner, R., Vaiana, G.S. and Serio, S., 1980, ``Magnetic
  fields and coronal heating'', {\it Astrophys. J.\/}, {\bf 238}, 343--348.
  {\small[\href{http://dx.doi.org/10.1086/157990}{DOI}]},
  {\small[\href{http://adsabs.harvard.edu/abs/1980ApJ...238..343G}{ADS}]}

\bibitem[Golub {\it et~al.\/}(1982)]{1982ApJ...259..359G}
Golub, L., Noci, G., Poletto, G. and Vaiana, G.S., 1982, ``Active region
  coronal evolution'', {\it Astrophys. J.\/}, {\bf 259}, 359--365.
  {\small[\href{http://dx.doi.org/10.1086/160172}{DOI}]},
  {\small[\href{http://adsabs.harvard.edu/abs/1982ApJ...259..359G}{ADS}]}

\bibitem[Golub {\it et~al.\/}(1989)]{1989SoPh..122..245G}
Golub, L., Hartquist, T.W. and Quillen, A.C., 1989, ``Comments on the
  observability of coronal variations'', {\it Solar Phys.\/}, {\bf 122},
  245--261. {\small[\href{http://dx.doi.org/10.1007/BF00912995}{DOI}]},
  {\small[\href{http://adsabs.harvard.edu/abs/1989SoPh..122..245G}{ADS}]}

\bibitem[Golub {\it et~al.\/}(2007)]{2007SoPh..243...63G}
Golub, L., Deluca, E., Austin, G., Bookbinder, J., Caldwell, D., Cheimets, P.,
  Cirtain, J., Cosmo, M., Reid, P., Sette, A., Weber, M., Sakao, T., Kano, R.,
  Shibasaki, K., Hara, H., Tsuneta, S., Kumagai, K., Tamura, T., Shimojo, M.,
  McCracken, J., Carpenter, J., Haight, H., Siler, R., Wright, E., Tucker, J.,
  Rutledge, H., Barbera, M., Peres, G. and Varisco, S., 2007, ``The X-Ray
  Telescope (XRT) for the Hinode Mission'', {\it Solar Phys.\/}, {\bf 243},
  63--86. {\small[\href{http://dx.doi.org/10.1007/s11207-007-0182-1}{DOI}]},
  {\small[\href{http://adsabs.harvard.edu/abs/2007SoPh..243...63G}{ADS}]}

\bibitem[Gomez {\it et~al.\/}(1993)]{1993ApJ...405..767G}
Gomez, D.O., Martens, P.C.H. and Golub, L., 1993, ``Normal incidence X-ray
  telescope power spectra of X-ray emission from solar active regions. I -
  Observations. II - Theory'', {\it Astrophys. J.\/}, {\bf 405}, 767--781.
  {\small[\href{http://dx.doi.org/10.1086/172405}{DOI}]},
  {\small[\href{http://adsabs.harvard.edu/abs/1993ApJ...405..767G}{ADS}]}

\bibitem[Gontikakis {\it et~al.\/}(2008)]{2008AA...489..441G}
Gontikakis, C., Contopoulos, I. and Dara, H.C., 2008, ``Distribution of coronal
  heating in a solar active region'', {\it Astron. Astrophys.\/}, {\bf 489},
  441--447. {\small[\href{http://dx.doi.org/10.1051/0004-6361:20079086}{DOI}]},
  {\small[\href{http://adsabs.harvard.edu/abs/2008A&A...489..441G}{ADS}]}

\bibitem[{Grappin} {\it et~al.\/}(2003)]{2003AIPC..679..750G}
{Grappin}, R., {L{\'e}orat}, J. and {Ofman}, L., 2003, ``{Flows in coronal
  loops driven by Alfv{\'e}n waves: 1.5 MHD simulations with transparent
  boundary conditions}'', in {\it Solar Wind Ten\/}, (Ed.) {M.~Velli, R.~Bruno,
  F.~Malara, \& B.~Bucci}, vol. 679 of American Institute of Physics Conference
  Series,  {\small[\href{http://dx.doi.org/10.1063/1.1618701}{DOI}]},
  {\small[\href{http://adsabs.harvard.edu/abs/2003AIPC..679..750G}{ADS}]}

\bibitem[Grappin {\it et~al.\/}(2005)]{2005AA...437.1081G}
Grappin, R., L{\'e}orat, J. and Habbal, S.R., 2005, ``Siphon flows and
  oscillations in long coronal loops due to Alfv{\'e}n waves'', {\it Astron.
  Astrophys.\/}, {\bf 437}, 1081--1092.
  {\small[\href{http://dx.doi.org/10.1051/0004-6361:20052745}{DOI}]},
  {\small[\href{http://adsabs.harvard.edu/abs/2005A&A...437.1081G}{ADS}]}

\bibitem[Griffiths {\it et~al.\/}(2000)]{2000ApJ...537..481G}
Griffiths, N.W., Fisher, G.H., Woods, D.T., Acton, L.W. and Siegmund, O.H.W.,
  2000, ``Simultaneous SOHO andYohkoh Observations of a Small Solar Active
  Region'', {\it Astrophys. J.\/}, {\bf 537}, 481--494.
  {\small[\href{http://dx.doi.org/10.1086/309005}{DOI}]},
  {\small[\href{http://adsabs.harvard.edu/abs/2000ApJ...537..481G}{ADS}]}

\bibitem[{Guarrasi} {\it et~al.\/}(2010)]{2010ApJ...719..576G}
{Guarrasi}, M., {Reale}, F. and {Peres}, G., 2010, ``{Coronal Fuzziness Modeled
  with Pulse-heated Multi-stranded Loop Systems}'', {\it Astrophys. J.\/}, {\bf
  719}, 576--582.
  {\small[\href{http://dx.doi.org/10.1088/0004-637X/719/1/576}{DOI}]},
  {\small[\href{http://adsabs.harvard.edu/abs/2010ApJ...719..576G}{ADS}]}

\bibitem[G{\"u}del(2004)]{2004AARv..12...71G}
G{\"u}del, M., 2004, ``X-ray astronomy of stellar coronae'', {\it Ann. Rev.
  Astr. Ap.\/}, {\bf 12}, 71--237.
  {\small[\href{http://dx.doi.org/10.1007/s00159-004-0023-2}{DOI}]},
  {\small[\href{http://adsabs.harvard.edu/abs/2004A&ARv..12...71G}{ADS}]},
  {\small[\href{http://arxiv.org/abs/arXiv:astro-ph/0406661}{{arXiv:astro-ph/0%
406661}}]}

\bibitem[Gudiksen and Nordlund(2005)]{2005ApJ...618.1020G}
Gudiksen, B.V. and Nordlund, {\AA}., 2005, ``An Ab Initio Approach to the Solar
  Coronal Heating Problem'', {\it Astrophys. J.\/}, {\bf 618}, 1020--1030.
  {\small[\href{http://dx.doi.org/10.1086/426063}{DOI}]},
  {\small[\href{http://adsabs.harvard.edu/abs/2005ApJ...618.1020G}{ADS}]},
  {\small[\href{http://arxiv.org/abs/arXiv:astro-ph/0407266}{{arXiv:astro-ph/0%
407266}}]}

\bibitem[Habbal {\it et~al.\/}(1985)]{1985SoPh...98..323H}
Habbal, S.R., Ronan, R. and Withbroe, G.L., 1985, ``Spatial and temporal
  variations of solar coronal loops'', {\it Solar Phys.\/}, {\bf 98}, 323--340.
  {\small[\href{http://dx.doi.org/10.1007/BF00152464}{DOI}]},
  {\small[\href{http://adsabs.harvard.edu/abs/1985SoPh...98..323H}{ADS}]}

\bibitem[Haisch {\it et~al.\/}(1988)]{1988ApJS...68..371H}
Haisch, B.M., Strong, K.T., Harrison, R.A. and Gary, G.A., 1988, ``Active
  region coronal loops - Structural and variability'', {\it Astrophys. J.
  Suppl. Ser.\/}, {\bf 68}, 371--405.
  {\small[\href{http://dx.doi.org/10.1086/191292}{DOI}]},
  {\small[\href{http://adsabs.harvard.edu/abs/1988ApJS...68..371H}{ADS}]}

\bibitem[Handy {\it et~al.\/}(1999)]{1999SoPh..187..229H}
Handy, B.N., Acton, L.W., Kankelborg, C.C., Wolfson, C.J., Akin, D.J., Bruner,
  M.E., Caravalho, R., Catura, R.C., Chevalier, R., Duncan, D.W., Edwards,
  C.G., Feinstein, C.N., Freeland, S.L., Friedlaender, F.M., Hoffmann, C.H.,
  Hurlburt, N.E., Jurcevich, B.K., Katz, N.L., Kelly, G.A., Lemen, J.R., Levay,
  M., Lindgren, R.W., Mathur, D.P., Meyer, S.B., Morrison, S.J., Morrison,
  M.D., Nightingale, R.W., Pope, T.P., Rehse, R.A., Schrijver, C.J., Shine,
  R.A., Shing, L., Strong, K.T., Tarbell, T.D., Title, A.M., Torgerson, D.D.,
  Golub, L., Bookbinder, J.A., Caldwell, D., Cheimets, P.N., Davis, W.N.,
  Deluca, E.E., McMullen, R.A., Warren, H.P., Amato, D., Fisher, R., Maldonado,
  H. and Parkinson, C., 1999, ``The transition region and coronal explorer'',
  {\it Solar Phys.\/}, {\bf 187}, 229--260.
  {\small[\href{http://dx.doi.org/10.1023/A:1005166902804}{DOI}]},
  {\small[\href{http://adsabs.harvard.edu/abs/1999SoPh..187..229H}{ADS}]}

\bibitem[Hansteen(1993)]{1993ApJ...402..741H}
Hansteen, V., 1993, ``A new interpretation of the redshift observed in
  optically thin transition region lines'', {\it Astrophys. J.\/}, {\bf 402},
  741--755. {\small[\href{http://dx.doi.org/10.1086/172174}{DOI}]},
  {\small[\href{http://adsabs.harvard.edu/abs/1993ApJ...402..741H}{ADS}]}

\bibitem[Hansteen {\it et~al.\/}(2007)]{2007ASPC..368..107H}
Hansteen, V.H., Carlsson, M. and Gudiksen, B., 2007, ``3D Numerical Models of
  the Chromosphere, Transition Region, and Corona'', in {\it The Physics of
  Chromospheric Plasmas\/}, (Eds.) Heinzel, P., Dorotovi{\v c}, I., Rutten,
  R.J., vol. 368 of Astronomical Society of the Pacific Conference Series,
  {\small[\href{http://adsabs.harvard.edu/abs/2007ASPC..368..107H}{ADS}]}

\bibitem[Hara {\it et~al.\/}(1992)]{1992PASJ...44L.135H}
Hara, H., Tsuneta, S., Lemen, J.R., Acton, L.W. and McTiernan, J.M., 1992,
  ``High-temperature plasmas in active regions observed with the Soft X-ray
  Telescope aboard YOHKOH'', {\it Publ. Astron. Soc. Japan\/}, {\bf 44},
  L135--L140.
  {\small[\href{http://adsabs.harvard.edu/abs/1992PASJ...44L.135H}{ADS}]}

\bibitem[Hara {\it et~al.\/}(2008)]{2008ApJ...678L..67H}
Hara, H., Watanabe, T., Harra, L.K., Culhane, J.L., Young, P.R., Mariska, J.T.
  and Doschek, G.A., 2008, ``Coronal Plasma Motions near Footpoints of Active
  Region Loops Revealed from Spectroscopic Observations with Hinode EIS'', {\it
  Astrophys. J. Lett.\/}, {\bf 678}, L67--L71.
  {\small[\href{http://dx.doi.org/10.1086/588252}{DOI}]},
  {\small[\href{http://adsabs.harvard.edu/abs/2008ApJ...678L..67H}{ADS}]}

\bibitem[Harrison {\it et~al.\/}(1995)]{1995SoPh..162..233H}
Harrison, R.A., Sawyer, E.C., Carter, M.K., Cruise, A.M., Cutler, R.M., Fludra,
  A., Hayes, R.W., Kent, B.J., Lang, J., Parker, D.J., Payne, J., Pike, C.D.,
  Peskett, S.C., Richards, A.G., Gulhane, J.L., Norman, K., Breeveld, A.A.,
  Breeveld, E.R., Al~Janabi, K.F., McCalden, A.J., Parkinson, J.H., Self, D.G.,
  Thomas, P.D., Poland, A.I., Thomas, R.J., Thompson, W.T., Kjeldseth-Moe, O.,
  Brekke, P., Karud, J., Maltby, P., Aschenbach, B., Br{\"a}uninger, H.,
  K{\"u}hne, M., Hollandt, J., Siegmund, O.H.W., Huber, M.C.E., Gabriel, A.H.,
  Mason, H.E. and Bromage, B.J.I., 1995, ``The Coronal Diagnostic Spectrometer
  for the Solar and Heliospheric Observatory'', {\it Solar Phys.\/}, {\bf 162},
  233--290. {\small[\href{http://dx.doi.org/10.1007/BF00733431}{DOI}]},
  {\small[\href{http://adsabs.harvard.edu/abs/1995SoPh..162..233H}{ADS}]}

\bibitem[Haynes {\it et~al.\/}(2008)]{2008AA...479..235H}
Haynes, M., Arber, T.D. and Verwichte, E., 2008, ``Coronal loop slow mode
  oscillations driven by the kink instability'', {\it Astron. Astrophys.\/},
  {\bf 479}, 235--239.
  {\small[\href{http://dx.doi.org/10.1051/0004-6361:20078534}{DOI}]},
  {\small[\href{http://adsabs.harvard.edu/abs/2008A&A...479..235H}{ADS}]}

\bibitem[Hollweg(1981)]{1981SoPh...70...25H}
Hollweg, J.V., 1981, ``Alfven waves in the solar atmosphere. II - Open and
  closed magnetic flux tubes'', {\it Solar Phys.\/}, {\bf 70}, 25--66.
  {\small[\href{http://dx.doi.org/10.1007/BF00154391}{DOI}]},
  {\small[\href{http://adsabs.harvard.edu/abs/1981SoPh...70...25H}{ADS}]}

\bibitem[Hollweg(1984)]{1984ApJ...277..392H}
Hollweg, J.V., 1984, ``Resonances of coronal loops'', {\it Astrophys. J.\/},
  {\bf 277}, 392--403. {\small[\href{http://dx.doi.org/10.1086/161706}{DOI}]},
  {\small[\href{http://adsabs.harvard.edu/abs/1984ApJ...277..392H}{ADS}]}

\bibitem[Hollweg(1985)]{1985JGR....90.7620H}
Hollweg, J.V., 1985, ``Viscosity in a magnetized plasma - Physical
  interpretation'', {\it J. Geophys. Res.\/}, {\bf 90}, 7620--7622.
  {\small[\href{http://dx.doi.org/10.1029/JA090iA08p07620}{DOI}]},
  {\small[\href{http://adsabs.harvard.edu/abs/1985JGR....90.7620H}{ADS}]}

\bibitem[Hollweg(1986)]{1986JGR....91.4111H}
Hollweg, J.V., 1986, ``Transition region, corona, and solar wind in coronal
  holes'', {\it J. Geophys. Res.\/}, {\bf 91}, 4111--4125.
  {\small[\href{http://dx.doi.org/10.1029/JA091iA04p04111}{DOI}]},
  {\small[\href{http://adsabs.harvard.edu/abs/1986JGR....91.4111H}{ADS}]}

\bibitem[Hollweg and Yang(1988)]{1988JGR....93.5423H}
Hollweg, J.V. and Yang, G., 1988, ``Resonance absorption of compressible
  magnetohydrodynamic waves at thin 'surfaces''', {\it J. Geoph. Res.\/}, {\bf
  93}, 5423--5436.
  {\small[\href{http://dx.doi.org/10.1029/JA093iA06p05423}{DOI}]},
  {\small[\href{http://adsabs.harvard.edu/abs/1988JGR....93.5423H}{ADS}]}

\bibitem[Hood and Priest(1979)]{1979AA....77..233H}
Hood, A.W. and Priest, E.R., 1979, ``The equilibrium of solar coronal magnetic
  loops'', {\it Astron. Astrophys.\/}, {\bf 77}, 233--251.
  {\small[\href{http://adsabs.harvard.edu/abs/1979A&A....77..233H}{ADS}]}

\bibitem[Hori {\it et~al.\/}(1997)]{1997ApJ...489..426H}
Hori, K., Yokoyama, T., Kosugi, T. and Shibata, K., 1997,
  ``Pseudo--Two-dimensional Hydrodynamic Modeling of Solar Flare Loops'', {\it
  Astrophys. J.\/}, {\bf 489}, 426.
  {\small[\href{http://dx.doi.org/10.1086/304754}{DOI}]},
  {\small[\href{http://adsabs.harvard.edu/abs/1997ApJ...489..426H}{ADS}]}

\bibitem[Hori {\it et~al.\/}(1998)]{1998ApJ...500..492H}
Hori, K., Yokoyama, T., Kosugi, T. and Shibata, K., 1998, ``Single and Multiple
  Solar Flare Loops: Hydrodynamics and Ca XIX Resonance Line Emission'', {\it
  Astrophys. J.\/}, {\bf 500}, 492.
  {\small[\href{http://dx.doi.org/10.1086/305725}{DOI}]},
  {\small[\href{http://adsabs.harvard.edu/abs/1998ApJ...500..492H}{ADS}]}

\bibitem[Hudson(1991)]{1991SoPh..133..357H}
Hudson, H.S., 1991, ``Solar flares, microflares, nanoflares, and coronal
  heating'', {\it Solar Phys.\/}, {\bf 133}, 357--369.
  {\small[\href{http://dx.doi.org/10.1007/BF00149894}{DOI}]},
  {\small[\href{http://adsabs.harvard.edu/abs/1991SoPh..133..357H}{ADS}]}

\bibitem[Ignatiev {\it et~al.\/}(1998)]{1998SPIE.3406...20I}
Ignatiev, A.P., Kolachevsky, N.N., Korneev, V.V., Krutov, V.V., Kuzin, S.V.,
  Mitrofanov, A.V., Pertsov, A., Ragozin, E.N., Slemzin, V.A., Tindo, I.P.,
  Zhitnik, I.A., Salashchenko, N.N. and Thomas, R.J., 1998, ``Manufacture and
  testing of x-ray optical elements for the TEREK-C and RES-C instruments on
  the CORONAS-I mission'', in {\it Society of Photo-Optical Instrumentation
  Engineers (SPIE) Conference Series\/}, (Ed.) V.A.Slemzin, I.I.Sobelman~\&,
  vol. 3406 of Presented at the Society of Photo-Optical Instrumentation
  Engineers (SPIE) Conference,
  {\small[\href{http://adsabs.harvard.edu/abs/1998SPIE.3406...20I}{ADS}]}

\bibitem[Ionson(1978)]{1978ApJ...226..650I}
Ionson, J.A., 1978, ``Resonant absorption of Alfvenic surface waves and the
  heating of solar coronal loops'', {\it Astrophys. J.\/}, {\bf 226}, 650--673.
  {\small[\href{http://dx.doi.org/10.1086/156648}{DOI}]},
  {\small[\href{http://adsabs.harvard.edu/abs/1978ApJ...226..650I}{ADS}]}

\bibitem[Ionson(1982)]{1982ApJ...254..318I}
Ionson, J.A., 1982, ``Resonant electrodynamic heating of stellar coronal loops
  - an LRC circuit analog'', {\it Astrophys. J.\/}, {\bf 254}, 318--334.
  {\small[\href{http://dx.doi.org/10.1086/159736}{DOI}]},
  {\small[\href{http://adsabs.harvard.edu/abs/1982ApJ...254..318I}{ADS}]}

\bibitem[Ionson(1983)]{1983ApJ...271..778I}
Ionson, J.A., 1983, ``Electrodynamic coupling in magnetically confined X-ray
  plasmas of astrophysical origin'', {\it Astrophys. J.\/}, {\bf 271},
  778--792. {\small[\href{http://dx.doi.org/10.1086/161244}{DOI}]},
  {\small[\href{http://adsabs.harvard.edu/abs/1983ApJ...271..778I}{ADS}]}

\bibitem[Jakimiec {\it et~al.\/}(1992)]{1992AA...253..269J}
Jakimiec, J., Sylwester, B., Sylwester, J., Serio, S., Peres, G. and Reale, F.,
  1992, ``Dynamics of flaring loops. II - Flare evolution in the
  density-temperature diagram'', {\it Astron. Astrophys.\/}, {\bf 253},
  269--276.
  {\small[\href{http://adsabs.harvard.edu/abs/1992A&A...253..269J}{ADS}]}

\bibitem[Jordan(1976)]{1976RSPTA.281..391J}
Jordan, C., 1976, ``The structure and energy balance of solar active regions'',
  {\it Royal Society of London Philosophical Transactions Series A\/}, {\bf
  281}, 391--404.
  {\small[\href{http://adsabs.harvard.edu/abs/1976RSPTA.281..391J}{ADS}]}

\bibitem[Jordan(1980)]{1980AA....86..355J}
Jordan, C., 1980, ``The energy balance of the solar transition region'', {\it
  Astron. Astrophys.\/}, {\bf 86}, 355--363.
  {\small[\href{http://adsabs.harvard.edu/abs/1980A&A....86..355J}{ADS}]}

\bibitem[Jordan {\it et~al.\/}(1987)]{1987MNRAS.225..903J}
Jordan, C., Ayres, T.R., Brown, A., Linsky, J.L. and Simon, T., 1987, ``The
  chromospheres and coronae of five G-K main-sequence stars'', {\it Mon. Not.
  R. Astron. Soc.\/}, {\bf 225}, 903--937.
  {\small[\href{http://adsabs.harvard.edu/abs/1987MNRAS.225..903J}{ADS}]}

\bibitem[Judge {\it et~al.\/}(1998)]{1998ApJ...502..981J}
Judge, P.G., Hansteen, V., Wikstol, O., Wilhelm, K., Schuehle, U. and Moran,
  T., 1998, ``Evidence in Support of the `Nanoflare' Picture of Coronal Heating
  from SUMER Data'', {\it Astrophys. J.\/}, {\bf 502}, 981.
  {\small[\href{http://dx.doi.org/10.1086/305915}{DOI}]},
  {\small[\href{http://adsabs.harvard.edu/abs/1998ApJ...502..981J}{ADS}]}

\bibitem[Kaiser {\it et~al.\/}(2008)]{2008SSRv..136....5K}
Kaiser, M.L., Kucera, T.A., Davila, J.M., St~Cyr, O.C., Guhathakurta, M. and
  Christian, E., 2008, ``The STEREO Mission: An Introduction'', {\it Space
  Science Reviews\/}, {\bf 136}, 5--16.
  {\small[\href{http://dx.doi.org/10.1007/s11214-007-9277-0}{DOI}]},
  {\small[\href{http://adsabs.harvard.edu/abs/2008SSRv..136....5K}{ADS}]}

\bibitem[Kano and Tsuneta(1995)]{1995ApJ...454..934K}
Kano, R. and Tsuneta, S., 1995, ``Scaling Law of Solar Coronal Loops Obtained
  with YOHKOH'', {\it Astrophys. J.\/}, {\bf 454}, 934.
  {\small[\href{http://dx.doi.org/10.1086/176547}{DOI}]},
  {\small[\href{http://adsabs.harvard.edu/abs/1995ApJ...454..934K}{ADS}]}

\bibitem[Kashyap and Drake(1998)]{1998ApJ...503..450K}
Kashyap, V. and Drake, J.J., 1998, ``Markov-Chain Monte Carlo Reconstruction of
  Emission Measure Distributions: Application to Solar Extreme-Ultraviolet
  Spectra'', {\it Astrophys. J.\/}, {\bf 503}, 450.
  {\small[\href{http://dx.doi.org/10.1086/305964}{DOI}]},
  {\small[\href{http://adsabs.harvard.edu/abs/1998ApJ...503..450K}{ADS}]}

\bibitem[Katsukawa and Tsuneta(2001)]{2001ApJ...557..343K}
Katsukawa, Y. and Tsuneta, S., 2001, ``Small Fluctuation of Coronal X-Ray
  Intensity and a Signature of Nanoflares'', {\it Astrophys. J.\/}, {\bf 557},
  343--350. {\small[\href{http://dx.doi.org/10.1086/321636}{DOI}]},
  {\small[\href{http://adsabs.harvard.edu/abs/2001ApJ...557..343K}{ADS}]}

\bibitem[Katsukawa and Tsuneta(2005)]{2005ApJ...621..498K}
Katsukawa, Y. and Tsuneta, S., 2005, ``Magnetic Properties at Footpoints of Hot
  and Cool Loops'', {\it Astrophys. J.\/}, {\bf 621}, 498--511.
  {\small[\href{http://dx.doi.org/10.1086/427488}{DOI}]},
  {\small[\href{http://adsabs.harvard.edu/abs/2005ApJ...621..498K}{ADS}]}

\bibitem[Klimchuk(1987)]{1987ApJ...323..368K}
Klimchuk, J.A., 1987, ``On the large-scale dynamics and magnetic structure of
  solar active regions'', {\it Astrophys. J.\/}, {\bf 323}, 368--379.
  {\small[\href{http://dx.doi.org/10.1086/165834}{DOI}]},
  {\small[\href{http://adsabs.harvard.edu/abs/1987ApJ...323..368K}{ADS}]}

\bibitem[Klimchuk(2000)]{2000SoPh..193...53K}
Klimchuk, J.A., 2000, ``Cross-Sectional Properties of Coronal Loops'', {\it
  Solar Phys.\/}, {\bf 193}, 53--75.
  {\small[\href{http://adsabs.harvard.edu/abs/2000SoPh..193...53K}{ADS}]}

\bibitem[Klimchuk(2006)]{2006SoPh..234...41K}
Klimchuk, J.A., 2006, ``On Solving the Coronal Heating Problem'', {\it Solar
  Phys.\/}, {\bf 234}, 41--77.
  {\small[\href{http://dx.doi.org/10.1007/s11207-006-0055-z}{DOI}]},
  {\small[\href{http://adsabs.harvard.edu/abs/2006SoPh..234...41K}{ADS}]},
  {\small[\href{http://arxiv.org/abs/arXiv:astro-ph/0511841}{{arXiv:astro-ph/0%
511841}}]}

\bibitem[Klimchuk and Gary(1995)]{1995ApJ...448..925K}
Klimchuk, J.A. and Gary, D.E., 1995, ``A Comparison of Active Region
  Temperatures and Emission Measures Observed in Soft X-Rays and Microwaves and
  Implications for Coronal Heating'', {\it Astrophys. J.\/}, {\bf 448}, 925.
  {\small[\href{http://dx.doi.org/10.1086/176021}{DOI}]},
  {\small[\href{http://adsabs.harvard.edu/abs/1995ApJ...448..925K}{ADS}]}

\bibitem[Klimchuk and Porter(1995)]{1995Natur.377..131K}
Klimchuk, J.A. and Porter, L.J., 1995, ``Scaling of heating rates in solar
  coronal loops'', {\it Nature\/}, {\bf 377}, 131--133.
  {\small[\href{http://dx.doi.org/10.1038/377131a0}{DOI}]},
  {\small[\href{http://adsabs.harvard.edu/abs/1995Natur.377..131K}{ADS}]}

\bibitem[Klimchuk {\it et~al.\/}(1992)]{1992PASJ...44L.181K}
Klimchuk, J.A., Lemen, J.R., Feldman, U., Tsuneta, S. and Uchida, Y., 1992,
  ``Thickness variations along coronal loops observed by the Soft X-ray
  Telescope on YOHKOH'', {\it Publ. Astron. Soc. Japan\/}, {\bf 44},
  L181--L185.
  {\small[\href{http://adsabs.harvard.edu/abs/1992PASJ...44L.181K}{ADS}]}

\bibitem[Klimchuk {\it et~al.\/}(2008)]{2008ApJ...682.1351K}
Klimchuk, J.A., Patsourakos, S. and Cargill, P.J., 2008, ``Highly Efficient
  Modeling of Dynamic Coronal Loops'', {\it Astrophys. J.\/}, {\bf 682},
  1351--1362. {\small[\href{http://dx.doi.org/10.1086/589426}{DOI}]},
  {\small[\href{http://adsabs.harvard.edu/abs/2008ApJ...682.1351K}{ADS}]},
  {\small[\href{http://arxiv.org/abs/0710.0185}{{arXiv:0710.0185}}]}

\bibitem[Ko {\it et~al.\/}(2009)]{2009ApJ...697.1956K}
Ko, Y.-K., Doschek, G.A., Warren, H.P. and Young, P.R., 2009, ``Hot Plasma in
  Nonflaring Active Regions Observed by the Extreme-Ultraviolet Imaging
  Spectrometer on Hinode'', {\it Astrophys. J.\/}, {\bf 697}, 1956--1970.
  {\small[\href{http://dx.doi.org/10.1088/0004-637X/697/2/1956}{DOI}]},
  {\small[\href{http://adsabs.harvard.edu/abs/2009ApJ...697.1956K}{ADS}]},
  {\small[\href{http://arxiv.org/abs/0903.3029}{{arXiv:0903.3029}}]}

\bibitem[Kopp and Poletto(1993)]{1993ApJ...418..496K}
Kopp, R.A. and Poletto, G., 1993, ``Coronal Heating by Nanoflares: Individual
  Events and Global Energetics'', {\it Astrophys. J.\/}, {\bf 418}, 496.
  {\small[\href{http://dx.doi.org/10.1086/173411}{DOI}]},
  {\small[\href{http://adsabs.harvard.edu/abs/1993ApJ...418..496K}{ADS}]}

\bibitem[Kopp {\it et~al.\/}(1985)]{1985SoPh...98...91K}
Kopp, R.A., Poletto, G., Noci, G. and Bruner, M., 1985, ``Analysis of loop
  flows observed on 27 March, 1980 by the UVSP instrument during the solar
  maximum mission'', {\it Solar Phys.\/}, {\bf 98}, 91--118.
  {\small[\href{http://dx.doi.org/10.1007/BF00177201}{DOI}]},
  {\small[\href{http://adsabs.harvard.edu/abs/1985SoPh...98...91K}{ADS}]}

\bibitem[Kosugi {\it et~al.\/}(2007)]{2007SoPh..243....3K}
Kosugi, T., Matsuzaki, K., Sakao, T., Shimizu, T., Sone, Y., Tachikawa, S.,
  Hashimoto, T., Minesugi, K., Ohnishi, A., Yamada, T., Tsuneta, S., Hara, H.,
  Ichimoto, K., Suematsu, Y., Shimojo, M., Watanabe, T., Shimada, S., Davis,
  J.M., Hill, L.D., Owens, J.K., Title, A.M., Culhane, J.L., Harra, L.K.,
  Doschek, G.A. and Golub, L., 2007, ``The Hinode (Solar-B) Mission: An
  Overview'', {\it Solar Phys.\/}, {\bf 243}, 3--17.
  {\small[\href{http://dx.doi.org/10.1007/s11207-007-9014-6}{DOI}]},
  {\small[\href{http://adsabs.harvard.edu/abs/2007SoPh..243....3K}{ADS}]}

\bibitem[{Kramar} {\it et~al.\/}(2009)]{2009SoPh..259..109K}
{Kramar}, M., {Jones}, S., {Davila}, J., {Inhester}, B. and {Mierla}, M., 2009,
  ``{On the Tomographic Reconstruction of the 3D Electron Density for the Solar
  Corona from STEREO COR1 Data}'', {\it Solar Phys.\/}, {\bf 259}, 109--121.
  {\small[\href{http://dx.doi.org/10.1007/s11207-009-9401-2}{DOI}]},
  {\small[\href{http://adsabs.harvard.edu/abs/2009SoPh..259..109K}{ADS}]}

\bibitem[Krieger {\it et~al.\/}(1972)]{1972SoPh...22..150K}
Krieger, A., Paolini, F., Vaiana, G.S. and Webb, D., 1972, ``Results from
  OSO-IV: the Long Term Behavior of X-Ray Emitting Regions'', {\it Solar
  Phys.\/}, {\bf 22}, 150--177.
  {\small[\href{http://dx.doi.org/10.1007/BF00145472}{DOI}]},
  {\small[\href{http://adsabs.harvard.edu/abs/1972SoPh...22..150K}{ADS}]}

\bibitem[Krieger(1978)]{1978SoPh...56..107K}
Krieger, A.S., 1978, ``The decay of coronal loops brightened by flares and
  transients'', {\it Solar Phys.\/}, {\bf 56}, 107--120.
  {\small[\href{http://dx.doi.org/10.1007/BF00152637}{DOI}]},
  {\small[\href{http://adsabs.harvard.edu/abs/1978SoPh...56..107K}{ADS}]}

\bibitem[Landi and Feldman(2004)]{2004ApJ...611..537L}
Landi, E. and Feldman, U., 2004, ``Models for Solar Magnetic Loops. IV. On the
  Relation between Coronal and Footpoint Plasma in Active Region Loops'', {\it
  Astrophys. J.\/}, {\bf 611}, 537--544.
  {\small[\href{http://dx.doi.org/10.1086/422169}{DOI}]},
  {\small[\href{http://adsabs.harvard.edu/abs/2004ApJ...611..537L}{ADS}]}

\bibitem[Landi and Feldman(2008)]{2008ApJ...672..674L}
Landi, E. and Feldman, U., 2008, ``The Thermal Structure of an Active Region
  Observed Outside the Solar Disk'', {\it Astrophys. J.\/}, {\bf 672},
  674--683. {\small[\href{http://dx.doi.org/10.1086/523629}{DOI}]},
  {\small[\href{http://adsabs.harvard.edu/abs/2008ApJ...672..674L}{ADS}]}

\bibitem[Landi and Landini(2004)]{2004ApJ...608.1133L}
Landi, E. and Landini, M., 2004, ``Models for Solar Magnetic Loops. III.
  Dynamic Models and Coronal Diagnostic Spectrometer Observations'', {\it
  Astrophys. J.\/}, {\bf 608}, 1133--1147.
  {\small[\href{http://dx.doi.org/10.1086/420813}{DOI}]},
  {\small[\href{http://adsabs.harvard.edu/abs/2004ApJ...608.1133L}{ADS}]}

\bibitem[Landi and Landini(2005)]{2005ApJ...618.1039L}
Landi, E. and Landini, M., 2005, ``Models for Solar Magnetic Loops. V. A New
  Diagnostic Technique to Compare Loop Models and Observations'', {\it
  Astrophys. J.\/}, {\bf 618}, 1039--1043.
  {\small[\href{http://dx.doi.org/10.1086/426015}{DOI}]},
  {\small[\href{http://adsabs.harvard.edu/abs/2005ApJ...618.1039L}{ADS}]}

\bibitem[Landi {\it et~al.\/}(2009)]{2009ApJ...695..221L}
Landi, E., Miralles, M.P., Curdt, W. and Hara, H., 2009, ``Physical Properties
  of Cooling Plasma in Quiescent Active Region Loops'', {\it Astrophys. J.\/},
  {\bf 695}, 221--237.
  {\small[\href{http://dx.doi.org/10.1088/0004-637X/695/1/221}{DOI}]},
  {\small[\href{http://adsabs.harvard.edu/abs/2009ApJ...695..221L}{ADS}]}

\bibitem[Landini and Landi(2002)]{2002AA...383..653L}
Landini, M. and Landi, E., 2002, ``Models for solar magnetic loops. I. A simple
  theoretical model and diagnostic procedure'', {\it Astron. Astrophys.\/},
  {\bf 383}, 653--660.
  {\small[\href{http://dx.doi.org/10.1051/0004-6361:20011759}{DOI}]},
  {\small[\href{http://adsabs.harvard.edu/abs/2002A&A...383..653L}{ADS}]}

\bibitem[Landini and Monsignori~Fossi(1975)]{1975AA....42..213L}
Landini, M. and Monsignori~Fossi, B.C., 1975, ``A loop model of active coronal
  regions'', {\it Astron. Astrophys.\/}, {\bf 42}, 213--220.
  {\small[\href{http://adsabs.harvard.edu/abs/1975A&A....42..213L}{ADS}]}

\bibitem[Lenz(2004)]{2004ApJ...604..433L}
Lenz, D.D., 2004, ``Effects of Flow on Structure and Abundances in Multispecies
  Solar Coronal Loops'', {\it Astrophys. J.\/}, {\bf 604}, 433--441.
  {\small[\href{http://dx.doi.org/10.1086/381879}{DOI}]},
  {\small[\href{http://adsabs.harvard.edu/abs/2004ApJ...604..433L}{ADS}]}

\bibitem[Lenz {\it et~al.\/}(1999)]{1999ApJ...517L.155L}
Lenz, D.D., Deluca, E.E., Golub, L., Rosner, R. and Bookbinder, J.A., 1999,
  ``Temperature and Emission-Measure Profiles along Long-lived Solar Coronal
  Loops Observed with the Transition Region and Coronal Explorer'', {\it
  Astrophys. J. Lett.\/}, {\bf 517}, L155--L158.
  {\small[\href{http://dx.doi.org/10.1086/312045}{DOI}]},
  {\small[\href{http://adsabs.harvard.edu/abs/1999ApJ...517L.155L}{ADS}]},
  {\small[\href{http://arxiv.org/abs/arXiv:astro-ph/9903491}{{arXiv:astro-ph/9%
903491}}]}

\bibitem[Litwin and Rosner(1998)]{1998ApJ...499..945L}
Litwin, C. and Rosner, R., 1998, ``Alfven Wave Transmission and Heating of
  Solar Coronal Loops'', {\it Astrophys. J.\/}, {\bf 499}, 945.
  {\small[\href{http://dx.doi.org/10.1086/305651}{DOI}]},
  {\small[\href{http://adsabs.harvard.edu/abs/1998ApJ...499..945L}{ADS}]}

\bibitem[L{\'o}pez~Fuentes {\it et~al.\/}(2006)]{2006ApJ...639..459L}
L{\'o}pez~Fuentes, M.C., Klimchuk, J.A. and D{\'e}moulin, P., 2006, ``The
  Magnetic Structure of Coronal Loops Observed by TRACE'', {\it Astrophys.
  J.\/}, {\bf 639}, 459--474.
  {\small[\href{http://dx.doi.org/10.1086/499155}{DOI}]},
  {\small[\href{http://adsabs.harvard.edu/abs/2006ApJ...639..459L}{ADS}]},
  {\small[\href{http://arxiv.org/abs/arXiv:astro-ph/0507462}{{arXiv:astro-ph/0%
507462}}]}

\bibitem[L{\'o}pez~Fuentes {\it et~al.\/}(2007)]{2007ApJ...657.1127L}
L{\'o}pez~Fuentes, M.C., Klimchuk, J.A. and Mandrini, C.H., 2007, ``The
  Temporal Evolution of Coronal Loops Observed by GOES SXI'', {\it Astrophys.
  J.\/}, {\bf 657}, 1127--1136.
  {\small[\href{http://dx.doi.org/10.1086/510662}{DOI}]},
  {\small[\href{http://adsabs.harvard.edu/abs/2007ApJ...657.1127L}{ADS}]},
  {\small[\href{http://arxiv.org/abs/arXiv:astro-ph/0611338}{{arXiv:astro-ph/0%
611338}}]}

\bibitem[Lundquist {\it et~al.\/}(2008{\natexlab{a}})]{2008ApJS..179..509L}
Lundquist, L.L., Fisher, G.H. and McTiernan, J.M., 2008{\natexlab{a}},
  ``Forward Modeling of Active Region Coronal Emissions. I. Methods and
  Testing'', {\it Astrophys. J. Suppl. Ser.\/}, {\bf 179}, 509--533.
  {\small[\href{http://dx.doi.org/10.1086/592775}{DOI}]},
  {\small[\href{http://adsabs.harvard.edu/abs/2008ApJS..179..509L}{ADS}]}

\bibitem[Lundquist {\it et~al.\/}(2008{\natexlab{b}})]{2008ApJ...689.1388L}
Lundquist, L.L., Fisher, G.H., Metcalf, T.R., Leka, K.D. and McTiernan, J.M.,
  2008{\natexlab{b}}, ``Forward Modeling of Active Region Coronal Emissions.
  II. Implications for Coronal Heating'', {\it Astrophys. J.\/}, {\bf 689},
  1388--1405. {\small[\href{http://dx.doi.org/10.1086/592760}{DOI}]},
  {\small[\href{http://adsabs.harvard.edu/abs/2008ApJ...689.1388L}{ADS}]}

\bibitem[MacNeice(1986)]{1986SoPh..103...47M}
MacNeice, P., 1986, ``A numerical hydrodynamic model of a heated coronal
  loop'', {\it Solar Phys.\/}, {\bf 103}, 47--66.
  {\small[\href{http://dx.doi.org/10.1007/BF00154858}{DOI}]},
  {\small[\href{http://adsabs.harvard.edu/abs/1986SoPh..103...47M}{ADS}]}

\bibitem[MacNeice {\it et~al.\/}(1985)]{1985SoPh...99..167M}
MacNeice, P., Pallavicini, R., Mason, H.E., Simnett, G.M., Antonucci, E.,
  Shine, R.A. and Dennis, B.R., 1985, ``Multiwavelength analysis of a well
  observed flare from SMM'', {\it Solar Phys.\/}, {\bf 99}, 167--188.
  {\small[\href{http://dx.doi.org/10.1007/BF00157307}{DOI}]},
  {\small[\href{http://adsabs.harvard.edu/abs/1985SoPh...99..167M}{ADS}]}

\bibitem[Maggio {\it et~al.\/}(2000)]{2000AA...356..627M}
Maggio, A., Pallavicini, R., Reale, F. and Tagliaferri, G., 2000, ``Twin X-ray
  flares and the active corona of AB Dor observed with BeppoSAX'', {\it Astron.
  Astrophys.\/}, {\bf 356}, 627--642.
  {\small[\href{http://adsabs.harvard.edu/abs/2000A&A...356..627M}{ADS}]}

\bibitem[Mandrini {\it et~al.\/}(2000)]{2000ApJ...530..999M}
Mandrini, C.H., D{\'e}moulin, P. and Klimchuk, J.A., 2000, ``Magnetic Field and
  Plasma Scaling Laws: Their Implications for Coronal Heating Models'', {\it
  Astrophys. J.\/}, {\bf 530}, 999--1015.
  {\small[\href{http://dx.doi.org/10.1086/308398}{DOI}]},
  {\small[\href{http://adsabs.harvard.edu/abs/2000ApJ...530..999M}{ADS}]}

\bibitem[Mariska {\it et~al.\/}(1980)]{1980ApJ...240..300M}
Mariska, J.T., Feldman, U. and Doschek, G.A., 1980, ``Physical conditions in
  the solar atmosphere above an active region'', {\it Astrophys. J.\/}, {\bf
  240}, 300--305. {\small[\href{http://dx.doi.org/10.1086/158233}{DOI}]},
  {\small[\href{http://adsabs.harvard.edu/abs/1980ApJ...240..300M}{ADS}]}

\bibitem[Martens {\it et~al.\/}(2000)]{2000ApJ...537..471M}
Martens, P.C.H., Kankelborg, C.C. and Berger, T.E., 2000, ``On the Nature of
  the `Moss' Observed by TRACE'', {\it Astrophys. J.\/}, {\bf 537}, 471--480.
  {\small[\href{http://dx.doi.org/10.1086/309000}{DOI}]},
  {\small[\href{http://adsabs.harvard.edu/abs/2000ApJ...537..471M}{ADS}]}

\bibitem[Martens {\it et~al.\/}(2002)]{2002ApJ...577L.115M}
Martens, P.C.H., Cirtain, J.W. and Schmelz, J.T., 2002, ``The Inadequacy of
  Temperature Measurements in the Solar Corona through Narrowband Filter and
  Line Ratios'', {\it Astrophys. J. Lett.\/}, {\bf 577}, L115--L117.
  {\small[\href{http://dx.doi.org/10.1086/344254}{DOI}]},
  {\small[\href{http://adsabs.harvard.edu/abs/2002ApJ...577L.115M}{ADS}]}

\bibitem[Mason {\it et~al.\/}(1999)]{1999SoPh..189..129M}
Mason, H.E., Landi, E., Pike, C.D. and Young, P.R., 1999, ``Electron density
  and temperature structure of two limb active regions observed by SOHO-CDS'',
  {\it Solar Phys.\/}, {\bf 189}, 129--146.
  {\small[\href{http://adsabs.harvard.edu/abs/1999SoPh..189..129M}{ADS}]}

\bibitem[{McLaughlin} and {Ofman}(2008)]{2008ApJ...682.1338M}
{McLaughlin}, J.~A. and {Ofman}, L., 2008, ``{Three-dimensional
  Magnetohydrodynamic Wave Behavior in Active Regions: Individual Loop Density
  Structure}'', {\it Astrophys. J.\/}, {\bf 682}, 1338--1350.
  {\small[\href{http://dx.doi.org/10.1086/588799}{DOI}]},
  {\small[\href{http://adsabs.harvard.edu/abs/2008ApJ...682.1338M}{ADS}]}

\bibitem[McTiernan(2009)]{2009ApJ...697...94M}
McTiernan, J.M., 2009, ``RHESSI/GOES Observations of the Nonflaring Sun from
  2002 to 2006'', {\it Astrophys. J.\/}, {\bf 697}, 94--99.
  {\small[\href{http://dx.doi.org/10.1088/0004-637X/697/1/94}{DOI}]},
  {\small[\href{http://adsabs.harvard.edu/abs/2009ApJ...697...94M}{ADS}]}

\bibitem[Mitra-Kraev and Benz(2001)]{2001AA...373..318M}
Mitra-Kraev, U. and Benz, A.O., 2001, ``A nanoflare heating model for the quiet
  solar corona'', {\it Astron. Astrophys.\/}, {\bf 373}, 318--328.
  {\small[\href{http://dx.doi.org/10.1051/0004-6361:20010524}{DOI}]},
  {\small[\href{http://adsabs.harvard.edu/abs/2001A&A...373..318M}{ADS}]},
  {\small[\href{http://arxiv.org/abs/arXiv:astro-ph/0104218}{{arXiv:astro-ph/0%
104218}}]}

\bibitem[Montesinos and Thomas(1989)]{1989ApJ...337..977M}
Montesinos, B. and Thomas, J.H., 1989, ``Siphon flows in isolated magnetic flux
  tubes. II - Adiabatic flows'', {\it Astrophys. J.\/}, {\bf 337}, 977--988.
  {\small[\href{http://dx.doi.org/10.1086/167169}{DOI}]},
  {\small[\href{http://adsabs.harvard.edu/abs/1989ApJ...337..977M}{ADS}]}

\bibitem[Montesinos and Thomas(1993)]{1993ApJ...402..314M}
Montesinos, B. and Thomas, J.H., 1993, ``Siphon flows in isolated magnetic flux
  tubes. V - Radiative flows with variable ionization'', {\it Astrophys. J.\/},
  {\bf 402}, 314--325. {\small[\href{http://dx.doi.org/10.1086/172135}{DOI}]},
  {\small[\href{http://adsabs.harvard.edu/abs/1993ApJ...402..314M}{ADS}]}

\bibitem[Morales and Charbonneau(2008)]{2008ApJ...682..654M}
Morales, L. and Charbonneau, P., 2008, ``Self-organized Critical Model of
  Energy Release in an Idealized Coronal Loop'', {\it Astrophys. J.\/}, {\bf
  682}, 654--666. {\small[\href{http://dx.doi.org/10.1086/588274}{DOI}]},
  {\small[\href{http://adsabs.harvard.edu/abs/2008ApJ...682..654M}{ADS}]}

\bibitem[M{\"u}ller {\it et~al.\/}(2003)]{2003AA...411..605M}
M{\"u}ller, D.A.N., Hansteen, V.H. and Peter, H., 2003, ``Dynamics of solar
  coronal loops. I. Condensation in cool loops and its effect on transition
  region lines'', {\it Astron. Astrophys.\/}, {\bf 411}, 605--613.
  {\small[\href{http://dx.doi.org/10.1051/0004-6361:20031328}{DOI}]},
  {\small[\href{http://adsabs.harvard.edu/abs/2003A&A...411..605M}{ADS}]}

\bibitem[M{\"u}ller {\it et~al.\/}(2004)]{2004AA...424..289M}
M{\"u}ller, D.A.N., Peter, H. and Hansteen, V.H., 2004, ``Dynamics of solar
  coronal loops. II. Catastrophic cooling and high-speed downflows'', {\it
  Astron. Astrophys.\/}, {\bf 424}, 289--300.
  {\small[\href{http://dx.doi.org/10.1051/0004-6361:20040403}{DOI}]},
  {\small[\href{http://adsabs.harvard.edu/abs/2004A&A...424..289M}{ADS}]},
  {\small[\href{http://arxiv.org/abs/arXiv:astro-ph/0405538}{{arXiv:astro-ph/0%
405538}}]}

\bibitem[M{\"u}ller {\it et~al.\/}(2005)]{2005AA...436.1067M}
M{\"u}ller, D.A.N., De~Groof, A., Hansteen, V.H. and Peter, H., 2005,
  ``High-speed coronal rain'', {\it Astron. Astrophys.\/}, {\bf 436},
  1067--1074.
  {\small[\href{http://dx.doi.org/10.1051/0004-6361:20042141}{DOI}]},
  {\small[\href{http://adsabs.harvard.edu/abs/2005A&A...436.1067M}{ADS}]}

\bibitem[Muller {\it et~al.\/}(1994)]{1994AA...283..232M}
Muller, R., Roudier, T., Vigneau, J. and Auffret, H., 1994, ``The proper motion
  of network bright points and the heating of the solar corona'', {\it Astron.
  Astrophys.\/}, {\bf 283}, 232--240.
  {\small[\href{http://adsabs.harvard.edu/abs/1994A&A...283..232M}{ADS}]}

\bibitem[Nagai(1980)]{1980SoPh...68..351N}
Nagai, F., 1980, ``A model of hot loops associated with solar flares. I -
  Gasdynamics in the loops'', {\it Solar Phys.\/}, {\bf 68}, 351--379.
  {\small[\href{http://adsabs.harvard.edu/abs/1980SoPh...68..351N}{ADS}]}

\bibitem[Nagai and Emslie(1984)]{1984ApJ...279..896N}
Nagai, F. and Emslie, A.G., 1984, ``Gas dynamics in the impulsive phase of
  solar flares. I Thick-target heating by nonthermal electrons'', {\it
  Astrophys. J.\/}, {\bf 279}, 896--908.
  {\small[\href{http://dx.doi.org/10.1086/161960}{DOI}]},
  {\small[\href{http://adsabs.harvard.edu/abs/1984ApJ...279..896N}{ADS}]}

\bibitem[Nagata {\it et~al.\/}(2003)]{2003ApJ...590.1095N}
Nagata, S., Hara, H., Kano, R., Kobayashi, K., Sakao, T., Shimizu, T., Tsuneta,
  S., Yoshida, T. and Gurman, J.B., 2003, ``Spatial and Temporal Properties of
  Hot and Cool Coronal Loops'', {\it Astrophys. J.\/}, {\bf 590}, 1095--1110.
  {\small[\href{http://dx.doi.org/10.1086/375127}{DOI}]},
  {\small[\href{http://adsabs.harvard.edu/abs/2003ApJ...590.1095N}{ADS}]}

\bibitem[{Nakariakov} and {Ofman}(2001)]{2001A&A...372L..53N}
{Nakariakov}, V.~M. and {Ofman}, L., 2001, ``{Determination of the coronal
  magnetic field by coronal loop oscillations}'', {\it Astron. Astrophys.\/},
  {\bf 372}, L53--L56.
  {\small[\href{http://dx.doi.org/10.1051/0004-6361:20010607}{DOI}]},
  {\small[\href{http://adsabs.harvard.edu/abs/2001A%26A...372L..53N}{ADS}]}

\bibitem[{Nakariakov} and {Verwichte}(2005)]{2005LRSP....2....3N}
{Nakariakov}, V.~M. and {Verwichte}, E., 2005, ``{Coronal Waves and
  Oscillations}'', {\it Living Reviews in Solar Physics\/}, {\bf 2}, 3--+.
  {\small[\href{http://adsabs.harvard.edu/abs/2005LRSP....2....3N}{ADS}]}

\bibitem[Nakariakov {\it et~al.\/}(1999)]{1999Sci...285..862N}
Nakariakov, V.M., Ofman, L., Deluca, E.E., Roberts, B. and Davila, J.M., 1999,
  ``TRACE observation of damped coronal loop oscillations: Implications for
  coronal heating'', {\it Science\/}, {\bf 285}, 862--864.
  {\small[\href{http://dx.doi.org/10.1126/science.285.5429.862}{DOI}]},
  {\small[\href{http://adsabs.harvard.edu/abs/1999Sci...285..862N}{ADS}]}

\bibitem[Nakariakov {\it et~al.\/}(2000)]{2000AA...362.1151N}
Nakariakov, V.M., Verwichte, E., Berghmans, D. and Robbrecht, E., 2000, ``Slow
  magnetoacoustic waves in coronal loops'', {\it Astron. Astrophys.\/}, {\bf
  362}, 1151--1157.
  {\small[\href{http://adsabs.harvard.edu/abs/2000A&A...362.1151N}{ADS}]}

\bibitem[Narain and Ulmschneider(1996)]{1996SSRv...75..453N}
Narain, U. and Ulmschneider, P., 1996, ``Chromospheric and Coronal Heating
  Mechanisms II'', {\it Space Science Reviews\/}, {\bf 75}, 453--509.
  {\small[\href{http://dx.doi.org/10.1007/BF00833341}{DOI}]},
  {\small[\href{http://adsabs.harvard.edu/abs/1996SSRv...75..453N}{ADS}]}

\bibitem[Nigro {\it et~al.\/}(2004)]{2004PhRvL..92s4501N}
Nigro, G., Malara, F., Carbone, V. and Veltri, P., 2004, ``Nanoflares and MHD
  Turbulence in Coronal Loops: A Hybrid Shell Model'', {\it Physical Review
  Letters\/}, {\bf 92}\penalty0 (19), 194\,501.
  {\small[\href{http://dx.doi.org/10.1103/PhysRevLett.92.194501}{DOI}]},
  {\small[\href{http://adsabs.harvard.edu/abs/2004PhRvL..92s4501N}{ADS}]}

\bibitem[Nitta(2000)]{2000SoPh..195..123N}
Nitta, N., 2000, ``The relation between hot and cool loops'', {\it Solar
  Phys.\/}, {\bf 195}, 123--133.
  {\small[\href{http://adsabs.harvard.edu/abs/2000SoPh..195..123N}{ADS}]}

\bibitem[Noci(1981)]{1981SoPh...69...63N}
Noci, G., 1981, ``Siphon flows in the solar corona'', {\it Solar Phys.\/}, {\bf
  69}, 63--76. {\small[\href{http://dx.doi.org/10.1007/BF00151256}{DOI}]},
  {\small[\href{http://adsabs.harvard.edu/abs/1981SoPh...69...63N}{ADS}]}

\bibitem[Noci {\it et~al.\/}(1989)]{1989ApJ...338.1131N}
Noci, G., Spadaro, D., Zappala, R.A. and Antiochos, S.K., 1989, ``Mass flows
  and the ionization states of coronal loops'', {\it Astrophys. J.\/}, {\bf
  338}, 1131--1138. {\small[\href{http://dx.doi.org/10.1086/167263}{DOI}]},
  {\small[\href{http://adsabs.harvard.edu/abs/1989ApJ...338.1131N}{ADS}]}

\bibitem[Noglik {\it et~al.\/}(2008)]{2008ApJ...674.1191N}
Noglik, J.B., Walsh, R.W. and Cirtain, J., 2008, ``Comparison of
  High-Resolution TRACE Data to Spectroscopic CDS Data for Temperature
  Determination'', {\it Astrophys. J.\/}, {\bf 674}, 1191--1200.
  {\small[\href{http://dx.doi.org/10.1086/525012}{DOI}]},
  {\small[\href{http://adsabs.harvard.edu/abs/2008ApJ...674.1191N}{ADS}]}

\bibitem[Ofman(2009)]{2009ApJ...694..502O}
Ofman, L., 2009, ``Three-Dimensional Magnetohydrodynamic Models of Twisted
  Multithreaded Coronal Loop Oscillations'', {\it Astrophys. J.\/}, {\bf 694},
  502--511.
  {\small[\href{http://dx.doi.org/10.1088/0004-637X/694/1/502}{DOI}]},
  {\small[\href{http://adsabs.harvard.edu/abs/2009ApJ...694..502O}{ADS}]}

\bibitem[Ofman and Aschwanden(2002)]{2002ApJ...576L.153O}
Ofman, L. and Aschwanden, M.J., 2002, ``Damping Time Scaling of Coronal Loop
  Oscillations Deduced from Transition Region and Coronal Explorer
  Observations'', {\it Astrophys. J.l\/}, {\bf 576}, L153--L156.
  {\small[\href{http://dx.doi.org/10.1086/343886}{DOI}]},
  {\small[\href{http://adsabs.harvard.edu/abs/2002ApJ...576L.153O}{ADS}]}

\bibitem[Ofman and Wang(2002)]{2002ApJ...580L..85O}
Ofman, L. and Wang, T., 2002, ``Hot Coronal Loop Oscillations Observed by
  SUMER: Slow Magnetosonic Wave Damping by Thermal Conduction'', {\it
  Astrophys. J. Lett.\/}, {\bf 580}, L85--L88.
  {\small[\href{http://dx.doi.org/10.1086/345548}{DOI}]},
  {\small[\href{http://adsabs.harvard.edu/abs/2002ApJ...580L..85O}{ADS}]}

\bibitem[Ofman and Wang(2008)]{2008AA...482L...9O}
Ofman, L. and Wang, T.J., 2008, ``Hinode observations of transverse waves with
  flows in coronal loops'', {\it Astron. Astrophys.\/}, {\bf 482}, L9--L12.
  {\small[\href{http://dx.doi.org/10.1051/0004-6361:20079340}{DOI}]},
  {\small[\href{http://adsabs.harvard.edu/abs/2008A&A...482L...9O}{ADS}]}

\bibitem[Ofman {\it et~al.\/}(1995)]{1995ApJ...444..471O}
Ofman, L., Davila, J.M. and Steinolfson, R.S., 1995, ``Coronal heating by the
  resonant absorption of Alfven waves: Wavenumber scaling laws.'', {\it
  Astrophys. J.\/}, {\bf 444}, 471--477.
  {\small[\href{http://dx.doi.org/10.1086/175621}{DOI}]},
  {\small[\href{http://adsabs.harvard.edu/abs/1995ApJ...444..471O}{ADS}]}

\bibitem[Ofman {\it et~al.\/}(1998)]{1998ApJ...493..474O}
Ofman, L., Klimchuk, J.A. and Davila, J.M., 1998, ``A Self-consistent Model for
  the Resonant Heating of Coronal Loops: The Effects of Coupling with the
  Chromosphere'', {\it Astrophys. J.\/}, {\bf 493}, 474--+.
  {\small[\href{http://dx.doi.org/10.1086/305109}{DOI}]},
  {\small[\href{http://adsabs.harvard.edu/abs/1998ApJ...493..474O}{ADS}]}

\bibitem[Ogawara {\it et~al.\/}(1991)]{1991SoPh..136....1O}
Ogawara, Y., Takano, T., Kato, T., Kosugi, T., Tsuneta, S., Watanabe, T.,
  Kondo, I. and Uchida, Y., 1991, ``The Solar-A Mission - an Overview'', {\it
  Solar Phys.\/}, {\bf 136}, 1--16.
  {\small[\href{http://dx.doi.org/10.1007/BF00151692}{DOI}]},
  {\small[\href{http://adsabs.harvard.edu/abs/1991SoPh..136....1O}{ADS}]}

\bibitem[O'Neill and Li(2005)]{2005AA...435.1159O}
O'Neill, I. and Li, X., 2005, ``Coronal loops heated by turbulence-driven
  Alfv{\'e}n waves: A two fluid model'', {\it Astron. Astrophys.\/}, {\bf 435},
  1159--1167.
  {\small[\href{http://dx.doi.org/10.1051/0004-6361:20041596}{DOI}]},
  {\small[\href{http://adsabs.harvard.edu/abs/2005A&A...435.1159O}{ADS}]}

\bibitem[Orayevskii and Sobelman(2002)]{2002russian}
Orayevskii, V.N. and Sobelman, I.~I., 2002, {\it Russian: Pis'ma v Astron.
  Zh.\/}, {\bf 28}, 457

\bibitem[Orlando and Peres(1999)]{1999PCEC...24..401O}
Orlando, S. and Peres, G., 1999, ``Effects on UV lines observations of
  stationary plasma flows confined in coronal loops'', {\it Physics and
  Chemistry of the Earth C\/}, {\bf 24}, 401--406.
  {\small[\href{http://dx.doi.org/10.1016/S1464-1917(99)00020-3}{DOI}]},
  {\small[\href{http://adsabs.harvard.edu/abs/1999PCEC...24..401O}{ADS}]}

\bibitem[Orlando {\it et~al.\/}(1995{\natexlab{a}})]{1995AA...294..861O}
Orlando, S., Peres, G. and Serio, S., 1995{\natexlab{a}}, ``Models of
  stationary siphon flows in stratified, thermally conducting coronal loops. 1:
  Regular solutions'', {\it Astron. Astrophys.\/}, {\bf 294}, 861--873.
  {\small[\href{http://adsabs.harvard.edu/abs/1995A&A...294..861O}{ADS}]}

\bibitem[Orlando {\it et~al.\/}(1995{\natexlab{b}})]{1995AA...300..549O}
Orlando, S., Peres, G. and Serio, S., 1995{\natexlab{b}}, ``Models of
  stationary siphon flows in stratified, thermally conducting coronal loops.
  II. Shocked solutions.'', {\it Astron. Astrophys.\/}, {\bf 300}, 549.
  {\small[\href{http://adsabs.harvard.edu/abs/1995A&A...300..549O}{ADS}]}

\bibitem[Orlando {\it et~al.\/}(2000)]{2000ApJ...528..524O}
Orlando, S., Peres, G. and Reale, F., 2000, ``The Sun as an X-Ray Star. I.
  Deriving the Emission Measure Distribution versus Temperature of the Whole
  Solar Corona from theYohkoh/Soft X-Ray Telescope Data'', {\it Astrophys.
  J.\/}, {\bf 528}, 524--536.
  {\small[\href{http://dx.doi.org/10.1086/308137}{DOI}]},
  {\small[\href{http://adsabs.harvard.edu/abs/2000ApJ...528..524O}{ADS}]}

\bibitem[Orlando {\it et~al.\/}(2001)]{2001ApJ...560..499O}
Orlando, S., Peres, G. and Reale, F., 2001, ``The Sun as an X-Ray Star. IV. The
  Contribution of Different Regions of the Corona to Its X-Ray Spectrum'', {\it
  Astrophys. J.\/}, {\bf 560}, 499--513.
  {\small[\href{http://dx.doi.org/10.1086/322333}{DOI}]},
  {\small[\href{http://adsabs.harvard.edu/abs/2001ApJ...560..499O}{ADS}]}

\bibitem[Orlando {\it et~al.\/}(2004)]{2004AA...424..677O}
Orlando, S., Peres, G. and Reale, F., 2004, ``The Sun as an X-ray star: Active
  region evolution, rotational modulation, and implications for stellar X-ray
  variability'', {\it Astron. Astrophys.\/}, {\bf 424}, 677--689.
  {\small[\href{http://dx.doi.org/10.1051/0004-6361:20040207}{DOI}]},
  {\small[\href{http://adsabs.harvard.edu/abs/2004A&A...424..677O}{ADS}]}

\bibitem[O'Shea {\it et~al.\/}(2007)]{2007AA...475L..25O}
O'Shea, E., Banerjee, D. and Doyle, J.G., 2007, ``Plasma condensation in
  coronal loops'', {\it Astron. Astrophys.\/}, {\bf 475}, L25--L28.
  {\small[\href{http://dx.doi.org/10.1051/0004-6361:20078617}{DOI}]},
  {\small[\href{http://adsabs.harvard.edu/abs/2007A&A...475L..25O}{ADS}]}

\bibitem[Parenti and Young(2008)]{2008AA...492..857P}
Parenti, S. and Young, P.R., 2008, ``On the ultraviolet signatures of small
  scale heating in coronal loops'', {\it Astron. Astrophys.\/}, {\bf 492},
  857--862.
  {\small[\href{http://dx.doi.org/10.1051/0004-6361:200809928}{DOI}]},
  {\small[\href{http://adsabs.harvard.edu/abs/2008A&A...492..857P}{ADS}]}

\bibitem[Parenti {\it et~al.\/}(2006)]{2006ApJ...651.1219P}
Parenti, S., Buchlin, E., Cargill, P.J., Galtier, S. and Vial, J.-C., 2006,
  ``Modeling the Radiative Signatures of Turbulent Heating in Coronal Loops'',
  {\it Astrophys. J.\/}, {\bf 651}, 1219--1228.
  {\small[\href{http://dx.doi.org/10.1086/507594}{DOI}]},
  {\small[\href{http://adsabs.harvard.edu/abs/2006ApJ...651.1219P}{ADS}]}

\bibitem[Parker(1988)]{1988ApJ...330..474P}
Parker, E.N., 1988, ``Nanoflares and the solar X-ray corona'', {\it Astrophys.
  J.\/}, {\bf 330}, 474--479.
  {\small[\href{http://dx.doi.org/10.1086/166485}{DOI}]},
  {\small[\href{http://adsabs.harvard.edu/abs/1988ApJ...330..474P}{ADS}]}

\bibitem[Parker(1991)]{1991ApJ...372..719P}
Parker, E.N., 1991, ``Heating solar coronal holes'', {\it Astrophys. J.\/},
  {\bf 372}, 719--727. {\small[\href{http://dx.doi.org/10.1086/170015}{DOI}]},
  {\small[\href{http://adsabs.harvard.edu/abs/1991ApJ...372..719P}{ADS}]}

\bibitem[Pascoe {\it et~al.\/}(2009)]{2009AA...505..319P}
Pascoe, D.J., de~Moortel, I. and McLaughlin, J.A., 2009, ``Impulsively
  generated oscillations in a 3D coronal loop'', {\it Astron. Astrophys.\/},
  {\bf 505}, 319--327.
  {\small[\href{http://dx.doi.org/10.1051/0004-6361/200912270}{DOI}]},
  {\small[\href{http://adsabs.harvard.edu/abs/2009A&A...505..319P}{ADS}]}

\bibitem[Patsourakos and Klimchuk(2005)]{2005ApJ...628.1023P}
Patsourakos, S. and Klimchuk, J.A., 2005, ``Coronal Loop Heating by Nanoflares:
  The Impact of the Field-aligned Distribution of the Heating on Loop
  Observations'', {\it Astrophys. J.\/}, {\bf 628}, 1023--1030.
  {\small[\href{http://dx.doi.org/10.1086/430662}{DOI}]},
  {\small[\href{http://adsabs.harvard.edu/abs/2005ApJ...628.1023P}{ADS}]}

\bibitem[Patsourakos and Klimchuk(2006)]{2006ApJ...647.1452P}
Patsourakos, S. and Klimchuk, J.A., 2006, ``Nonthermal Spectral Line Broadening
  and the Nanoflare Model'', {\it Astrophys. J.\/}, {\bf 647}, 1452--1465.
  {\small[\href{http://dx.doi.org/10.1086/505517}{DOI}]},
  {\small[\href{http://adsabs.harvard.edu/abs/2006ApJ...647.1452P}{ADS}]}

\bibitem[Patsourakos and Klimchuk(2007)]{2007ApJ...667..591P}
Patsourakos, S. and Klimchuk, J.A., 2007, ``The Cross-Field Thermal Structure
  of Coronal Loops from Triple-Filter TRACE Observations'', {\it Astrophys.
  J.\/}, {\bf 667}, 591--601.
  {\small[\href{http://dx.doi.org/10.1086/520713}{DOI}]},
  {\small[\href{http://adsabs.harvard.edu/abs/2007ApJ...667..591P}{ADS}]}

\bibitem[Patsourakos and Klimchuk(2008)]{2008ApJ...689.1406P}
Patsourakos, S. and Klimchuk, J.A., 2008, ``Static and Impulsive Models of
  Solar Active Regions'', {\it Astrophys. J.\/}, {\bf 689}, 1406--1411.
  {\small[\href{http://dx.doi.org/10.1086/592683}{DOI}]},
  {\small[\href{http://adsabs.harvard.edu/abs/2008ApJ...689.1406P}{ADS}]},
  {\small[\href{http://arxiv.org/abs/0808.2745}{{arXiv:0808.2745}}]}

\bibitem[Patsourakos and Klimchuk(2009)]{2009ApJ...696..760P}
Patsourakos, S. and Klimchuk, J.A., 2009, ``Spectroscopic Observations of Hot
  Lines Constraining Coronal Heating in Solar Active Regions'', {\it Astrophys.
  J.\/}, {\bf 696}, 760--765.
  {\small[\href{http://dx.doi.org/10.1088/0004-637X/696/1/760}{DOI}]},
  {\small[\href{http://adsabs.harvard.edu/abs/2009ApJ...696..760P}{ADS}]},
  {\small[\href{http://arxiv.org/abs/0903.3880}{{arXiv:0903.3880}}]}

\bibitem[Patsourakos {\it et~al.\/}(2004)]{2004ApJ...603..322P}
Patsourakos, S., Klimchuk, J.A. and MacNeice, P.J., 2004, ``The Inability of
  Steady-Flow Models to Explain the Extreme-Ultraviolet Coronal Loops'', {\it
  Astrophys. J.\/}, {\bf 603}, 322--329.
  {\small[\href{http://dx.doi.org/10.1086/381426}{DOI}]},
  {\small[\href{http://adsabs.harvard.edu/abs/2004ApJ...603..322P}{ADS}]}

\bibitem[Pek{\"u}nl{\"u} {\it et~al.\/}(2001)]{2001MNRAS.326..675P}
Pek{\"u}nl{\"u}, E.R., {\c C}ak{\i}rl{\i}, {\"O}. and {\"O}zetken, E., 2001,
  ``Solar coronal heating by magnetosonic waves'', {\it Mon. Not. R. Astron.
  Soc.\/}, {\bf 326}, 675--685.
  {\small[\href{http://dx.doi.org/10.1046/j.1365-8711.2001.04639.x}{DOI}]},
  {\small[\href{http://adsabs.harvard.edu/abs/2001MNRAS.326..675P}{ADS}]}

\bibitem[Peres and Vaiana(1990)]{1990MmSAI..61..401P}
Peres, G. and Vaiana, G.S., 1990, ``X-ray observations, scaling laws and
  magnetic fields'', {\it Memorie della Societa Astronomica Italiana\/}, {\bf
  61}, 401--430.
  {\small[\href{http://adsabs.harvard.edu/abs/1990MmSAI..61..401P}{ADS}]}

\bibitem[Peres {\it et~al.\/}(1982)]{1982ApJ...252..791P}
Peres, G., Serio, S., Vaiana, G.S. and Rosner, R., 1982, ``Coronal closed
  structures. IV - Hydrodynamical stability and response to heating
  perturbations'', {\it Astrophys. J.\/}, {\bf 252}, 791--799.
  {\small[\href{http://dx.doi.org/10.1086/159601}{DOI}]},
  {\small[\href{http://adsabs.harvard.edu/abs/1982ApJ...252..791P}{ADS}]}

\bibitem[Peres {\it et~al.\/}(1992)]{1992ApJ...389..777P}
Peres, G., Spadaro, D. and Noci, G., 1992, ``Steady siphon flows in closed
  coronal structures - Comparison with extreme-ultraviolet observations'', {\it
  Astrophys. J.\/}, {\bf 389}, 777--783.
  {\small[\href{http://dx.doi.org/10.1086/171250}{DOI}]},
  {\small[\href{http://adsabs.harvard.edu/abs/1992ApJ...389..777P}{ADS}]}

\bibitem[Peres {\it et~al.\/}(1993)]{1993pssc.symp..151P}
Peres, G., Reale, F. and Serio, S., 1993, ``Hydrodynamics and Diagnostics of
  Coronal Loops Subject to Dynamic Heating'', in {\it Physics of Solar and
  Stellar Coronae: G.S. Vaiana Memorial Symposuim\/}, (Eds.) Linsky, J.L.,
  Serio, S., Proceedings of a conference of the IAU, held in Palermo, Italy,
  22\,--\,26 June, vol. 183 of Astrophysics and Space Science Library, p. 151,
  Kluwer, Dordrecht; Boston.
  {\small[\href{http://adsabs.harvard.edu/abs/1993pssc.symp..151P}{ADS}]}

\bibitem[Peres {\it et~al.\/}(1994)]{1994ApJ...422..412P}
Peres, G., Reale, F. and Golub, L., 1994, ``Loop models of low coronal
  structures observed by the Normal Incidence X-Ray Telescope (NIXT)'', {\it
  Astrophys. J.\/}, {\bf 422}, 412--415.
  {\small[\href{http://dx.doi.org/10.1086/173736}{DOI}]},
  {\small[\href{http://adsabs.harvard.edu/abs/1994ApJ...422..412P}{ADS}]}

\bibitem[Peres {\it et~al.\/}(2000)]{2000ApJ...528..537P}
Peres, G., Orlando, S., Reale, F., Rosner, R. and Hudson, H., 2000, ``The Sun
  as an X-Ray Star. II. Using theYohkoh/Soft X-Ray Telescope-derived Solar
  Emission Measure versus Temperature to Interpret Stellar X-Ray
  Observations'', {\it Astrophys. J.\/}, {\bf 528}, 537--551.
  {\small[\href{http://dx.doi.org/10.1086/308136}{DOI}]},
  {\small[\href{http://adsabs.harvard.edu/abs/2000ApJ...528..537P}{ADS}]}

\bibitem[Peres {\it et~al.\/}(2001)]{2001ApJ...563.1045P}
Peres, G., Orlando, S., Reale, F. and Rosner, R., 2001, ``The Distribution of
  the Emission Measure, and of the Heating Budget, among the Loops in the
  Corona'', {\it Astrophys. J.\/}, {\bf 563}, 1045--1054.
  {\small[\href{http://dx.doi.org/10.1086/323769}{DOI}]},
  {\small[\href{http://adsabs.harvard.edu/abs/2001ApJ...563.1045P}{ADS}]},
  {\small[\href{http://arxiv.org/abs/arXiv:astro-ph/0111192}{{arXiv:astro-ph/0%
111192}}]}

\bibitem[Peres {\it et~al.\/}(2004)]{2004ApJ...612..472P}
Peres, G., Orlando, S. and Reale, F., 2004, ``Are Coronae of Late-Type Stars
  Made of Solar-like Structures? The X-Ray Surface Flux versus Hardness Ratio
  Diagram and the Pressure-Temperature Correlation'', {\it Astrophys. J.\/},
  {\bf 612}, 472--480. {\small[\href{http://dx.doi.org/10.1086/422461}{DOI}]},
  {\small[\href{http://adsabs.harvard.edu/abs/2004ApJ...612..472P}{ADS}]},
  {\small[\href{http://arxiv.org/abs/arXiv:astro-ph/0405281}{{arXiv:astro-ph/0%
405281}}]}

\bibitem[Peter(1999)]{1999ApJ...516..490P}
Peter, H., 1999, ``Analysis of Transition-Region Emission-Line Profiles from
  Full-Disk Scans of the Sun Using the SUMER Instrument on SOHO'', {\it
  Astrophys. J.\/}, {\bf 516}, 490--504.
  {\small[\href{http://dx.doi.org/10.1086/307102}{DOI}]},
  {\small[\href{http://adsabs.harvard.edu/abs/1999ApJ...516..490P}{ADS}]}

\bibitem[Poletto {\it et~al.\/}(1975)]{1975SoPh...44...83P}
Poletto, G., Vaiana, G.S., Zombeck, M.V., Krieger, A.S. and Timothy, A.F.,
  1975, ``A comparison of coronal X-ray structures of active regions with
  magnetic fields computed from photospheric observations'', {\it Solar
  Phys.\/}, {\bf 44}, 83--99.
  {\small[\href{http://dx.doi.org/10.1007/BF00156848}{DOI}]},
  {\small[\href{http://adsabs.harvard.edu/abs/1975SoPh...44...83P}{ADS}]}

\bibitem[Porter and Klimchuk(1995)]{1995ApJ...454..499P}
Porter, L.J. and Klimchuk, J.A., 1995, ``Soft X-Ray Loops and Coronal
  Heating'', {\it Astrophys. J.\/}, {\bf 454}, 499.
  {\small[\href{http://dx.doi.org/10.1086/176501}{DOI}]},
  {\small[\href{http://adsabs.harvard.edu/abs/1995ApJ...454..499P}{ADS}]}

\bibitem[Porter {\it et~al.\/}(1994)]{1994ApJ...435..482P}
Porter, L.J., Klimchuk, J.A. and Sturrock, P.A., 1994, ``The possible role of
  MHD waves in heating the solar corona'', {\it Astrophys. J.\/}, {\bf 435},
  482--501. {\small[\href{http://dx.doi.org/10.1086/174830}{DOI}]},
  {\small[\href{http://adsabs.harvard.edu/abs/1994ApJ...435..482P}{ADS}]}

\bibitem[Priest(1978)]{1978SoPh...58...57P}
Priest, E.R., 1978, ``The structure of coronal loops'', {\it Solar Phys.\/},
  {\bf 58}, 57--87. {\small[\href{http://dx.doi.org/10.1007/BF00152555}{DOI}]},
  {\small[\href{http://adsabs.harvard.edu/abs/1978SoPh...58...57P}{ADS}]}

\bibitem[Priest(1981)]{1981sars.work..213P}
Priest, E.R., 1981, ``Theory of loop flows and instability'', in {\it Solar
  Active Regions: A monograph from Skylab Solar Workshop III\/}, (Ed.) Orrall,
  F.Q.,
  {\small[\href{http://adsabs.harvard.edu/abs/1981sars.work..213P}{ADS}]}

\bibitem[Priest {\it et~al.\/}(2000)]{2000ApJ...539.1002P}
Priest, E.R., Foley, C.R., Heyvaerts, J., Arber, T.D., Mackay, D., Culhane,
  J.L. and Acton, L.W., 2000, ``A Method to Determine the Heating Mechanisms of
  the Solar Corona'', {\it Astrophys. J.\/}, {\bf 539}, 1002--1022.
  {\small[\href{http://dx.doi.org/10.1086/309238}{DOI}]},
  {\small[\href{http://adsabs.harvard.edu/abs/2000ApJ...539.1002P}{ADS}]}

\bibitem[Rappazzo {\it et~al.\/}(2007)]{2007ApJ...657L..47R}
Rappazzo, A.F., Velli, M., Einaudi, G. and Dahlburg, R.B., 2007, ``Coronal
  Heating, Weak MHD Turbulence, and Scaling Laws'', {\it Astrophys. J.
  Lett.\/}, {\bf 657}, L47--L51.
  {\small[\href{http://dx.doi.org/10.1086/512975}{DOI}]},
  {\small[\href{http://adsabs.harvard.edu/abs/2007ApJ...657L..47R}{ADS}]},
  {\small[\href{http://arxiv.org/abs/arXiv:astro-ph/0701872}{{arXiv:astro-ph/0%
701872}}]}

\bibitem[Raymond {\it et~al.\/}(1976)]{1976ApJ...204..290R}
Raymond, J.C., Cox, D.P. and Smith, B.W., 1976, ``Radiative cooling of a
  low-density plasma'', {\it Astrophys. J.\/}, {\bf 204}, 290--292.
  {\small[\href{http://dx.doi.org/10.1086/154170}{DOI}]},
  {\small[\href{http://adsabs.harvard.edu/abs/1976ApJ...204..290R}{ADS}]}

\bibitem[Reale(1999)]{1999SoPh..190..139R}
Reale, F., 1999, ``Inclination of large coronal loops observed by TRACE'', {\it
  Solar Phys.\/}, {\bf 190}, 139--144.
  {\small[\href{http://dx.doi.org/10.1023/A:1005278823545}{DOI}]},
  {\small[\href{http://adsabs.harvard.edu/abs/1999SoPh..190..139R}{ADS}]}

\bibitem[Reale(2002{\natexlab{a}})]{2002ASPC..277..103R}
Reale, F., 2002{\natexlab{a}}, ``Stellar Flare Modeling'', in {\it Stellar
  Coronae in the Chandra and XMM-NEWTON Era\/}, (Eds.) Favata, F., Drake, J.J.,
  vol. 277 of Astronomical Society of the Pacific Conference Series,
  {\small[\href{http://adsabs.harvard.edu/abs/2002ASPC..277..103R}{ADS}]}

\bibitem[Reale(2002{\natexlab{b}})]{2002ApJ...580..566R}
Reale, F., 2002{\natexlab{b}}, ``More on the Determination of the Coronal
  Heating Function from Yohkoh Data'', {\it Astrophys. J.\/}, {\bf 580},
  566--573. {\small[\href{http://dx.doi.org/10.1086/343123}{DOI}]},
  {\small[\href{http://adsabs.harvard.edu/abs/2002ApJ...580..566R}{ADS}]},
  {\small[\href{http://arxiv.org/abs/arXiv:astro-ph/0207550}{{arXiv:astro-ph/0%
207550}}]}

\bibitem[Reale(2003)]{2003AdSpR..32.1057R}
Reale, F., 2003, ``Modeling solar and stellar flares'', {\it Advances in Space
  Research\/}, {\bf 32}, 1057--1066.
  {\small[\href{http://dx.doi.org/10.1016/S0273-1177(03)00309-0}{DOI}]},
  {\small[\href{http://adsabs.harvard.edu/abs/2003AdSpR..32.1057R}{ADS}]}

\bibitem[Reale(2005)]{2005ESASP.600E..27R}
Reale, F., 2005, ``Sub-Structuring Dynamics and Heating in Dense Coronal
  Structures'', in {\it The Dynamic Sun: Challenges for Theory and
  Observations\/},
  {\small[\href{http://adsabs.harvard.edu/abs/2005ESASP.600E..27R}{ADS}]}

\bibitem[Reale(2007)]{2007AA...471..271R}
Reale, F., 2007, ``Diagnostics of stellar flares from X-ray observations: from
  the decay to the rise phase'', {\it Astron. Astrophys.\/}, {\bf 471},
  271--279. {\small[\href{http://dx.doi.org/10.1051/0004-6361:20077223}{DOI}]},
  {\small[\href{http://adsabs.harvard.edu/abs/2007A&A...471..271R}{ADS}]},
  {\small[\href{http://arxiv.org/abs/arXiv:0705.3254}{{arXiv:0705.3254}}]}

\bibitem[Reale and Ciaravella(2006)]{2006AA...449.1177R}
Reale, F. and Ciaravella, A., 2006, ``Analysis of a multi-wavelength
  time-resolved observation of a coronal loop'', {\it Astron. Astrophys.\/},
  {\bf 449}, 1177--1192.
  {\small[\href{http://dx.doi.org/10.1051/0004-6361:20054314}{DOI}]},
  {\small[\href{http://adsabs.harvard.edu/abs/2006A&A...449.1177R}{ADS}]},
  {\small[\href{http://arxiv.org/abs/arXiv:astro-ph/0512397}{{arXiv:astro-ph/0%
512397}}]}

\bibitem[Reale and Orlando(2008)]{2008ApJ...684..715R}
Reale, F. and Orlando, S., 2008, ``Nonequilibrium of Ionization and the
  Detection of Hot Plasma in Nanoflare-heated Coronal Loops'', {\it Astrophys.
  J.\/}, {\bf 684}, 715--724.
  {\small[\href{http://dx.doi.org/10.1086/590338}{DOI}]},
  {\small[\href{http://adsabs.harvard.edu/abs/2008ApJ...684..715R}{ADS}]},
  {\small[\href{http://arxiv.org/abs/0805.3512}{{arXiv:0805.3512}}]}

\bibitem[Reale and Peres(2000)]{2000ApJ...528L..45R}
Reale, F. and Peres, G., 2000, ``TRACE-derived Temperature and Emission Measure
  Profiles along Long-lived Coronal Loops: The Role of Filamentation'', {\it
  Astrophys. J. Lett.\/}, {\bf 528}, L45--L48.
  {\small[\href{http://dx.doi.org/10.1086/312414}{DOI}]},
  {\small[\href{http://adsabs.harvard.edu/abs/2000ApJ...528L..45R}{ADS}]},
  {\small[\href{http://arxiv.org/abs/arXiv:astro-ph/9911096}{{arXiv:astro-ph/9%
911096}}]}

\bibitem[Reale {\it et~al.\/}(1988)]{1988ApJ...328..256R}
Reale, F., Peres, G., Serio, S., Rosner, R. and Schmitt, J.H.M.M., 1988,
  ``Hydrodynamic modeling of an X-ray flare on Proxima Centauri observed by the
  Einstein telescope'', {\it Astrophys. J.\/}, {\bf 328}, 256--264.
  {\small[\href{http://dx.doi.org/10.1086/166288}{DOI}]},
  {\small[\href{http://adsabs.harvard.edu/abs/1988ApJ...328..256R}{ADS}]}

\bibitem[Reale {\it et~al.\/}(1993)]{1993AA...272..486R}
Reale, F., Serio, S. and Peres, G., 1993, ``Dynamics of the decay of confined
  stellar X-ray flares'', {\it Astron. Astrophys.\/}, {\bf 272}, 486.
  {\small[\href{http://adsabs.harvard.edu/abs/1993A&A...272..486R}{ADS}]}

\bibitem[Reale {\it et~al.\/}(1996)]{1996AA...316..215R}
Reale, F., Peres, G. and Serio, S., 1996, ``Radiatively-driven downdrafts and
  redshifts in transition region lines. I. Reference model.'', {\it Astron.
  Astrophys.\/}, {\bf 316}, 215--228.
  {\small[\href{http://adsabs.harvard.edu/abs/1996A&A...316..215R}{ADS}]}

\bibitem[Reale {\it et~al.\/}(1997{\natexlab{a}})]{1997AA...325..782R}
Reale, F., Betta, R., Peres, G., Serio, S. and McTiernan, J.,
  1997{\natexlab{a}}, ``Determination of the length of coronal loops from the
  decay of X-ray flares I. Solar flares observed with YOHKOH SXT.'', {\it
  Astron. Astrophys.\/}, {\bf 325}, 782--790.
  {\small[\href{http://adsabs.harvard.edu/abs/1997A&A...325..782R}{ADS}]}

\bibitem[Reale {\it et~al.\/}(1997{\natexlab{b}})]{1997AA...318..506R}
Reale, F., Peres, G. and Serio, S., 1997{\natexlab{b}}, ``Radiatively driven
  downdrafts and redshifts in transition region lines. II. Exploring the
  parameter space.'', {\it Astron. Astrophys.\/}, {\bf 318}, 506--520.
  {\small[\href{http://adsabs.harvard.edu/abs/1997A&A...318..506R}{ADS}]}

\bibitem[Reale {\it et~al.\/}(2000{\natexlab{a}})]{2000ApJ...535..423R}
Reale, F., Peres, G., Serio, S., Betta, R.M., DeLuca, E.E. and Golub, L.,
  2000{\natexlab{a}}, ``A Brightening Coronal Loop Observed by TRACE. II. Loop
  Modeling and Constraints on Heating'', {\it Astrophys. J.\/}, {\bf 535},
  423--437. {\small[\href{http://dx.doi.org/10.1086/308817}{DOI}]},
  {\small[\href{http://adsabs.harvard.edu/abs/2000ApJ...535..423R}{ADS}]}

\bibitem[Reale {\it et~al.\/}(2000{\natexlab{b}})]{2000ApJ...535..412R}
Reale, F., Peres, G., Serio, S., DeLuca, E.E. and Golub, L.,
  2000{\natexlab{b}}, ``A Brightening Coronal Loop Observed by TRACE. I.
  Morphology and Evolution'', {\it Astrophys. J.\/}, {\bf 535}, 412--422.
  {\small[\href{http://dx.doi.org/10.1086/308816}{DOI}]},
  {\small[\href{http://adsabs.harvard.edu/abs/2000ApJ...535..412R}{ADS}]}

\bibitem[Reale {\it et~al.\/}(2001)]{2001ApJ...557..906R}
Reale, F., Peres, G. and Orlando, S., 2001, ``The Sun as an X-Ray Star. III.
  Flares'', {\it Astrophys. J.\/}, {\bf 557}, 906--920.
  {\small[\href{http://dx.doi.org/10.1086/321598}{DOI}]},
  {\small[\href{http://adsabs.harvard.edu/abs/2001ApJ...557..906R}{ADS}]},
  {\small[\href{http://arxiv.org/abs/arXiv:astro-ph/0104021}{{arXiv:astro-ph/0%
104021}}]}

\bibitem[Reale {\it et~al.\/}(2002)]{2002AA...383..952R}
Reale, F., Bocchino, F. and Peres, G., 2002, ``Modeling non-confined coronal
  flares: Dynamics and X-ray diagnostics'', {\it Astron. Astrophys.\/}, {\bf
  383}, 952--971.
  {\small[\href{http://dx.doi.org/10.1051/0004-6361:20011792}{DOI}]},
  {\small[\href{http://adsabs.harvard.edu/abs/2002A&A...383..952R}{ADS}]},
  {\small[\href{http://arxiv.org/abs/arXiv:astro-ph/0112333}{{arXiv:astro-ph/0%
112333}}]}

\bibitem[Reale {\it et~al.\/}(2004)]{2004AA...416..733R}
Reale, F., G{\"u}del, M., Peres, G. and Audard, M., 2004, ``Modeling an X-ray
  flare on Proxima Centauri: Evidence of two flaring loop components and of two
  heating mechanisms at work'', {\it Astron. Astrophys.\/}, {\bf 416},
  733--747. {\small[\href{http://dx.doi.org/10.1051/0004-6361:20034027}{DOI}]},
  {\small[\href{http://adsabs.harvard.edu/abs/2004A&A...416..733R}{ADS}]},
  {\small[\href{http://arxiv.org/abs/arXiv:astro-ph/0312267}{{arXiv:astro-ph/0%
312267}}]}

\bibitem[Reale {\it et~al.\/}(2005)]{2005ApJ...633..489R}
Reale, F., Nigro, G., Malara, F., Peres, G. and Veltri, P., 2005, ``Modeling a
  Coronal Loop Heated by Magnetohydrodynamic Turbulence Nanoflares'', {\it
  Astrophys. J.\/}, {\bf 633}, 489--498.
  {\small[\href{http://dx.doi.org/10.1086/444409}{DOI}]},
  {\small[\href{http://adsabs.harvard.edu/abs/2005ApJ...633..489R}{ADS}]},
  {\small[\href{http://arxiv.org/abs/arXiv:astro-ph/0506694}{{arXiv:astro-ph/0%
506694}}]}

\bibitem[Reale {\it et~al.\/}(2007)]{2007Sci...318.1582R}
Reale, F., Parenti, S., Reeves, K.K., Weber, M., Bobra, M.G., Barbera, M.,
  Kano, R., Narukage, N., Shimojo, M., Sakao, T., Peres, G. and Golub, L.,
  2007, ``Fine Thermal Structure of a Coronal Active Region'', {\it Science\/},
  {\bf 318}, 1582--.
  {\small[\href{http://dx.doi.org/10.1126/science.1146590}{DOI}]},
  {\small[\href{http://adsabs.harvard.edu/abs/2007Sci...318.1582R}{ADS}]}

\bibitem[Reale {\it et~al.\/}(2009{\natexlab{a}})]{2009ApJ...704L..58R}
Reale, F., Mc~Tiernan, J.M. and Testa, P., 2009{\natexlab{a}}, ``Comparison of
  Hinode/XRT and RHESSI Detection of Hot Plasma in the Non-Flaring Solar
  Corona'', {\it Astrophys. J. Lett.\/}, {\bf 704}, L58--L61.
  {\small[\href{http://dx.doi.org/10.1088/0004-637X/704/1/L58}{DOI}]},
  {\small[\href{http://adsabs.harvard.edu/abs/2009ApJ...704L..58R}{ADS}]},
  {\small[\href{http://arxiv.org/abs/0909.2529}{{arXiv:0909.2529}}]}

\bibitem[Reale {\it et~al.\/}(2009{\natexlab{b}})]{2009ApJ...698..756R}
Reale, F., Testa, P., Klimchuk, J.A. and Parenti, S., 2009{\natexlab{b}},
  ``Evidence of Widespread Hot Plasma in a Nonflaring Coronal Active Region
  from Hinode/X-Ray Telescope'', {\it Astrophys. J.\/}, {\bf 698}, 756--765.
  {\small[\href{http://dx.doi.org/10.1088/0004-637X/698/1/756}{DOI}]},
  {\small[\href{http://adsabs.harvard.edu/abs/2009ApJ...698..756R}{ADS}]},
  {\small[\href{http://arxiv.org/abs/0904.0878}{{arXiv:0904.0878}}]}

\bibitem[Reeves {\it et~al.\/}(1977)]{1977ApOpt..16..837R}
Reeves, E.M., Timothy, J.G. and Huber, M.C.E., 1977, ``Extreme UV
  spectroheliometer on the Apollo Telescope Mount'', {\it Appl. Optics\/}, {\bf
  16}, 837--848.
  {\small[\href{http://adsabs.harvard.edu/abs/1977ApOpt..16..837R}{ADS}]}

\bibitem[Reeves and Warren(2002)]{2002ApJ...578..590R}
Reeves, K.K. and Warren, H.P., 2002, ``Modeling the Cooling of Postflare
  Loops'', {\it Astrophys. J.\/}, {\bf 578}, 590--597.
  {\small[\href{http://dx.doi.org/10.1086/342310}{DOI}]},
  {\small[\href{http://adsabs.harvard.edu/abs/2002ApJ...578..590R}{ADS}]}

\bibitem[Reidy {\it et~al.\/}(1968)]{1968ApJ...151..333R}
Reidy, W.P., Vaiana, G.S., Zehnpfennig, T. and Giacconi, R., 1968, ``Study of
  X-Ray Images of the Sun at Solar Minimum'', {\it Astrophys. J.\/}, {\bf 151},
  333. {\small[\href{http://dx.doi.org/10.1086/149440}{DOI}]},
  {\small[\href{http://adsabs.harvard.edu/abs/1968ApJ...151..333R}{ADS}]}

\bibitem[Rosner {\it et~al.\/}(1978)]{1978ApJ...220..643R}
Rosner, R., Tucker, W.H. and Vaiana, G.S., 1978, ``Dynamics of the quiescent
  solar corona'', {\it Astrophys. J.\/}, {\bf 220}, 643--645.
  {\small[\href{http://dx.doi.org/10.1086/155949}{DOI}]},
  {\small[\href{http://adsabs.harvard.edu/abs/1978ApJ...220..643R}{ADS}]}

\bibitem[Rosner {\it et~al.\/}(1985)]{1985ARAA..23..413R}
Rosner, R., Golub, L. and Vaiana, G.S., 1985, ``On stellar X-ray emission'',
  {\it Annu. Rev. Astron. Astrophys.\/}, {\bf 23}, 413--452.
  {\small[\href{http://dx.doi.org/10.1146/annurev.aa.23.090185.002213}{DOI}]},
  {\small[\href{http://adsabs.harvard.edu/abs/1985ARA&A..23..413R}{ADS}]}

\bibitem[Rottman {\it et~al.\/}(1990)]{1990ApJ...358..693R}
Rottman, G.J., Hassler, D.D., Jones, M.D. and Orrall, F.Q., 1990, ``The
  systematic radial downflow in the transition region of the quiet sun from
  limb-to-limb observations of the C IV resonance lines'', {\it Astrophys.
  J.\/}, {\bf 358}, 693--697.
  {\small[\href{http://dx.doi.org/10.1086/169023}{DOI}]},
  {\small[\href{http://adsabs.harvard.edu/abs/1990ApJ...358..693R}{ADS}]}

\bibitem[Saito and Billings(1964)]{1964ApJ...140..760S}
Saito, K. and Billings, D.E., 1964, ``Polarimetric Observations of a Coronal
  Condensation.'', {\it Astrophys. J.\/}, {\bf 140}, 760--+.
  {\small[\href{http://dx.doi.org/10.1086/147970}{DOI}]},
  {\small[\href{http://adsabs.harvard.edu/abs/1964ApJ...140..760S}{ADS}]}

\bibitem[Sakamoto {\it et~al.\/}(2008)]{2008ApJ...689.1421S}
Sakamoto, Y., Tsuneta, S. and Vekstein, G., 2008, ``Observational Appearance of
  Nanoflares with SXT and TRACE'', {\it Astrophys. J.\/}, {\bf 689},
  1421--1432. {\small[\href{http://dx.doi.org/10.1086/592488}{DOI}]},
  {\small[\href{http://adsabs.harvard.edu/abs/2008ApJ...689.1421S}{ADS}]}

\bibitem[Sakamoto {\it et~al.\/}(2009)]{2009ApJ...703.2118S}
Sakamoto, Y., Tsuneta, S. and Vekstein, G., 2009, ``A Nanoflare Heating Model
  and Comparison with Observations'', {\it Astrophys. J.\/}, {\bf 703},
  2118--2130.
  {\small[\href{http://dx.doi.org/10.1088/0004-637X/703/2/2118}{DOI}]},
  {\small[\href{http://adsabs.harvard.edu/abs/2009ApJ...703.2118S}{ADS}]}

\bibitem[Sakurai(1981)]{1981SoPh...69..343S}
Sakurai, T., 1981, ``Calculation of force-free magnetic field with non-constant
  {$\alpha$}'', {\it Solar Phys.\/}, {\bf 69}, 343--359.
  {\small[\href{http://dx.doi.org/10.1007/BF00149999}{DOI}]},
  {\small[\href{http://adsabs.harvard.edu/abs/1981SoPh...69..343S}{ADS}]}

\bibitem[Sandman {\it et~al.\/}(2009)]{2009SoPh..259....1S}
Sandman, A.W., Aschwanden, M.J., Derosa, M.L., W{\"u}lser, J.P. and Alexander,
  D., 2009, ``Comparison of STEREO/EUVI Loops with Potential Magnetic Field
  Models'', {\it Solar Phys.\/}, {\bf 259}, 1--11.
  {\small[\href{http://dx.doi.org/10.1007/s11207-009-9383-0}{DOI}]},
  {\small[\href{http://adsabs.harvard.edu/abs/2009SoPh..259....1S}{ADS}]}

\bibitem[Sanz-Forcada {\it et~al.\/}(2003)]{2003ApJS..145..147S}
Sanz-Forcada, J., Brickhouse, N.S. and Dupree, A.K., 2003, ``The Structure of
  Stellar Coronae in Active Binary Systems'', {\it Astrophys. J. Suppl.
  Ser.\/}, {\bf 145}, 147--179.
  {\small[\href{http://dx.doi.org/10.1086/345815}{DOI}]},
  {\small[\href{http://adsabs.harvard.edu/abs/2003ApJS..145..147S}{ADS}]},
  {\small[\href{http://arxiv.org/abs/arXiv:astro-ph/0210652}{{arXiv:astro-ph/0%
210652}}]}

\bibitem[Scelsi {\it et~al.\/}(2005)]{2005AA...432..671S}
Scelsi, L., Maggio, A., Peres, G. and Pallavicini, R., 2005, ``Coronal
  properties of G-type stars in different evolutionary phases'', {\it Astron.
  Astrophys.\/}, {\bf 432}, 671--685.
  {\small[\href{http://dx.doi.org/10.1051/0004-6361:20041739}{DOI}]},
  {\small[\href{http://adsabs.harvard.edu/abs/2005A&A...432..671S}{ADS}]},
  {\small[\href{http://arxiv.org/abs/arXiv:astro-ph/0501631}{{arXiv:astro-ph/0%
501631}}]}

\bibitem[Schmelz(2002)]{2002ApJ...578L.161S}
Schmelz, J.T., 2002, ``Are Coronal Loops Isothermal?'', {\it Astrophys. J.
  Lett.\/}, {\bf 578}, L161--L164.
  {\small[\href{http://dx.doi.org/10.1086/344715}{DOI}]},
  {\small[\href{http://adsabs.harvard.edu/abs/2002ApJ...578L.161S}{ADS}]}

\bibitem[Schmelz {\it et~al.\/}(2001)]{2001ApJ...556..896S}
Schmelz, J.T., Scopes, R.T., Cirtain, J.W., Winter, H.D. and Allen, J.D., 2001,
  ``Observational Constraints on Coronal Heating Models Using Coronal
  Diagnostics Spectrometer and Soft X-Ray Telescope Data'', {\it Astrophys.
  J.\/}, {\bf 556}, 896--904.
  {\small[\href{http://dx.doi.org/10.1086/321588}{DOI}]},
  {\small[\href{http://adsabs.harvard.edu/abs/2001ApJ...556..896S}{ADS}]}

\bibitem[Schmelz {\it et~al.\/}(2003)]{2003ApJ...599..604S}
Schmelz, J.T., Beene, J.E., Nasraoui, K., Blevins, H.T., Martens, P.C.H. and
  Cirtain, J.W., 2003, ``The Effect of Background Subtraction on the
  Temperature of EIT Coronal Loops'', {\it Astrophys. J.\/}, {\bf 599},
  604--614. {\small[\href{http://dx.doi.org/10.1086/379212}{DOI}]},
  {\small[\href{http://adsabs.harvard.edu/abs/2003ApJ...599..604S}{ADS}]}

\bibitem[Schmelz {\it et~al.\/}(2005)]{2005ApJ...627L..81S}
Schmelz, J.T., Nasraoui, K., Richardson, V.L., Hubbard, P.J., Nevels, C.R. and
  Beene, J.E., 2005, ``All Coronal Loops Are the Same: Evidence to the
  Contrary'', {\it Astrophys. J. Lett.\/}, {\bf 627}, L81--L84.
  {\small[\href{http://dx.doi.org/10.1086/431950}{DOI}]},
  {\small[\href{http://adsabs.harvard.edu/abs/2005ApJ...627L..81S}{ADS}]},
  {\small[\href{http://arxiv.org/abs/arXiv:astro-ph/0505593}{{arXiv:astro-ph/0%
505593}}]}

\bibitem[Schmelz {\it et~al.\/}(2007{\natexlab{a}})]{2007ApJ...660L.157S}
Schmelz, J.T., Kashyap, V.L. and Weber, M.A., 2007{\natexlab{a}}, ``Coronal
  Heat: Solar Loop Temperatures from TRACE Triple-Filter Data'', {\it
  Astrophys. J. Lett.\/}, {\bf 660}, L157--L160.
  {\small[\href{http://dx.doi.org/10.1086/518363}{DOI}]},
  {\small[\href{http://adsabs.harvard.edu/abs/2007ApJ...660L.157S}{ADS}]}

\bibitem[Schmelz {\it et~al.\/}(2007{\natexlab{b}})]{2007ApJ...658L.119S}
Schmelz, J.T., Nasraoui, K., Del~Zanna, G., Cirtain, J.W., DeLuca, E.E. and
  Mason, H.E., 2007{\natexlab{b}}, ``Coronal Diagnostic Spectrometer
  Observations of Isothermal and Multithermal Coronal Loops'', {\it Astrophys.
  J. Lett.\/}, {\bf 658}, L119--L122.
  {\small[\href{http://dx.doi.org/10.1086/514815}{DOI}]},
  {\small[\href{http://adsabs.harvard.edu/abs/2007ApJ...658L.119S}{ADS}]}

\bibitem[Schmelz {\it et~al.\/}(2008)]{2008ApJ...684L.115S}
Schmelz, J.T., Scott, J. and Rightmire, L.A., 2008, ``May Day! Coronal Loop
  Temperatures from the Hinode EUV Imaging Spectrometer'', {\it Astrophys. J.
  Lett.\/}, {\bf 684}, L115--L118.
  {\small[\href{http://dx.doi.org/10.1086/592215}{DOI}]},
  {\small[\href{http://adsabs.harvard.edu/abs/2008ApJ...684L.115S}{ADS}]}

\bibitem[Schmelz {\it et~al.\/}(2009{\natexlab{a}})]{2009ApJ...691..503S}
Schmelz, J.T., Nasraoui, K., Rightmire, L.A., Kimble, J.A., Del~Zanna, G.,
  Cirtain, J.W., DeLuca, E.E. and Mason, H.E., 2009{\natexlab{a}}, ``Are
  Coronal Loops Isothermal or Multithermal?'', {\it Astrophys. J.\/}, {\bf
  691}, 503--515.
  {\small[\href{http://dx.doi.org/10.1088/0004-637X/691/1/503}{DOI}]},
  {\small[\href{http://adsabs.harvard.edu/abs/2009ApJ...691..503S}{ADS}]},
  {\small[\href{http://arxiv.org/abs/0901.3281}{{arXiv:0901.3281}}]}

\bibitem[Schmelz {\it et~al.\/}(2009{\natexlab{b}})]{2009ApJ...693L.131S}
Schmelz, J.T., Saar, S.H., DeLuca, E.E., Golub, L., Kashyap, V.L., Weber, M.A.
  and Klimchuk, J.A., 2009{\natexlab{b}}, ``Hinode X-Ray Telescope Detection of
  Hot Emission from Quiescent Active Regions: A Nanoflare Signature?'', {\it
  Astrophys. J. Lett.\/}, {\bf 693}, L131--L135.
  {\small[\href{http://dx.doi.org/10.1088/0004-637X/693/2/L131}{DOI}]},
  {\small[\href{http://adsabs.harvard.edu/abs/2009ApJ...693L.131S}{ADS}]},
  {\small[\href{http://arxiv.org/abs/0901.3122}{{arXiv:0901.3122}}]}

\bibitem[Schmidt(1964)]{1964NASSP..50..107S}
Schmidt, H.U., 1964, ``On the Observable Effects of Magnetic Energy Storage and
  Release Connected With Solar Flares'', {\it NASA Special Publication\/}, {\bf
  50}, 107.
  {\small[\href{http://adsabs.harvard.edu/abs/1964NASSP..50..107S}{ADS}]}

\bibitem[Schmieder {\it et~al.\/}(2004)]{2004ApJ...601..530S}
Schmieder, B., Rust, D.M., Georgoulis, M.K., D{\'e}moulin, P. and Bernasconi,
  P.N., 2004, ``Emerging Flux and the Heating of Coronal Loops'', {\it
  Astrophys. J.\/}, {\bf 601}, 530--545.
  {\small[\href{http://dx.doi.org/10.1086/380199}{DOI}]},
  {\small[\href{http://adsabs.harvard.edu/abs/2004ApJ...601..530S}{ADS}]}

\bibitem[Schmitt {\it et~al.\/}(1990)]{1990ApJ...365..704S}
Schmitt, J.H.M.M., Collura, A., Sciortino, S., Vaiana, G.S., Harnden~Jr, F.R.
  and Rosner, R., 1990, ``Einstein Observatory coronal temperatures of
  late-type stars'', {\it Astrophys. J.\/}, {\bf 365}, 704--728.
  {\small[\href{http://dx.doi.org/10.1086/169525}{DOI}]},
  {\small[\href{http://adsabs.harvard.edu/abs/1990ApJ...365..704S}{ADS}]}

\bibitem[Schrijver(2007)]{2007ApJ...662L.119S}
Schrijver, C.J., 2007, ``Braiding-induced Interchange Reconnection of the
  Magnetic Field and the Width of Solar Coronal Loops'', {\it Astrophys. J.
  Lett.\/}, {\bf 662}, L119--L122.
  {\small[\href{http://dx.doi.org/10.1086/519455}{DOI}]},
  {\small[\href{http://adsabs.harvard.edu/abs/2007ApJ...662L.119S}{ADS}]}

\bibitem[Schrijver {\it et~al.\/}(2004)]{2004ApJ...615..512S}
Schrijver, C.J., Sandman, A.W., Aschwanden, M.J. and De~Rosa, M.L., 2004, ``The
  Coronal Heating Mechanism as Identified by Full-Sun Visualizations'', {\it
  Astrophys. J.\/}, {\bf 615}, 512--525.
  {\small[\href{http://dx.doi.org/10.1086/424028}{DOI}]},
  {\small[\href{http://adsabs.harvard.edu/abs/2004ApJ...615..512S}{ADS}]}

\bibitem[Seaton {\it et~al.\/}(2001)]{2001ApJ...563L.173S}
Seaton, D.B., Winebarger, A.R., DeLuca, E.E., Golub, L., Reeves, K.K. and
  Gallagher, P.T., 2001, ``Active Region Transient Events Observed with
  TRACE'', {\it Astrophys. J. Lett.\/}, {\bf 563}, L173--L177.
  {\small[\href{http://dx.doi.org/10.1086/338737}{DOI}]},
  {\small[\href{http://adsabs.harvard.edu/abs/2001ApJ...563L.173S}{ADS}]}

\bibitem[{Selwa} and {Ofman}(2009)]{2009AnGeo..27.3899S}
{Selwa}, M. and {Ofman}, L., 2009, ``{3-D numerical simulations of coronal
  loops oscillations}'', {\it Annales Geophysicae\/}, {\bf 27}, 3899--3908.
  {\small[\href{http://dx.doi.org/10.5194/angeo-27-3899-2009}{DOI}]},
  {\small[\href{http://adsabs.harvard.edu/abs/2009AnGeo..27.3899S}{ADS}]}

\bibitem[Serio {\it et~al.\/}(1981)]{1981ApJ...243..288S}
Serio, S., Peres, G., Vaiana, G.S., Golub, L. and Rosner, R., 1981, ``Closed
  coronal structures. II - Generalized hydrostatic model'', {\it Astrophys.
  J.\/}, {\bf 243}, 288--300.
  {\small[\href{http://dx.doi.org/10.1086/158597}{DOI}]},
  {\small[\href{http://adsabs.harvard.edu/abs/1981ApJ...243..288S}{ADS}]}

\bibitem[Serio {\it et~al.\/}(1991)]{1991AA...241..197S}
Serio, S., Reale, F., Jakimiec, J., Sylwester, B. and Sylwester, J., 1991,
  ``Dynamics of flaring loops. I - Thermodynamic decay scaling laws'', {\it
  Astron. Astrophys.\/}, {\bf 241}, 197--202.
  {\small[\href{http://adsabs.harvard.edu/abs/1991A&A...241..197S}{ADS}]}

\bibitem[Sheeley~Jr(1980)]{1980SoPh...66...79S}
Sheeley~Jr, N.R., 1980, ``Temporal variations of loop structures in the solar
  atmosphere'', {\it Solar Phys.\/}, {\bf 66}, 79--87.
  {\small[\href{http://dx.doi.org/10.1007/BF00150520}{DOI}]},
  {\small[\href{http://adsabs.harvard.edu/abs/1980SoPh...66...79S}{ADS}]}

\bibitem[Shimizu(1995)]{1995PASJ...47..251S}
Shimizu, T., 1995, ``Energetics and Occurrence Rate of Active-Region Transient
  Brightenings and Implications for the Heating of the Active-Region Corona'',
  {\it Publ. Astron. Soc. Japan\/}, {\bf 47}, 251--263.
  {\small[\href{http://adsabs.harvard.edu/abs/1995PASJ...47..251S}{ADS}]}

\bibitem[Shimizu and Tsuneta(1997)]{1997ApJ...486.1045S}
Shimizu, T. and Tsuneta, S., 1997, ``Deep Survey of Solar Nanoflares with
  YOHKOH'', {\it Astrophys. J.\/}, {\bf 486}, 1045.
  {\small[\href{http://dx.doi.org/10.1086/304542}{DOI}]},
  {\small[\href{http://adsabs.harvard.edu/abs/1997ApJ...486.1045S}{ADS}]}

\bibitem[Shimizu {\it et~al.\/}(1994)]{1994ApJ...422..906S}
Shimizu, T., Tsuneta, S., Acton, L.W., Lemen, J.R., Ogawara, Y. and Uchida, Y.,
  1994, ``Morphology of active region transient brightenings with the YOHKOH
  Soft X-ray Telescope'', {\it Astrophys. J.\/}, {\bf 422}, 906--911.
  {\small[\href{http://dx.doi.org/10.1086/173782}{DOI}]},
  {\small[\href{http://adsabs.harvard.edu/abs/1994ApJ...422..906S}{ADS}]}

\bibitem[Shimojo {\it et~al.\/}(2002)]{2002SoPh..206..133S}
Shimojo, M., Kurokawa, H. and Yoshimura, K., 2002, ``Dynamical Features and
  Evolutional Characteristics of Brightening Coronal Loops'', {\it Solar
  Phys.\/}, {\bf 206}, 133--142.
  {\small[\href{http://dx.doi.org/10.1023/A:1014939525194}{DOI}]},
  {\small[\href{http://adsabs.harvard.edu/abs/2002SoPh..206..133S}{ADS}]}

\bibitem[Spadaro {\it et~al.\/}(1990)]{1990ApJ...355..342S}
Spadaro, D., Noci, G., Zappala, R.A. and Antiochos, S.K., 1990, ``The effect of
  nonequilibrium ionization on ultraviolet line shifts in the solar transition
  region'', {\it Astrophys. J.\/}, {\bf 355}, 342--347.
  {\small[\href{http://dx.doi.org/10.1086/168768}{DOI}]},
  {\small[\href{http://adsabs.harvard.edu/abs/1990ApJ...355..342S}{ADS}]}

\bibitem[Spadaro {\it et~al.\/}(2003)]{2003ApJ...582..486S}
Spadaro, D., Lanza, A.F., Lanzafame, A.C., Karpen, J.T., Antiochos, S.K.,
  Klimchuk, J.A. and MacNeice, P.J., 2003, ``A Transient Heating Model for
  Coronal Structure and Dynamics'', {\it Astrophys. J.\/}, {\bf 582}, 486--494.
  {\small[\href{http://dx.doi.org/10.1086/344508}{DOI}]},
  {\small[\href{http://adsabs.harvard.edu/abs/2003ApJ...582..486S}{ADS}]}

\bibitem[Spitzer(1962)]{1962pfig.book.....S}
Spitzer, L., 1962, {\it Physics of Fully Ionized Gases\/}, vol.~3 of
  Interscience Tracts on Physics and Astronomy, Interscience, New York, 2nd
  edn.  {\small[\href{http://adsabs.harvard.edu/abs/1962pfig.book.....S}{ADS}]}

\bibitem[Steinolfson and Davila(1993)]{1993ApJ...415..354S}
Steinolfson, R.S. and Davila, J.M., 1993, ``Coronal heating by the resonant
  absorption of Alfven waves - Importance of the global mode and scaling
  laws'', {\it Astrophys. J.\/}, {\bf 415}, 354--363.
  {\small[\href{http://dx.doi.org/10.1086/173169}{DOI}]},
  {\small[\href{http://adsabs.harvard.edu/abs/1993ApJ...415..354S}{ADS}]}

\bibitem[Stelzer {\it et~al.\/}(2002)]{2002AA...392..585S}
Stelzer, B., Burwitz, V., Audard, M., G{\"u}del, M., Ness, J.-U., Grosso, N.,
  Neuh{\"a}user, R., Schmitt, J.H.M.M., Predehl, P. and Aschenbach, B., 2002,
  ``Simultaneous X-ray spectroscopy of YY Gem with Chandra and XMM-Newton'',
  {\it Astron. Astrophys.\/}, {\bf 392}, 585--598.
  {\small[\href{http://dx.doi.org/10.1051/0004-6361:20021188}{DOI}]},
  {\small[\href{http://adsabs.harvard.edu/abs/2002A&A...392..585S}{ADS}]},
  {\small[\href{http://arxiv.org/abs/arXiv:astro-ph/0206429}{{arXiv:astro-ph/0%
206429}}]}

\bibitem[Strong {\it et~al.\/}(1992)]{1992PASJ...44L.161S}
Strong, K.T., Harvey, K., Hirayama, T., Nitta, N., Shimizu, T. and Tsuneta, S.,
  1992, ``Observations of the variability of coronal bright points by the Soft
  X-ray Telescope on YOHKOH'', {\it Publ. Astron. Soc. Japan\/}, {\bf 44},
  L161--L166.
  {\small[\href{http://adsabs.harvard.edu/abs/1992PASJ...44L.161S}{ADS}]}

\bibitem[Sylwester {\it et~al.\/}(1993)]{1993AA...267..586S}
Sylwester, B., Sylwester, J., Serio, S., Reale, F., Bentley, R.D. and Fludra,
  A., 1993, ``Dynamics of flaring loops. III - Interpretation of flare
  evolution in the emission measure-temperature diagram'', {\it Astron.
  Astrophys.\/}, {\bf 267}, 586--594.
  {\small[\href{http://adsabs.harvard.edu/abs/1993A&A...267..586S}{ADS}]}

\bibitem[Sylwester {\it et~al.\/}(1998)]{1998ESASP.417..313S}
Sylwester, J., Gaicki, I., Kordylewski, Z., Nowak, M., Kowalinski, S.,
  Sjarkowski, M., Bentley, W., Trzebinski, R.D., Whyndham, M.W., Guttridge,
  P.R., Culhane, J.L., Lang, J., Phillips, K.J.H., Brown, C.M., Doschek, G.A.,
  Oraevsky, V.N., Boldyrev, S.I., Kopaev, I.M., Stepanov, A.I. and Klepikov,
  V.Y., 1998, ``RESIK: High Sensitivity Soft X-ray Spectrometer for the Study
  of Solar Flare Plasma'', in {\it Crossroads for European Solar and
  Heliospheric Physics. Recent Achievements and Future Mission
  Possibilities\/}, vol. 417 of ESA Special Publication,
  {\small[\href{http://adsabs.harvard.edu/abs/1998ESASP.417..313S}{ADS}]}

\bibitem[Sylwester {\it et~al.\/}(2008)]{2008JApA...29..339S}
Sylwester, J., Kuzin, S., Kotov, Y.D., Farnik, F. and Reale, F., 2008,
  ``SphinX: A fast solar Photometer in X-rays'', {\it Journal of Astrophysics
  and Astronomy\/}, {\bf 29}, 339--343.
  {\small[\href{http://dx.doi.org/10.1007/s12036-008-0044-8}{DOI}]},
  {\small[\href{http://adsabs.harvard.edu/abs/2008JApA...29..339S}{ADS}]}

\bibitem[Tanaka(1983)]{1983SoPh...86....3T}
Tanaka, Y., 1983, ``Introduction to HINOTORI'', {\it Solar Phys.\/}, {\bf 86},
  3--6. {\small[\href{http://dx.doi.org/10.1007/BF00157168}{DOI}]},
  {\small[\href{http://adsabs.harvard.edu/abs/1983SoPh...86....3T}{ADS}]}

\bibitem[Taroyan(2009)]{2009ApJ...694...69T}
Taroyan, Y., 2009, ``Alfv{\'e}n Instability in Coronal Loops With Siphon
  Flows'', {\it Astrophys. J.\/}, {\bf 694}, 69--75.
  {\small[\href{http://dx.doi.org/10.1088/0004-637X/694/1/69}{DOI}]},
  {\small[\href{http://adsabs.harvard.edu/abs/2009ApJ...694...69T}{ADS}]}

\bibitem[Taroyan {\it et~al.\/}(2005)]{2005AA...438..713T}
Taroyan, Y., Erd{\'e}lyi, R., Doyle, J.G. and Bradshaw, S.J., 2005, ``Footpoint
  excitation of standing acoustic waves in coronal loops'', {\it Astron.
  Astrophys.\/}, {\bf 438}, 713--720.
  {\small[\href{http://dx.doi.org/10.1051/0004-6361:20052794}{DOI}]},
  {\small[\href{http://adsabs.harvard.edu/abs/2005A&A...438..713T}{ADS}]}

\bibitem[Telleschi {\it et~al.\/}(2005)]{2005ApJ...622..653T}
Telleschi, A., G{\"u}del, M., Briggs, K., Audard, M., Ness, J.-U. and Skinner,
  S.L., 2005, ``Coronal Evolution of the Sun in Time: High-Resolution X-Ray
  Spectroscopy of Solar Analogs with Different Ages'', {\it Astrophys. J.\/},
  {\bf 622}, 653--679. {\small[\href{http://dx.doi.org/10.1086/428109}{DOI}]},
  {\small[\href{http://adsabs.harvard.edu/abs/2005ApJ...622..653T}{ADS}]},
  {\small[\href{http://arxiv.org/abs/arXiv:astro-ph/0503546}{{arXiv:astro-ph/0%
503546}}]}

\bibitem[Teriaca {\it et~al.\/}(1999{\natexlab{a}})]{1999AA...349..636T}
Teriaca, L., Banerjee, D. and Doyle, J.G., 1999{\natexlab{a}}, ``SUMER
  observations of Doppler shift in the quiet Sun and in an active region'',
  {\it Astron. Astrophys.\/}, {\bf 349}, 636--648.
  {\small[\href{http://adsabs.harvard.edu/abs/1999A&A...349..636T}{ADS}]}

\bibitem[Teriaca {\it et~al.\/}(1999{\natexlab{b}})]{1999AA...352L..99T}
Teriaca, L., Doyle, J.G., Erd{\'e}lyi, R. and Sarro, L.M., 1999{\natexlab{b}},
  ``New insight into transition region dynamics via SUMER observations and
  numerical modelling'', {\it Astron. Astrophys.\/}, {\bf 352}, L99--L102.
  {\small[\href{http://adsabs.harvard.edu/abs/1999A&A...352L..99T}{ADS}]}

\bibitem[Teriaca {\it et~al.\/}(2002)]{2002AA...392..309T}
Teriaca, L., Madjarska, M.S. and Doyle, J.G., 2002, ``Transition region
  explosive events: Do they have a coronal counterpart?'', {\it Astron.
  Astrophys.\/}, {\bf 392}, 309--317.
  {\small[\href{http://dx.doi.org/10.1051/0004-6361:20020795}{DOI}]},
  {\small[\href{http://adsabs.harvard.edu/abs/2002A&A...392..309T}{ADS}]}

\bibitem[Teriaca {\it et~al.\/}(2004)]{2004AA...427.1065T}
Teriaca, L., Banerjee, D., Falchi, A., Doyle, J.G. and Madjarska, M.S., 2004,
  ``Transition region small-scale dynamics as seen by SUMER on SOHO'', {\it
  Astron. Astrophys.\/}, {\bf 427}, 1065--1074.
  {\small[\href{http://dx.doi.org/10.1051/0004-6361:20040503}{DOI}]},
  {\small[\href{http://adsabs.harvard.edu/abs/2004A&A...427.1065T}{ADS}]}

\bibitem[Terzo and Reale(2010)]{2010AA...515A...7T}
Terzo, S. and Reale, F., 2010, ``On the importance of background subtraction in
  the analysis of coronal loops observed with TRACE'', {\it Astron.
  Astrophys.\/}, {\bf 515}, A7+.
  {\small[\href{http://dx.doi.org/10.1051/0004-6361/200913469}{DOI}]},
  {\small[\href{http://adsabs.harvard.edu/abs/2010A&A...515A...7T}{ADS}]},
  {\small[\href{http://arxiv.org/abs/1002.2121}{{arXiv:1002.2121
  {\small[astro-ph.SR]}}}]}

\bibitem[Testa {\it et~al.\/}(2002)]{2002ApJ...580.1159T}
Testa, P., Peres, G., Reale, F. and Orlando, S., 2002, ``Temperature and
  Density Structure of Hot and Cool Loops Derived from the Analysis of TRACE
  Data'', {\it Astrophys. J.\/}, {\bf 580}, 1159--1171.
  {\small[\href{http://dx.doi.org/10.1086/343732}{DOI}]},
  {\small[\href{http://adsabs.harvard.edu/abs/2002ApJ...580.1159T}{ADS}]}

\bibitem[Testa {\it et~al.\/}(2005)]{2005ApJ...622..695T}
Testa, P., Peres, G. and Reale, F., 2005, ``Emission Measure Distribution in
  Loops Impulsively Heated at the Footpoints'', {\it Astrophys. J.\/}, {\bf
  622}, 695--703. {\small[\href{http://dx.doi.org/10.1086/427900}{DOI}]},
  {\small[\href{http://adsabs.harvard.edu/abs/2005ApJ...622..695T}{ADS}]},
  {\small[\href{http://arxiv.org/abs/arXiv:astro-ph/0412482}{{arXiv:astro-ph/0%
412482}}]}

\bibitem[Thomas(1988)]{1988ApJ...333..407T}
Thomas, J.H., 1988, ``Siphon flows in isolated magnetic flux tubes'', {\it
  Astrophys. J.\/}, {\bf 333}, 407--419.
  {\small[\href{http://dx.doi.org/10.1086/166756}{DOI}]},
  {\small[\href{http://adsabs.harvard.edu/abs/1988ApJ...333..407T}{ADS}]}

\bibitem[Thomas and Montesinos(1990)]{1990ApJ...359..550T}
Thomas, J.H. and Montesinos, B., 1990, ``Siphon flows in isolated magnetic flux
  tubes. III - The equilibrium path of the flux-tube arch'', {\it Astrophys.
  J.\/}, {\bf 359}, 550--559.
  {\small[\href{http://dx.doi.org/10.1086/169086}{DOI}]},
  {\small[\href{http://adsabs.harvard.edu/abs/1990ApJ...359..550T}{ADS}]}

\bibitem[Thomas and Montesinos(1991)]{1991ApJ...375..404T}
Thomas, J.H. and Montesinos, B., 1991, ``Siphon flows in isolated magnetic flux
  tubes. IV - Critical flows with standing tube shocks'', {\it Astrophys.
  J.\/}, {\bf 375}, 404--413.
  {\small[\href{http://dx.doi.org/10.1086/170198}{DOI}]},
  {\small[\href{http://adsabs.harvard.edu/abs/1991ApJ...375..404T}{ADS}]}

\bibitem[{Title}(2010)]{2010AAS...21630803T}
{Title}, A.~M., 2010, ``{AIA on SDO}'', in {\it American Astronomical Society
  Meeting Abstracts\/}, vol. 216 of American Astronomical Society Meeting
  Abstracts,
  {\small[\href{http://adsabs.harvard.edu/abs/2010AAS...21630803T}{ADS}]}

\bibitem[Tomczyk {\it et~al.\/}(2007)]{2007Sci...317.1192T}
Tomczyk, S., McIntosh, S.W., Keil, S.L., Judge, P.G., Schad, T., Seeley, D.H.
  and Edmondson, J., 2007, ``Alfv{\'e}n Waves in the Solar Corona'', {\it
  Science\/}, {\bf 317}, 1192--.
  {\small[\href{http://dx.doi.org/10.1126/science.1143304}{DOI}]},
  {\small[\href{http://adsabs.harvard.edu/abs/2007Sci...317.1192T}{ADS}]}

\bibitem[Tousey {\it et~al.\/}(1977)]{1977ApOpt..16..870T}
Tousey, R., Bartoe, J.-D.F., Brueckner, G.E. and Purcell, J.D., 1977, ``Extreme
  ultraviolet spectroheliograph ATM experiment S082A'', {\it Appl. Optics\/},
  {\bf 16}, 870--878.
  {\small[\href{http://adsabs.harvard.edu/abs/1977ApOpt..16..870T}{ADS}]}

\bibitem[Tripathi {\it et~al.\/}(2008)]{2008AA...481L..53T}
Tripathi, D., Mason, H.E., Young, P.R. and Del~Zanna, G., 2008, ``Density
  structure of an active region and associated moss using Hinode/EIS'', {\it
  Astron. Astrophys.\/}, {\bf 481}, L53--L56.
  {\small[\href{http://dx.doi.org/10.1051/0004-6361:20079034}{DOI}]},
  {\small[\href{http://adsabs.harvard.edu/abs/2008A&A...481L..53T}{ADS}]},
  {\small[\href{http://arxiv.org/abs/0802.3311}{{arXiv:0802.3311}}]}

\bibitem[Tripathi {\it et~al.\/}(2009)]{2009ApJ...694.1256T}
Tripathi, D., Mason, H.E., Dwivedi, B.N., Del~Zanna, G. and Young, P.R., 2009,
  ``Active Region Loops: Hinode/Extreme-Ultraviolet Imaging Spectrometer
  Observations'', {\it Astrophys. J.\/}, {\bf 694}, 1256--1265.
  {\small[\href{http://dx.doi.org/10.1088/0004-637X/694/2/1256}{DOI}]},
  {\small[\href{http://adsabs.harvard.edu/abs/2009ApJ...694.1256T}{ADS}]},
  {\small[\href{http://arxiv.org/abs/0901.0095}{{arXiv:0901.0095}}]}

\bibitem[Tsuneta {\it et~al.\/}(1991)]{1991SoPh..136...37T}
Tsuneta, S., Acton, L., Bruner, M., Lemen, J., Brown, W., Caravalho, R.,
  Catura, R., Freeland, S., Jurcevich, B. and Owens, J., 1991, ``The soft X-ray
  telescope for the SOLAR-A mission'', {\it Solar Phys.\/}, {\bf 136}, 37--67.
  {\small[\href{http://dx.doi.org/10.1007/BF00151694}{DOI}]},
  {\small[\href{http://adsabs.harvard.edu/abs/1991SoPh..136...37T}{ADS}]}

\bibitem[Tsuneta {\it et~al.\/}(1992)]{1992PASJ...44L..63T}
Tsuneta, S., Hara, H., Shimizu, T., Acton, L.W., Strong, K.T., Hudson, H.S. and
  Ogawara, Y., 1992, ``Observation of a solar flare at the limb with the YOHKOH
  Soft X-ray Telescope'', {\it Publ. Astron. Soc. Japan\/}, {\bf 44}, L63--L69.
  {\small[\href{http://adsabs.harvard.edu/abs/1992PASJ...44L..63T}{ADS}]}

\bibitem[{Uchida}(1970)]{1970PASJ...22..341U}
{Uchida}, Y., 1970, ``{Diagnosis of Coronal Magnetic Structure by
  Flare-Associated Hydromagnetic Disturbances}'', {\it Publ. Astron. Soc.
  Japan\/}, {\bf 22}, 341--+.
  {\small[\href{http://adsabs.harvard.edu/abs/1970PASJ...22..341U}{ADS}]}

\bibitem[Ugarte-Urra {\it et~al.\/}(2005)]{2005AA...439..351U}
Ugarte-Urra, I., Doyle, J.G., Walsh, R.W. and Madjarska, M.S., 2005, ``Electron
  density along a coronal loop observed with CDS/SOHO'', {\it Astron.
  Astrophys.\/}, {\bf 439}, 351--359.
  {\small[\href{http://dx.doi.org/10.1051/0004-6361:20042560}{DOI}]},
  {\small[\href{http://adsabs.harvard.edu/abs/2005A&A...439..351U}{ADS}]}

\bibitem[Ugarte-Urra {\it et~al.\/}(2009)]{2009ApJ...695..642U}
Ugarte-Urra, I., Warren, H.P. and Brooks, D.H., 2009, ``Active Region
  Transition Region Loop Populations and Their Relationship to the Corona'',
  {\it Astrophys. J.\/}, {\bf 695}, 642--651.
  {\small[\href{http://dx.doi.org/10.1088/0004-637X/695/1/642}{DOI}]},
  {\small[\href{http://adsabs.harvard.edu/abs/2009ApJ...695..642U}{ADS}]},
  {\small[\href{http://arxiv.org/abs/0901.1075}{{arXiv:0901.1075}}]}

\bibitem[Ulrich(1996)]{1996ApJ...465..436U}
Ulrich, R.K., 1996, ``Observations of Magnetohydrodynamic Oscillations in the
  Solar Atmosphere with Properties of Alfv{\'e}n Waves'', {\it Astrophys.
  J.\/}, {\bf 465}, 436--+.
  {\small[\href{http://dx.doi.org/10.1086/177431}{DOI}]},
  {\small[\href{http://adsabs.harvard.edu/abs/1996ApJ...465..436U}{ADS}]}

\bibitem[Uzdensky(2007)]{2007ApJ...671.2139U}
Uzdensky, D.A., 2007, ``The Fast Collisionless Reconnection Condition and the
  Self-Organization of Solar Coronal Heating'', {\it Astrophys. J.\/}, {\bf
  671}, 2139--2153. {\small[\href{http://dx.doi.org/10.1086/522915}{DOI}]},
  {\small[\href{http://adsabs.harvard.edu/abs/2007ApJ...671.2139U}{ADS}]},
  {\small[\href{http://arxiv.org/abs/0707.1316}{{arXiv:0707.1316}}]}

\bibitem[Vaiana and Rosner(1978)]{1978ARAA..16..393V}
Vaiana, G.S. and Rosner, R., 1978, ``Recent advances in coronal physics'', {\it
  Annu. Rev. Astron. Astrophys.\/}, {\bf 16}, 393--428.
  {\small[\href{http://dx.doi.org/10.1146/annurev.aa.16.090178.002141}{DOI}]},
  {\small[\href{http://adsabs.harvard.edu/abs/1978ARA&A..16..393V}{ADS}]}

\bibitem[Vaiana {\it et~al.\/}(1968)]{1968Sci...161..564V}
Vaiana, G.S., Reidy, W.P., Zehnpfennig, T., Vanspeybroeck, L. and Giacconi, R.,
  1968, ``X-ray Structures of the Sun during the Importance 1N Flare of 8 June
  1968'', {\it Science\/}, {\bf 161}, 564--567.
  {\small[\href{http://dx.doi.org/10.1126/science.161.3841.564}{DOI}]},
  {\small[\href{http://adsabs.harvard.edu/abs/1968Sci...161..564V}{ADS}]}

\bibitem[Vaiana {\it et~al.\/}(1973)]{1973SoPh...32...81V}
Vaiana, G.S., Krieger, A.S. and Timothy, A.F., 1973, ``Identification and
  Analysis of Structures in the Corona from X-Ray Photography'', {\it Solar
  Phys.\/}, {\bf 32}, 81--116.
  {\small[\href{http://dx.doi.org/10.1007/BF00152731}{DOI}]},
  {\small[\href{http://adsabs.harvard.edu/abs/1973SoPh...32...81V}{ADS}]}

\bibitem[van Ballegooijen(1986)]{1986ApJ...311.1001V}
van Ballegooijen, A.A., 1986, ``Cascade of magnetic energy as a mechanism of
  coronal heating'', {\it Astrophys. J.\/}, {\bf 311}, 1001--1014.
  {\small[\href{http://dx.doi.org/10.1086/164837}{DOI}]},
  {\small[\href{http://adsabs.harvard.edu/abs/1986ApJ...311.1001V}{ADS}]}

\bibitem[van~den Oord and Mewe(1989)]{1989AA...213..245V}
van~den Oord, G.H.J. and Mewe, R., 1989, ``The X-ray flare and the quiescent
  emission from Algol as detected by EXOSAT'', {\it Astron. Astrophys.\/}, {\bf
  213}, 245--260.
  {\small[\href{http://adsabs.harvard.edu/abs/1989A&A...213..245V}{ADS}]}

\bibitem[van~den Oord {\it et~al.\/}(1988)]{1988AA...205..181V}
van~den Oord, G.H.J., Mewe, R. and Brinkman, A.C., 1988, ``An EXOSAT
  observation of an X-ray flare and quiescent emission from the RS CVn binary
  sigma2 CrB'', {\it Astron. Astrophys.\/}, {\bf 205}, 181--196.
  {\small[\href{http://adsabs.harvard.edu/abs/1988A&A...205..181V}{ADS}]}

\bibitem[Vekstein(2009)]{2009AA...499L...5V}
Vekstein, G., 2009, ``Probing nanoflares with observed fluctuations of the
  coronal EUV emission'', {\it Astron. Astrophys.\/}, {\bf 499}, L5--L8.
  {\small[\href{http://dx.doi.org/10.1051/0004-6361/200911872}{DOI}]},
  {\small[\href{http://adsabs.harvard.edu/abs/2009A&A...499L...5V}{ADS}]}

\bibitem[Vekstein and Katsukawa(2000)]{2000ApJ...541.1096V}
Vekstein, G. and Katsukawa, Y., 2000, ``Scaling Laws for a Nanoflare-Heated
  Solar Corona'', {\it Astrophys. J.\/}, {\bf 541}, 1096--1103.
  {\small[\href{http://dx.doi.org/10.1086/309480}{DOI}]},
  {\small[\href{http://adsabs.harvard.edu/abs/2000ApJ...541.1096V}{ADS}]}

\bibitem[Vekstein and Jain(2003)]{2003PPCF...45..535V}
Vekstein, G.E. and Jain, R., 2003, ``Signatures of a nanoflare heated solar
  corona'', {\it Plasma Physics and Controlled Fusion\/}, {\bf 45}, 535--545.
  {\small[\href{http://dx.doi.org/10.1088/0741-3335/45/5/302}{DOI}]},
  {\small[\href{http://adsabs.harvard.edu/abs/2003PPCF...45..535V}{ADS}]}

\bibitem[Vernazza {\it et~al.\/}(1981)]{1981ApJS...45..635V}
Vernazza, J.E., Avrett, E.H. and Loeser, R., 1981, ``Structure of the solar
  chromosphere. III - Models of the EUV brightness components of the
  quiet-sun'', {\it Astrophys. J. Suppl. Ser.\/}, {\bf 45}, 635--725.
  {\small[\href{http://dx.doi.org/10.1086/190731}{DOI}]},
  {\small[\href{http://adsabs.harvard.edu/abs/1981ApJS...45..635V}{ADS}]}

\bibitem[Vesecky {\it et~al.\/}(1979)]{1979ApJ...233..987V}
Vesecky, J.F., Antiochos, S.K. and Underwood, J.H., 1979, ``Numerical modeling
  of quasi-static coronal loops. I - Uniform energy input'', {\it Astrophys.
  J.\/}, {\bf 233}, 987--997.
  {\small[\href{http://dx.doi.org/10.1086/157462}{DOI}]},
  {\small[\href{http://adsabs.harvard.edu/abs/1979ApJ...233..987V}{ADS}]}

\bibitem[Wang {\it et~al.\/}(1997)]{1997ApJ...478L..41W}
Wang, J., Shibata, K., Nitta, N., Slater, G.L., Savy, S.K. and Ogawara, Y.,
  1997, ``Shrinkage of Coronal X-Ray Loops'', {\it Astrophys. J. Lett.\/}, {\bf
  478}, L41+. {\small[\href{http://dx.doi.org/10.1086/310539}{DOI}]},
  {\small[\href{http://adsabs.harvard.edu/abs/1997ApJ...478L..41W}{ADS}]}

\bibitem[Wang {\it et~al.\/}(2009)]{2009ApJ...696.1448W}
Wang, T.J., Ofman, L. and Davila, J.M., 2009, ``Propagating Slow
  Magnetoacoustic Waves in Coronal Loops Observed by Hinode/EIS'', {\it
  Astrophys. J.\/}, {\bf 696}, 1448--1460.
  {\small[\href{http://dx.doi.org/10.1088/0004-637X/696/2/1448}{DOI}]},
  {\small[\href{http://adsabs.harvard.edu/abs/2009ApJ...696.1448W}{ADS}]},
  {\small[\href{http://arxiv.org/abs/0902.4480}{{arXiv:0902.4480
  {\small[astro-ph.SR]}}}]}

\bibitem[{Warren} {\it et~al.\/}(2010)]{2010ApJ...711..228W}
{Warren}, H.~P., {Winebarger}, A.~R. and {Brooks}, D.~H., 2010, ``{Evidence for
  Steady Heating: Observations of an Active Region Core with Hinode and
  TRACE}'', {\it Astrophys. J.\/}, {\bf 711}, 228--238.
  {\small[\href{http://dx.doi.org/10.1088/0004-637X/711/1/228}{DOI}]},
  {\small[\href{http://adsabs.harvard.edu/abs/2010ApJ...711..228W}{ADS}]},
  {\small[\href{http://arxiv.org/abs/0910.0458}{{arXiv:0910.0458
  {\small[astro-ph.SR]}}}]}

\bibitem[Warren(2006)]{2006ApJ...637..522W}
Warren, H.P., 2006, ``Multithread Hydrodynamic Modeling of a Solar Flare'',
  {\it Astrophys. J.\/}, {\bf 637}, 522--530.
  {\small[\href{http://dx.doi.org/10.1086/497904}{DOI}]},
  {\small[\href{http://adsabs.harvard.edu/abs/2006ApJ...637..522W}{ADS}]},
  {\small[\href{http://arxiv.org/abs/arXiv:astro-ph/0507328}{{arXiv:astro-ph/0%
507328}}]}

\bibitem[Warren and Winebarger(2006)]{2006ApJ...645..711W}
Warren, H.P. and Winebarger, A.R., 2006, ``Hydrostatic Modeling of the
  Integrated Soft X-Ray and Extreme Ultraviolet Emission in Solar Active
  Regions'', {\it Astrophys. J.\/}, {\bf 645}, 711--719.
  {\small[\href{http://dx.doi.org/10.1086/504075}{DOI}]},
  {\small[\href{http://adsabs.harvard.edu/abs/2006ApJ...645..711W}{ADS}]},
  {\small[\href{http://arxiv.org/abs/arXiv:astro-ph/0602052}{{arXiv:astro-ph/0%
602052}}]}

\bibitem[Warren and Winebarger(2007)]{2007ApJ...666.1245W}
Warren, H.P. and Winebarger, A.R., 2007, ``Static and Dynamic Modeling of a
  Solar Active Region'', {\it Astrophys. J.\/}, {\bf 666}, 1245--1255.
  {\small[\href{http://dx.doi.org/10.1086/519943}{DOI}]},
  {\small[\href{http://adsabs.harvard.edu/abs/2007ApJ...666.1245W}{ADS}]},
  {\small[\href{http://arxiv.org/abs/arXiv:astro-ph/0609023}{{arXiv:astro-ph/0%
609023}}]}

\bibitem[Warren {\it et~al.\/}(2002)]{2002ApJ...579L..41W}
Warren, H.P., Winebarger, A.R. and Hamilton, P.S., 2002, ``Hydrodynamic
  Modeling of Active Region Loops'', {\it Astrophys. J. Lett.\/}, {\bf 579},
  L41--L44. {\small[\href{http://dx.doi.org/10.1086/344921}{DOI}]},
  {\small[\href{http://adsabs.harvard.edu/abs/2002ApJ...579L..41W}{ADS}]}

\bibitem[Warren {\it et~al.\/}(2003)]{2003ApJ...593.1174W}
Warren, H.P., Winebarger, A.R. and Mariska, J.T., 2003, ``Evolving Active
  Region Loops Observed with the Transition Region and Coronal explorer. II.
  Time-dependent Hydrodynamic Simulations'', {\it Astrophys. J.\/}, {\bf 593},
  1174--1186. {\small[\href{http://dx.doi.org/10.1086/376678}{DOI}]},
  {\small[\href{http://adsabs.harvard.edu/abs/2003ApJ...593.1174W}{ADS}]}

\bibitem[Warren {\it et~al.\/}(2008{\natexlab{a}})]{2008ApJ...686L.131W}
Warren, H.P., Ugarte-Urra, I., Doschek, G.A., Brooks, D.H. and Williams, D.R.,
  2008{\natexlab{a}}, ``Observations of Active Region Loops with the EUV
  Imaging Spectrometer on Hinode'', {\it Astrophys. J. Lett.\/}, {\bf 686},
  L131--L134. {\small[\href{http://dx.doi.org/10.1086/592960}{DOI}]},
  {\small[\href{http://adsabs.harvard.edu/abs/2008ApJ...686L.131W}{ADS}]},
  {\small[\href{http://arxiv.org/abs/0808.3227}{{arXiv:0808.3227}}]}

\bibitem[Warren {\it et~al.\/}(2008{\natexlab{b}})]{2008ApJ...677.1395W}
Warren, H.P., Winebarger, A.R., Mariska, J.T., Doschek, G.A. and Hara, H.,
  2008{\natexlab{b}}, ``Observation and Modeling of Coronal `Moss' With the EUV
  Imaging Spectrometer on Hinode'', {\it Astrophys. J.\/}, {\bf 677},
  1395--1400. {\small[\href{http://dx.doi.org/10.1086/529186}{DOI}]},
  {\small[\href{http://adsabs.harvard.edu/abs/2008ApJ...677.1395W}{ADS}]},
  {\small[\href{http://arxiv.org/abs/0709.0396}{{arXiv:0709.0396}}]}

\bibitem[Watanabe {\it et~al.\/}(2009)]{2009ApJ...692.1294W}
Watanabe, T., Hara, H., Yamamoto, N., Kato, D., Sakaue, H.A., Murakami, I.,
  Kato, T., Nakamura, N. and Young, P.R., 2009, ``Fe XIII Density Diagnostics
  in the EIS Observing Wavelengths'', {\it Astrophys. J.\/}, {\bf 692},
  1294--1304.
  {\small[\href{http://dx.doi.org/10.1088/0004-637X/692/2/1294}{DOI}]},
  {\small[\href{http://adsabs.harvard.edu/abs/2009ApJ...692.1294W}{ADS}]}

\bibitem[Weber {\it et~al.\/}(2005)]{2005ApJ...635L.101W}
Weber, M.A., Schmelz, J.T., DeLuca, E.E. and Roames, J.K., 2005, ``Isothermal
  Bias of the `Filter Ratio' Method for Observations of Multithermal Plasma'',
  {\it Astrophys. J. Lett.\/}, {\bf 635}, L101--L104.
  {\small[\href{http://dx.doi.org/10.1086/499125}{DOI}]},
  {\small[\href{http://adsabs.harvard.edu/abs/2005ApJ...635L.101W}{ADS}]}

\bibitem[West {\it et~al.\/}(2008)]{2008SoPh..252...89W}
West, M.J., Bradshaw, S.J. and Cargill, P.J., 2008, ``On the Lifetime of Hot
  Coronal Plasmas Arising from Nanoflares'', {\it Solar Phys.\/}, {\bf 252},
  89--100. {\small[\href{http://dx.doi.org/10.1007/s11207-008-9243-3}{DOI}]},
  {\small[\href{http://adsabs.harvard.edu/abs/2008SoPh..252...89W}{ADS}]}

\bibitem[White {\it et~al.\/}(1991)]{1991ApJ...366L..43W}
White, S.M., Kundu, M.R. and Gopalswamy, N., 1991, ``Strong magnetic fields and
  inhomogeneity in the solar corona'', {\it Astrophys. J. Lett.\/}, {\bf 366},
  L43--L46. {\small[\href{http://dx.doi.org/10.1086/185905}{DOI}]},
  {\small[\href{http://adsabs.harvard.edu/abs/1991ApJ...366L..43W}{ADS}]}

\bibitem[Wilhelm {\it et~al.\/}(1995)]{1995SoPh..162..189W}
Wilhelm, K., Curdt, W., Marsch, E., Sch{\"u}hle, U., Lemaire, P., Gabriel, A.,
  Vial, J.-C., Grewing, M., Huber, M.C.E., Jordan, S.D., Poland, A.I., Thomas,
  R.J., K{\"u}hne, M., Timothy, J.G., Hassler, D.M. and Siegmund, O.H.W., 1995,
  ``SUMER - Solar Ultraviolet Measurements of Emitted Radiation'', {\it Solar
  Phys.\/}, {\bf 162}, 189--231.
  {\small[\href{http://dx.doi.org/10.1007/BF00733430}{DOI}]},
  {\small[\href{http://adsabs.harvard.edu/abs/1995SoPh..162..189W}{ADS}]}

\bibitem[Winebarger and Warren(2004)]{2004ApJ...610L.129W}
Winebarger, A.R. and Warren, H.P., 2004, ``Can TRACE Extreme-Ultraviolet
  Observations of Cooling Coronal Loops Be Used to Determine the Heating
  Parameters?'', {\it Astrophys. J. Lett.\/}, {\bf 610}, L129--L132.
  {\small[\href{http://dx.doi.org/10.1086/423304}{DOI}]},
  {\small[\href{http://adsabs.harvard.edu/abs/2004ApJ...610L.129W}{ADS}]}

\bibitem[Winebarger and Warren(2005)]{2005ApJ...626..543W}
Winebarger, A.R. and Warren, H.P., 2005, ``Cooling Active Region Loops Observed
  with SXT and TRACE'', {\it Astrophys. J.\/}, {\bf 626}, 543--550.
  {\small[\href{http://dx.doi.org/10.1086/429817}{DOI}]},
  {\small[\href{http://adsabs.harvard.edu/abs/2005ApJ...626..543W}{ADS}]},
  {\small[\href{http://arxiv.org/abs/arXiv:astro-ph/0502270}{{arXiv:astro-ph/0%
502270}}]}

\bibitem[Winebarger {\it et~al.\/}(1999)]{1999ApJ...526..471W}
Winebarger, A.R., Emslie, A.G., Mariska, J.T. and Warren, H.P., 1999,
  ``Analyzing the Energetics of Explosive Events Observed by SUMER on SOHO'',
  {\it Astrophys. J.\/}, {\bf 526}, 471--477.
  {\small[\href{http://dx.doi.org/10.1086/307966}{DOI}]},
  {\small[\href{http://adsabs.harvard.edu/abs/1999ApJ...526..471W}{ADS}]}

\bibitem[Winebarger {\it et~al.\/}(2001)]{2001ApJ...553L..81W}
Winebarger, A.R., DeLuca, E.E. and Golub, L., 2001, ``Apparent Flows above an
  Active Region Observed with the Transition Region and Coronal Explorer'',
  {\it Astrophys. J. Lett.\/}, {\bf 553}, L81--L84.
  {\small[\href{http://dx.doi.org/10.1086/320496}{DOI}]},
  {\small[\href{http://adsabs.harvard.edu/abs/2001ApJ...553L..81W}{ADS}]}

\bibitem[Winebarger {\it et~al.\/}(2002{\natexlab{a}})]{2002ApJ...565.1298W}
Winebarger, A.R., Emslie, A.G., Mariska, J.T. and Warren, H.P.,
  2002{\natexlab{a}}, ``Energetics of Explosive Events Observed with SUMER'',
  {\it Astrophys. J.\/}, {\bf 565}, 1298--1311.
  {\small[\href{http://dx.doi.org/10.1086/324714}{DOI}]},
  {\small[\href{http://adsabs.harvard.edu/abs/2002ApJ...565.1298W}{ADS}]}

\bibitem[Winebarger {\it et~al.\/}(2002{\natexlab{b}})]{2002ApJ...570L.105W}
Winebarger, A.R., Updike, A.C. and Reeves, K.K., 2002{\natexlab{b}},
  ``Correlating Transition Region Explosive Events with Extreme-Ultraviolet
  Brightenings'', {\it Astrophys. J. Lett.\/}, {\bf 570}, L105--L108.
  {\small[\href{http://dx.doi.org/10.1086/341121}{DOI}]},
  {\small[\href{http://adsabs.harvard.edu/abs/2002ApJ...570L.105W}{ADS}]}

\bibitem[Winebarger {\it et~al.\/}(2002{\natexlab{c}})]{2002ApJ...567L..89W}
Winebarger, A.R., Warren, H., van Ballegooijen, A., DeLuca, E.E. and Golub, L.,
  2002{\natexlab{c}}, ``Steady Flows Detected in Extreme-Ultraviolet Loops'',
  {\it Astrophys. J. Lett.\/}, {\bf 567}, L89--L92.
  {\small[\href{http://dx.doi.org/10.1086/339796}{DOI}]},
  {\small[\href{http://adsabs.harvard.edu/abs/2002ApJ...567L..89W}{ADS}]}

\bibitem[Winebarger {\it et~al.\/}(2003{\natexlab{a}})]{2003ApJ...587..439W}
Winebarger, A.R., Warren, H.P. and Mariska, J.T., 2003{\natexlab{a}},
  ``Transition Region and Coronal Explorer and Soft X-Ray Telescope Active
  Region Loop Observations: Comparisons with Static Solutions of the
  Hydrodynamic Equations'', {\it Astrophys. J.\/}, {\bf 587}, 439--449.
  {\small[\href{http://dx.doi.org/10.1086/368017}{DOI}]},
  {\small[\href{http://adsabs.harvard.edu/abs/2003ApJ...587..439W}{ADS}]}

\bibitem[Winebarger {\it et~al.\/}(2003{\natexlab{b}})]{2003ApJ...593.1164W}
Winebarger, A.R., Warren, H.P. and Seaton, D.B., 2003{\natexlab{b}}, ``Evolving
  Active Region Loops Observed with the Transition Region and Coronal Explorer.
  I. Observations'', {\it Astrophys. J.\/}, {\bf 593}, 1164--1173.
  {\small[\href{http://dx.doi.org/10.1086/376679}{DOI}]},
  {\small[\href{http://adsabs.harvard.edu/abs/2003ApJ...593.1164W}{ADS}]}

\bibitem[Winebarger {\it et~al.\/}(2008)]{2008ApJ...676..672W}
Winebarger, A.R., Warren, H.P. and Falconer, D.A., 2008, ``Modeling X-Ray Loops
  and EUV `Moss' in an Active Region Core'', {\it Astrophys. J.\/}, {\bf 676},
  672--679. {\small[\href{http://dx.doi.org/10.1086/527291}{DOI}]},
  {\small[\href{http://adsabs.harvard.edu/abs/2008ApJ...676..672W}{ADS}]},
  {\small[\href{http://arxiv.org/abs/0712.0756}{{arXiv:0712.0756}}]}

\bibitem[Wragg and Priest(1981)]{1981SoPh...70..293W}
Wragg, M.A. and Priest, E.R., 1981, ``The temperature-density structure of
  coronal loops in hydrostatic equilibrium'', {\it Solar Phys.\/}, {\bf 70},
  293--313. {\small[\href{http://dx.doi.org/10.1007/BF00151335}{DOI}]},
  {\small[\href{http://adsabs.harvard.edu/abs/1981SoPh...70..293W}{ADS}]}

\bibitem[Yokoyama and Shibata(2001)]{2001ApJ...549.1160Y}
Yokoyama, T. and Shibata, K., 2001, ``Magnetohydrodynamic Simulation of a Solar
  Flare with Chromospheric Evaporation Effect Based on the Magnetic
  Reconnection Model'', {\it Astrophys. J.\/}, {\bf 549}, 1160--1174.
  {\small[\href{http://dx.doi.org/10.1086/319440}{DOI}]},
  {\small[\href{http://adsabs.harvard.edu/abs/2001ApJ...549.1160Y}{ADS}]}

\bibitem[Yoshida and Tsuneta(1996)]{1996ApJ...459..342Y}
Yoshida, T. and Tsuneta, S., 1996, ``Temperature Structure of Solar Active
  Regions'', {\it Astrophys. J.\/}, {\bf 459}, 342--+.
  {\small[\href{http://dx.doi.org/10.1086/176897}{DOI}]},
  {\small[\href{http://adsabs.harvard.edu/abs/1996ApJ...459..342Y}{ADS}]}

\bibitem[Yoshida {\it et~al.\/}(1995)]{1995PASJ...47L..15Y}
Yoshida, T., Tsuneta, S., Golub, L., Strong, K. and Ogawara, Y., 1995,
  ``Temperature Structure of the Solar Corona: Comparison of the NIXT and
  YOHKOH X-Ray Images'', {\it Publ. Astron. Soc. Japan\/}, {\bf 47}, L15--L19.
  {\small[\href{http://adsabs.harvard.edu/abs/1995PASJ...47L..15Y}{ADS}]}

\bibitem[Young {\it et~al.\/}(2009)]{2009AA...495..587Y}
Young, P.R., Watanabe, T., Hara, H. and Mariska, J.T., 2009, ``High-precision
  density measurements in the solar corona. I. Analysis methods and results for
  Fe XII and Fe XIII'', {\it Astron. Astrophys.\/}, {\bf 495}, 587--606.
  {\small[\href{http://dx.doi.org/10.1051/0004-6361:200810143}{DOI}]},
  {\small[\href{http://adsabs.harvard.edu/abs/2009A&A...495..587Y}{ADS}]},
  {\small[\href{http://arxiv.org/abs/0805.0958}{{arXiv:0805.0958}}]}

\bibitem[Zhitnik {\it et~al.\/}(2003)]{2003MNRAS.338...67Z}
Zhitnik, I.A., Bugaenko, O.I., {Ignat'ev}, A.P., Krutov, V.V., Kuzin, S.V.,
  Mitrofanov, A.V., Oparin, S.N., Pertsov, A.A., Slemzin, V.A., Stepanov, A.I.
  and Urnov, A.M., 2003, ``Dynamic 10 MK plasma structures observed in
  monochromatic full-Sun images by the SPIRIT spectroheliograph on the
  CORONAS-F mission'', {\it Mon. Not. R. Astron. Soc.\/}, {\bf 338}, 67--71.
  {\small[\href{http://dx.doi.org/10.1046/j.1365-8711.2003.06014.x}{DOI}]},
  {\small[\href{http://adsabs.harvard.edu/abs/2003MNRAS.338...67Z}{ADS}]}

\bibitem[Zirker(1993)]{1993SoPh..148...43Z}
Zirker, J.B., 1993, ``Coronal heating'', {\it Solar Phys.\/}, {\bf 148},
  43--60. {\small[\href{http://dx.doi.org/10.1007/BF00675534}{DOI}]},
  {\small[\href{http://adsabs.harvard.edu/abs/1993SoPh..148...43Z}{ADS}]}

\end{thebibliography}

\end{document}